\documentclass[final,onefignum,onetabnum]{siamart171218}

\usepackage{lipsum}
\usepackage{amsfonts}
\usepackage{graphicx}
\usepackage{epstopdf}
\usepackage{algorithmic}
\ifpdf
  \DeclareGraphicsExtensions{.eps,.pdf,.png,.jpg}
\else
  \DeclareGraphicsExtensions{.eps}
\fi

\usepackage{nicefrac}
\usepackage{bm}
\usepackage{tikz}
\usetikzlibrary{decorations.pathmorphing}
\usetikzlibrary{matrix}
\newcommand{\veq}{\mathrel{\rotatebox{90}{$=$}}}
\usepackage{mathtools}
\usepackage{multirow, makecell}
\usepackage{subcaption}
\usepackage{accents}

\usepackage{enumitem}
\setlist[enumerate]{leftmargin=.5in}
\setlist[itemize]{leftmargin=.5in}

\newsiamremark{remark}{Remark}
\newsiamremark{hypothesis}{Hypothesis}
\crefname{hypothesis}{Hypothesis}{Hypotheses}
\newsiamthm{claim}{Claim}

\headers{Dynamic Ecological System Analysis}{Huseyin Coskun}

\title{Dynamic Ecological System Analysis} 
\subtitle{A Holistic Analysis of Compartmental Systems} 

\author{Huseyin Coskun\thanks{Department of Mathematics, University of Georgia, Athens, GA 30602 (\email{hcoskun@uga.edu}).} }

\usepackage{amsopn}
\DeclareMathOperator{\diag}{diag}
\DeclareMathOperator{\sgn}{sgn}

\ifpdf
\hypersetup{
  pdftitle={Dynamic Ecological System Analysis},
  pdfauthor={Huseyin Coskun}
}
\fi

\begin{document}

\maketitle

\begin{abstract}
This article develops a new mathematical method for holistic analysis of nonlinear dynamic compartmental systems through the system decomposition theory. The method is based on the novel dynamic system and subsystem partitioning methodologies through which compartmental systems are decomposed to the utmost level. The dynamic system and subsystem partitioning enable tracking the evolution of the initial stocks, environmental inputs, and intercompartmental system flows, as well as the associated storages derived from these stocks, inputs, and flows individually and separately within the system. Moreover, the transient and the dynamic direct, indirect, acyclic, cycling, and transfer (\texttt{diact}) flows and associated storages transmitted along a given flow path or from one compartment, directly or indirectly, to any other are analytically characterized, systematically classified, and mathematically formulated. Further, the article develops a dynamic technique based on the \texttt{diact} transactions for the quantitative classification of interspecific interactions and the determination of their strength within food webs. Major concepts and quantities of the current static network analyses are also extended to nonlinear dynamic settings and integrated with the proposed dynamic measures and indices within the proposed unifying mathematical framework. Therefore, the proposed methodology enables a holistic view and analysis of ecological systems. We consider that this methodology brings a novel complex system theory to the service of urgent and challenging environmental problems of the day and has the potential to lead the way to a more formalistic ecological science.
\end{abstract}

\begin{keywords}
system decomposition theory, complex systems theory, dynamic ecological network analysis, nonlinear dynamic compartmental systems, dynamic system and subsystem partitioning, transient flows and storages, \texttt{diact} flows and storages, food webs, interspecific interactions, dynamic input-output economics, socio-economic systems, dynamic input-output analysis, epidemiology, infectious diseases, toxicology, pharmacokinetics, neural networks, chemical and biological systems, control theory, information theory, information diffusion, social networks, computer networks, malware propagation, graph theory, traffic flow
\end{keywords}


\section{Introduction}
\label{sec:intro}

Compartmental systems are mathematical abstractions of networks that model behaviors of continuous physical systems composed of discrete living and nonliving homogeneous components. Based on conservation principles, system compartments are interconnected through the flow of energy, matter, or currency between them and their environment. Therefore, formulating flows and associated storages accurately and explicitly is critically important in quantifying compartmental system function. Various mathematical aspects of compartmental systems are studied in the literature \cite{Jacquez1993,Anderson2013}. While many fields utilize compartmental modeling, this approach proves particularly well-suited for analysis of ecological systems to address environmental phenomena.

Due to the current technological advancements as well as scientific understandings of population and industrial growth and resource demands, environmental issues have assumed center stage in human communities. On the other hand, in spite of this increased attention to the environment, traditional ecology has an applied nature and is still in the empirical stage of development. In the mainstream framework of traditional ecology, a first principles-based formal theory has yet to emerge. This disconnect narrows the scope of applicability of the field and reduces its ability to deal with complex organism-environment relationships. To that extent, ecology and environmental science are limited in their capacity to realistically model and analyze complex systems. Mathematical theories and modeling have significant potential to lead the way to a more formalistic and theoretical ecoscience devoted to the discovery of basic scientific laws. More exact, precise, and incisive environmental applications can then be materialized based on this understanding.

Sound rationales have been offered in the literature for ecological network analysis, but these are for special cases, such as linear and static models. One such static approach called the {\em environ theory} has been developed by \cite{Patten1978,Matis1981} based on economic {\em input-output analysis} of \cite{Leontief1936, Leontief1966} introduced into ecology by \cite{Hannon1973}. Ecological networks and complexity in living systems are analyzed also in the context of {\em information theory}, {\em thermodynamics} \cite{Ulanowicz1972,Hirata1985,Ulanowicz2004,Ulanowicz2013}, and {\em hierarchy theory} \cite{Allen2014}, yet only for static systems. Several software have been developped to computerize these static methods \cite{Ulanowicz1991,Christensen1992, Fath2006, Kazanci2009a, Schramski2011, Borrett2014c}.

Although the steady-state analysis is well-established, dynamic analysis of nonlinear compartmental systems has remained a long-standing, open problem. For example, Finn's cycling index\textemdash a celebrated ecosystem measure that quantifies cycling system flows defined in static ecological network analyses over four decades ago\textemdash has still not been made applicable to ecosystem models that change over time \cite{Finn1976}. The indirect effects in ecosystems have also long been a well-established empirical fact \cite{Paine1966,Strauss1991,Wootton1993,Menge1995,Menge1997,Wootton2002}. Theoretical explorations of the concept began as early as the 1970s, and it has been a topic of scholarly conversation for the past five decades \cite{Holt1977,Patten1992,Fath1999,Ma2013,Coskun2019ITR}. Despite the urgent need, the indirect flow and storage transfers have never been formulated before. There are earlier approaches in the literature for the analysis of dynamic ecosystems, but these are either essentially closed-form abstract formulations \cite{Hallam1985}, or designed for special cases, such as linear systems with time-dependent inputs \cite{Hippe1983}. In addition, there are also agent-based techniques for dynamic compartmental system analysis \cite{Shevtsov2009,Kazanci2009,Kazanci2012}. These are, however, computational methods that rely on network particle tracking simulations.

In ecosystem ecology, food webs provide a framework to link community structure with flows of energy and material through trophic interactions and, therefore, relate biodiversity with ecosystem function. Temporal variation in web architecture and nonlinearity are discussed in the literature \cite{Dell2006,Zanden2016}. It is suggested that the dynamic nature of food webs is affecting ecosystem attributes. Nonlinearity and dynamic behavior, such as extinction in food webs, however, has yet to be addressed methodologically. Not only food webs, but today's major environmental and ecological phenomena and problems--human impact, climate change, biodiversity loss, etc.-- all involve change, which demonstrates that the need for dynamic methods for nonlinear system analysis is not only appropriate, but also urgent \cite{Chen2001,Hastings2004}. 

This is the first manuscript in the literature that potentially addresses the disconnect between the current static and computational methods and applied ecological needs. We consider that the methodology proposed herein, in effect, brings a novel complex system theory to the service of pressing and challenging environmental problems of the day. Due to its theoretical and mathematical nature, it has the potential to lead the way to a more formalistic ecological science. The proposed methodology is a comprehensive approach in the sense that the major concepts and quantities in the current static ecological network analyses are extended from static to nonlinear dynamic settings, as well as integrated effectively with the proposed dynamic measures in this unifying mathematical framework. This novel and unifying approach leads to a {\em holistic analysis of ecosystems}.

Aligned with the mathematical theory, called the \textit{system decomposition theory}, introduced recently by~\cite{Coskun2017NDP}, the proposed comprehensive method is composed of the novel {\em dynamic system} and {\em subsystem partitioning methodologies}. The system partitioning methodology yields the {\em subthroughflow} and {\em substorage vectors} and {\em matrices} that represent the flows and storages generated by the initial stocks and individual environmental inputs in each compartment separately. Therefore, the system partitioning enables dynamically decomposing composite compartmental flows and storages into subcompartmental segments based on their constituent sources from the initial stocks and environmental inputs. In other words, this methodology enables dynamically tracking the evolution of the initial stocks and environmental inputs, as well as the associated storages derived from these stocks and inputs individually and separately within the system. 

The {\em transient flows} transmitted along a given flow path and the associated {\em storages} generated by these flows in each compartment on the path are then formulated through the subsystem partitioning methodology. Therefore, this methodology allows for the dynamic decomposition of arbitrary composite intercompartmental flows and associated storages into the transient subflow and substorage segments along a given set of subflow paths. Consequently, the subsystem partitioning enables dynamically tracking the fate of arbitrary intercompartmental flows and associated storages within the subsystems. Moreover, the spread of an arbitrary flow or storage segment from one compartment to the entire system can be determined and monitored. For the quantification of intercompartmental flow and storage transfer dynamics, the {\em direct}, {\em indirect}, {\em acyclic}, {\em cycling}, and {\em transfer} ($\texttt{diact}$) flows and associated storages transmitted from one compartment, directly or indirectly, to any other are also analytically characterized, systematically classified, and mathematically formulated.

In a nutshell, the system and subsystem partitioning methodologies dynamically determine the distribution of the initial stocks, environmental inputs, and arbitrary intercompartmental flows, as well as the organization of the associated storages derived from these stocks, inputs, and flows individually and separately within the system. In other words, the proposed method as a whole enables tracking the evolution of the initial stocks, environment inputs, and arbitrary intercmopartmental system flows, as well as associated storages individually and separately. The dynamic quantities such as the subthroughflows, substorages, and transient and \texttt{diact} flows and storages are systematically introduced through the proposed method for the first time in the literature. Equipped with these measures, the proposed methodology serves as a quantitative platform for testing empirical hypotheses, ecological inferences, and, potentially, theoretical developments. The method also constructs a foundation for the development of new mathematical system analysis tools as quantitative ecological indicators. Multiple such dynamic \texttt{diact} measures and indices of matrix, vector, and scalar types which may prove useful for environmental assessment and management were systematically introduced by \cite{Coskun2017DESM}.

The temporal variations of trophic interactions in food webs is an important topic in ecology as outlined above \cite{Dell2006,Zanden2016}. The conditions or states of communities in food webs, such as extinction, can be dynamically regulated by the temporal variations and seasonal shifts. The present manuscript develops also a novel mathematical technique based on the \texttt{diact} transactions for the dynamic classification of interspecific interactions, and notably, for the determination of their strength within food webs. This technique effectively addresses the nonlinearity in and dynamic architecture of the food chains and webs.

The proposed methodology is applicable to any conservative compartmental system of naturogenic or anthropogenic nature. The method can be used, for example, to analyze models designed for material flows in industry \cite{Bailey2004}. It can also be used to analyze mass or energy transfers between species of different trophic levels in a complex network or along a given food chain of a food web in nonlinear dynamic settings \cite{Hastings1996,Belgrano2005,Franks2002}. Although the motivating applications are ecological and environmental for this paper, the applicability of the proposed method extends to other realms, such as economics, pharmacokinetics, chemical reaction kinetics, epidemiology, biomedical systems, neural networks, social networks, and information science\textemdash in fact, wherever dynamical compartmental models of conserved quantities can be constructed. An input-output analysis in economics was developed several decades ago, but only for static systems \cite{Leontief1936,Leontief1966}. The proposed methodology in the context of economics, in particular, can be considered as the mathematical foundation of the {\em dynamic input-output economics}.

The proposed method is applied to two models in Section~\ref{sec:results} to illustrate its efficiency and wide applicability. In the first case study, a linear ecosystem model introduced by \cite{Hippe1983} is analyzed. The second case study concerns nutrient transfer within a nutrient-producer-consumer ecosystem \cite{Hallam1985}. Analytical and numerical solutions for the substorages, subthroughflows, transient and \texttt{diact} transactions, and residence times are presented for both models. The interspecific interactions in the nonlinear model and their strength are also analyzed through the proposed mathematical classification technique.

This paper is organized as follows: the mathematical method is introduced in Section~\ref{sec:sc}, the transient and \texttt{diact} flows and storages are formulated in Section~\ref{sec:dsp}, system analysis and measures are discussed in Section~\ref{sec:dsami}, results and case studies are presented in Section~\ref{sec:results}, and discussion and conclusions follow in Section~\ref{sec:disc} and~\ref{sec:conc}.

\section{Methods}
\label{sec:method}

A new mathematical theory for the dynamic decomposition of nonlinear compartmental systems has recently been introduced by \cite{Coskun2017NDP}. In line with this theory, a mathematical method for the dynamic analysis of nonlinear ecological systems is developed in the present paper. 

The proposed theory is based on the novel {\em system} and {\em subsystem partitioning methodologies}. The {\em system} and {\em subsystem} partitioning determine the distribution of the initial stocks, environmental inputs, and intercompartmental flows, as well as the organization of the associated storages derived from these stocks, inputs, and flows individually and separately within compartmental systems. The proposed method, therefore, as a whole, yields the decomposition of all system flows and storages to the utmost level. The method together with the corresponding concepts and quantities will be introduced in this section.

The terminology and notations used in this paper are adopted from \cite{Coskun2017NDP} as follows:
\begin{center} 
    \begin{tabular}[h]{ l p{.7\textwidth} }
\hfill \\    
    $n$ & number of compartments \\
    $t$ & time $[\mathrm{t}]$ \\
    $x_i(t)$ & total material (mass) $[\mathrm{m}]$  (or energy, currency) in compartment $i$, $i = 1,\ldots,n$, at time $t$ \\
    $f_{ij}(t,x)$ & nonnegative flow from compartment $j$ to $i$, at time $t$ $[\mathrm{m}/\mathrm{t}]$ \\    
    $y_{i}(t,x) = f_{0{i}}(t,x)$ & environmental ($j=0$) output from compartment $i$ at time $t$ \\ 
     $z_{i}(t,x) = f_{{i}0}(t,x)$ & environmental input into compartment $i$ at time $t$ \\
     \hfill
    \end{tabular}
\end{center} 

The governing equations for the compartmental dynamics are
\begin{equation}
\label{eq:model}
\dot{x}_i(t) = \check{\tau}_i(t,x) - \hat{\tau}_i(t,x)
\end{equation}
for $i=1,\ldots,n$. The state vector $x(t)=[x_1(t),\ldots,x_n(t)]^T$ is a differentiable function of compartmental storages with the initial conditions of $x(t_0) = x_0 = [x_{1,0},\ldots,x_{n,0}]^T$ where the superscript $T$ represents the matrix transpose. The total inflow, $\check{\tau}_i(t,x)$, and outflow, $\hat{\tau}_i(t,x)$, are called the {\em inward} and {\em outward throughflows} at compartment $i$, respectively, and formulated as
\begin{equation}
\label{eq:in_out_flows}
\check{\tau}_i(t,x) = \sum_{\substack{j=0}}^n f_{ij}(t,x) \quad \mbox{and} \quad  \hat{\tau}_i(t,x) = \sum_{\substack{j=0}}^n f_{ji}(t,x)
\end{equation} 
for $i=1,\ldots,n$. The nonlinear differentiable function $f_{ij}(t,x) \geq 0$ represents nonnegative flow rate from compartment $j$ to $i$ at time $t$. In general, it is assumed that $f_{ii}(t,x) = 0$, but the following analysis is also valid for nonnegative flow from a compartment into itself. Index $j=0$ stands for the environment. We further assume that $f_{ij}(t,x)$ has the following special form:
\begin{equation}
\label{eq:sf}
f_{ij}(t,x) = q^x_{ij}(t,x) \, {x}_j (t)
\end{equation}
where $q^x_{ij}(t,x)$ is a nonlinear function of ${x}$ and $t$, and has the same properties as $f_{ij}(t,x)$. We will call $q^x_{ij}(t,x) = { f_{ij}(t,x) }/{x_j(t)}$ the {\em flow intensity} directed from compartment $j$ to $i$ per unit storage or the {\em flow distribution factor} for system storages in the context of the proposed methodology \cite{Coskun2017SCSA}.

Combining Eqs.~\ref{eq:model} and~\ref{eq:in_out_flows} and separating environmental inputs and outputs, the system of governing equations takes the following standard form:
\begin{equation}
\label{eq:model_c}
\dot{x}_i(t) = \left( z_i(t,x) + \sum_{\substack{j=1}}^n f_{ij}(t,x)  \right) - \left( y_i(t,x) + \sum_{\substack{j=1 }}^n f_{ji}(t,x) \right)
\end{equation} 
with the initial conditions $x_i(t_0) = x_{i,0}$, for $i=1,\ldots,n$. There are $n$ equations; one for each compartment. The condition, Eq.~\ref{eq:sf}, guaranties non-negativity of the compartmental storages, that is $x_i(t) \geq 0$ for all $i$. If an environmental input or initial condition is positive, that is, $z_i(t,x)>0$ or $x_{i,0} > 0$, the corresponding storage value is always strictly positive, $x_i(t) >0$.

The proposed methodology is designed for {\em conservative compartmental systems}. A dynamical system is called {\em compartmental} if it can be expressed in the form of Eq.~\ref{eq:model_c}. The compartmental systems will be called {\em conservative} if all internal flow rates add up to zero when the system is closed, that is, when there is neither environmental input nor output:
\begin{equation}
\label{eq:consv}
\sum_{i=1}^{n} \dot{x}_ i(t) = 0 \quad \mbox{when} \quad {z}(t,x) = {y}(t,x) = \bm{0} \quad \mbox{for all } t
\end{equation}  
where $\mathbf{0}$ is used for both the $n \times n$ zero matrix and zero vector of size $n$ \cite{Coskun2017NDP}.
 
For notational convenience, we define a {\em direct flow matrix} function $F$ of size $n \times n$ as
\begin{equation}
\begin{aligned}
\label{eq:Fmatrix_1}
F(t,x) = \left(f_{ij}(t,x) \right)
\end{aligned}
\nonumber
\end{equation}
and the {\em inward} and {\em outward throughflow vector} functions as
\begin{equation}
\begin{aligned}
\label{eq:Tvector_1}
\check{\tau}(t,x) &= \left[ \check{\tau}_1(t,x), \ldots,\check{\tau}_n(t,x) \right]^T = z(t,x) + F(t,x) \, \mathbf{1}  \quad  \mbox{and} \\ 
\hat{\tau}(t,x) &= \left[ \hat{\tau}_1(t,x),\ldots,\hat{\tau}_n(t,x) \right]^T = y(t,x) + F^T(t,x) \, \mathbf{1} , 
\end{aligned}
\end{equation}
respectively, where 
\[ z(t,x)=[z_1(t,x),\ldots,z_n(t,x)]^T \quad \mbox{and} \quad y(t,x)=[y_1(t,x),\ldots,y_n(t,x)]^T \]
are the {\em input} and {\em output vector} functions, and ${\mathbf 1}$ denotes the column vector of size $n$ whose entries are all one.

\subsection{System partitioning methodology}
\label{sec:sc}

In this section, we introduce the {\em dynamic system partitioning methodology} for analytically partitioning the governing system into mutually exclusive and exhaustive {\em subsystems}, as a simplified version of the \emph{system decomposition methodology} recently proposed by \cite{Coskun2017NDP}. By {\em mutual exclusiveness}, we mean that transactions are possible only among corresponding {\em subcompartments} belong to the same subsystem. By \emph{exhaustiveness}, we mean that all generated subsystems sum to the entire system so partitioned. The system partitioning enables dynamically partitioning composite compartmental flows and storages into subcompartmental segments based on their constituent sources from the initial stocks and environmental inputs of the same conserved quantity. The system partitioning methodology, consequently, yields the subthroughflow and substorage matrices representing the distribution of the initial stocks and environmental inputs, as well as the organization of the associated storages derived from these stocks and inputs individually and separately within the system. In other words, this methodology enables tracking the evolution of the initial stocks and environmental inputs, as well as associated storages individually and separately within the system.

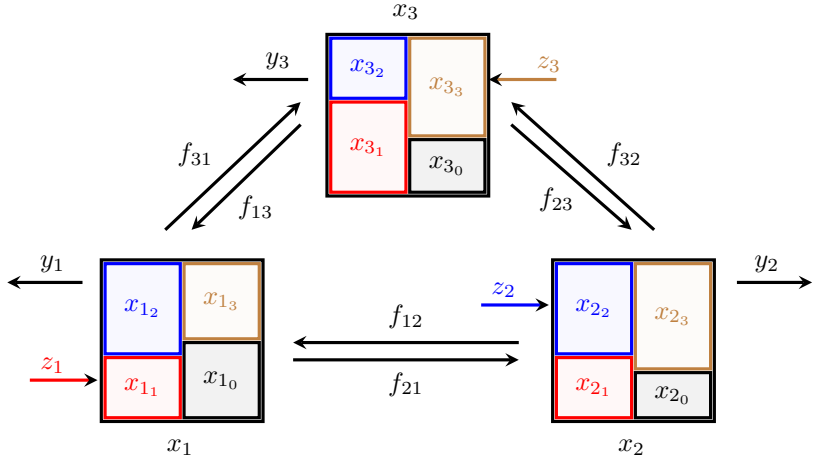
\begin{figure}[t]
\begin{center}
\begin{tikzpicture}
\centering
   \draw[very thick, draw=black] (-.05,-.05) rectangle node(R1) [pos=.5] { } (2.1,2.1) ;
   \draw[very thick, fill=red!3, draw=red, text=red] (0,0) rectangle node(R1) [pos=.5] {$x_{1_1}$} (1,0.8) ;
   \draw[very thick, fill=gray!10, draw=black, text=black] (1.05,0) rectangle node(R2) [pos=.5] {${x}_{1_0}$} (2.05,1) ;
   \draw[very thick, fill=brown!3, draw=brown, text=brown] (1.05,1.05) rectangle node(R3) [pos=.5] {$x_{1_3}$} (2.05,2.05) ;
   \draw[very thick, fill=blue!3, draw=blue, text=blue] (0,0.85) rectangle node(R4) [pos=.5] {$x_{1_2}$} (1,2.05) ;      
    \draw[very thick,-stealth,draw=red]  (-1,.5) -- (-.1,.5) ;     
    \node (z) [text=red] at (-.7,0.7) {$z_1$};                   
    \draw[very thick,-stealth,draw=black]  (-.3,1.8) -- (-1.3,1.8) ;     
    \node (z) [text=black] at (-.7,2.05) {$y_1$};                                                 
    \node (x) at (1,-.4) {${x}_{1}$};      
   \draw[very thick, draw=black] (5.95,-.05) rectangle node(R1) [pos=.5] { } (8.1,2.1) ;
   \draw[very thick, fill=red!3, draw=red, text=red] (6,0) rectangle node(R1) [pos=.5] {$x_{2_1}$} (7,0.8) ;
   \draw[very thick, fill=gray!10, draw=black, text=black] (7.05,0) rectangle node(R2) [pos=.5] {${x}_{2_0}$} (8.05,.6) ;
   \draw[very thick, fill=brown!3, draw=brown, text=brown] (7.05,2.05) rectangle node(R3) [pos=.5] {$x_{2_3}$} (8.05,0.65) ;
   \draw[very thick, fill=blue!3, draw=blue, text=blue] (6,0.85) rectangle node(R4) [pos=.5] {$x_{2_2}$} (7,2.05) ;      
    \draw[very thick,-stealth,draw=blue]  (5,1.5) -- (5.9,1.5) ;     
    \node (z) [text=blue] at (5.3,1.7) {$z_2$};       
    \draw[very thick,-stealth,draw=black]  (8.4,1.8) -- (9.4,1.8) ;     
    \node (z) [text=black] at (8.8,2.05) {$y_2$};                                                 
    \node (x) at (7,-.4) {${x}_{2}$};      
   \draw[very thick, draw=black] (2.95,2.95) rectangle node(R1) [pos=.5] { } (5.1,5.1) ;
   \draw[very thick, fill=red!3, draw=red, text=red] (3,3) rectangle node(R1) [pos=.5] {$x_{3_1}$} (4,4.2) ;
   \draw[very thick, fill=gray!10, draw=black, text=black] (4.05,3) rectangle node(R2) [pos=.5] {${x}_{3_0}$} (5.05,3.7) ;
   \draw[very thick, fill=brown!3, draw=brown, text=brown] (4.05,3.75) rectangle node(R3) [pos=.5] {$x_{3_3}$} (5.05,5.05) ;
   \draw[very thick, fill=blue!3, draw=blue, text=blue] (3,4.25) rectangle node(R4) [pos=.5] {$x_{3_2}$} (4,5.05) ;      
    \draw[very thick,-stealth,draw=brown]  (6,4.5) -- (5.1,4.5) ;     
    \node (z) [text=brown] at (5.9,4.7) {$z_3$};                        
    \draw[very thick,-stealth,draw=black]  (2.7,4.5) -- (1.7,4.5) ;     
    \node (z) [text=black] at (2.3,4.75) {$y_3$};                            
    \node (x) at (4,5.4) {${x}_{3}$};        
    \draw[very thick,-stealth]  (0.8,2.5) -- (2.6,4.2) ;  
    \draw[very thick,-stealth]  (2.6,3.9)  -- (1.15,2.5) ;
    \node (x) at (2,2.8) {${f}_{13}$};  
    \node (x) at (1.2,3.5) {${f}_{31}$};                                           
    \draw[very thick,-stealth]  (7.3,2.5) -- (5.4,4.2) ;  
    \draw[very thick,-stealth]  (5.4,3.9) -- (7,2.5) ;  
    \node (x) at (6,2.9) {${f}_{23}$};  
    \node (x) at (6.9,3.5) {${f}_{32}$};                                           
    \draw[very thick,-stealth]  (2.5,.77) -- (5.5,0.77) ;  
    \draw[very thick,-stealth]  (5.5,1) -- (2.5,1) ;  
    \node (x) at (4,1.35) {${f}_{12}$};  
    \node (x) at (4,.45) {${f}_{21}$};                                           
\end{tikzpicture}
\end{center}
\caption{Schematic representation of the dynamic subcompartmentalization in a three-compartment model system. Each subsystem is colored differently; the second subsystem ($k=2$) is blue, for example. Only the subcompartments in the same subsystem ($x_{1_2}(t)$, $x_{2_2}(t)$, and $x_{3_2}(t)$ in the second subsystem, for example) interact with each other. Subsystem $k$ receives environmental input only at subcompartment ${k_k}$. The initial subsystem receives no environmental input. The dynamic flow partitioning is not represented in this figure. Compare this figure with Fig.~\ref{fig:sc}, in which the subcompartmentalization and corresponding flow partitioning are illustrated for $x_1(t)$ only.}
\label{fig:sd}
\end{figure}

\begin{figure}[t]
\begin{center}
\begin{tikzpicture}[scale=.7]
\centering
    \draw[fill=gray!2]    (4.7,0) -- ++(2.3,0) -- ++(1,2) -- ++(-1,2) -- ++(-2.3,0);           
    \draw[draw=none,fill=red!7]  (4.7,1.2) -- ++ (2.8,0) -- ++ (.4,.8) -- ++ (-3.2,0);             
    \draw[draw=none,fill=gray!7]  (4.7,.1) -- ++ (2.25,0) -- ++ (.5,1.05) -- ++ (-2.75,0);                 
    \draw[draw=none,fill=blue!7]  (4.7,2.05) -- ++ (3.2,0) -- ++ (-.5,.9) -- ++ (-2.7,0);                     
    \draw[draw=none,fill=brown!7]  (4.7,3) -- ++ (2.7,0) -- ++ (-.5,.9) -- ++ (-2.2,0);                         
   \draw[very thick, draw=black] (-.05,-.05) rectangle node(R1) [pos=.5] { } (4.1,4.1) ;
   \draw[very thick, fill=red!3, draw=red, text=red] (0,0) rectangle node(R1) [pos=.5] {\small $x_{1_1}$} (2,1.8) ;
   \draw[very thick, fill=gray!10, draw=black, text=black] (2.05,0) rectangle node(R2) [pos=.5] {\small ${x}_{1_0}$} (4.05,2.5) ;
   \draw[very thick, fill=brown!3, draw=brown, text=brown] (2.05,2.55) rectangle node(R3) [pos=.5] {\small $x_{1_3}$} (4.05,4.05) ;
   \draw[very thick, fill=blue!3, draw=blue, text=blue] (0,1.85) rectangle node(R4) [pos=.5] {\small $x_{1_2}$} (2,4.05) ;      
    \draw[very thick]  (4.7,0) -- (7,0) ;
    \draw[very thick]  (7,0) -- (8,2) ;   
    \draw[very thick]  (8,2) -- (7,4) ;        
    \draw[very thick]  (7,4) -- (4.7,4) ; 
    \draw[very thick,-stealth,draw=red]  (-1.5,.5) -- (-.1,.5) ;     
    \node (z) [text=red] at (-1,0.9) {$z_1$};                        
    \draw[very thick,-stealth,draw=black]  (5,.4) -- (6.8,.4) ;     
    \node (f0) at (6,.8) {\small ${f}_{j_0 1_0}$};   
    \draw[very thick,-stealth,draw=red]  (5,1.3) -- (7.4,1.3) ;         
    \node (f1) [text=red] at (6,1.7) {\small ${f}_{j_1 1_1}$};
    \draw[very thick,-stealth,draw=blue]  (5,2.2) -- (7.4,2.2) ;     
    \node (f2) [text=blue] at (6,2.6) {\small ${f}_{j_2 1_2}$};   
    \draw[very thick,-stealth,draw=brown]  (5,3.1) -- (6.8,3.1) ;     
    \node (f3) [text=brown] at (6,3.5) {\small ${f}_{j_3 1_3}$};                                                     
    \node (x) at (2,4.7) {${x}_{1}$};      
    \node (x) at (6,4.7) {${f}_{j1}$};               
\end{tikzpicture}
\end{center}
\caption{Schematic representation of the dynamic flow partitioning in a three-compartment model system. The figure illustrates subcompartmentalization of compartment $i=1$ and the corresponding dynamic flow partitioning from this compartment to others, $j$.}
\label{fig:sc}
\end{figure}
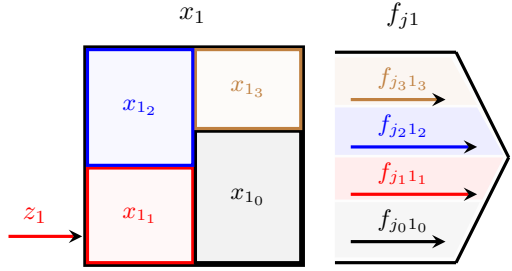

The system partitioning involves the dynamic {\em subcompartmentalization} and {\em flow partitioning} components, whose mechanisms are explained in this section (Figs.~\ref{fig:sd} and~\ref{fig:sc}). The related concepts and notations are summarized below: 
\begin{center}
    \begin{tabular}[h]{ l p{.70\textwidth} }
    $x_{i_k}(t)$ & storage in subcompartment $k$ of compartment $i$, that is, in subcompartment $i_k$, $k = 0,\ldots,n$, at time $t$, generated by environmental input $z_k(t,{x})$ during $[t_0,t]$ \\ 
    $f_{{i_k}{j_k}}(t,{\rm x})$ & nonnegative flow from subcompartment $j_k$ to $i_k$ at time $t$ \\    
    $y_{i_k}(t,{\rm x}) = f_{0{i_k}}(t,{\rm x})$ & environmental ($j=0$) output from subcompartment $i_k$ at time $t$ \\     
    $z_{i_k}(t,{\rm x}) = \delta_{ik} \, z_{i}(t,x)$ & environmental input into subcompartment $i_k$ at time $t$, where $\delta_{ik}$ is the discrete delta function \\  
    \hfill  
    \end{tabular}
\end{center}

The system is partitioned explicitly and analytically into mutually exclusive and exhaustive {\em subsystems} as follows: Each compartment is partitioned into $n+1$ subcompartments; $n$ initially empty subcompartments for $n$ environmental inputs and $1$ subcompartment for the initial stock of the compartment. The notation $i_k$ is used to represent the $k^{th}$ subcompartment of the $i^{th}$ compartment for $i=1,\ldots,n$ and $k=0,\ldots,n$. The subscript index $k=0$ represents the initial subcompartment of compartment $i$ (see Fig.~\ref{fig:sd}).

The storage in subcompartment $i_k$ will be called the {\em substorage} in $i_k$ and denoted by $x_{i_k}(t)$. More specifically, the substorage $x_{i_k}(t)$ is defined as the storage in compartment $i$ at time $t$ that is generated by the environmental input into compartment $k \neq 0$, $z_k(t)$, during the time interval $[t_0,t ]$ (see Fig.~\ref{fig:sc}). Consequently, due to the exhaustiveness of the system partitioning, we have 
\begin{equation}
\label{eq:partition}
x_i(t) = \sum \limits_{k=0}^n x_{i_k}(t), \quad i=1,\ldots,n.
\end{equation}
We define a new vector variable for the substorages as
\[ {\rm x}(t) = \left[ x_{1_0}(t), \ldots, x_{n_0}(t), x_{1_1}(t), \ldots, x_{n_1}(t), \ldots, x_{1_n}(t), \ldots, x_{n_n}(t) \right]^T .\]

We assume that environmental input $z_k(t,\rm{x})$ enters the system at subcompartment $k_k$, for all $k$. Moreover, no other $k^{th}$ subcompartment of any other compartment $i$, that is, subcompartment $i_k$, receives environmental input. This {\em input partitioning} can be expressed as
\begin{equation}
\label{eq:Z}
z_{i_k}(t,{\rm x}) = \delta_{ik} \, z_{k}(t,x) = \left \{
\begin{aligned}
& z_{k_k}(t,{\rm x}) = z_{k}(t,x), \quad i = k \\
& 0. \quad \quad \quad \quad \quad \quad \quad \quad \, \, \, i \neq k
\end{aligned}
\right .
\nonumber
\end{equation}

The intercompartmental flows are also partitioned in line with the subcompartmentalization (see Fig.~\ref{fig:sc}). The composite intercompartmental direct flow, $f_{ij}(t,x)$, is partitioned based on the assumption that the subcompartmental flow segments, $f_{{i_k}{j_k}}(t,{\rm x})$, $k = 0,\ldots,n$, are proportional to the corresponding substorages, $x_{j_k}(t)$, with the proportionality factor of the flow intensity in the flow direction, $q^{x}_{ij}(t,{x})$. The subcompartmental flow $f_{{i_k}{j_k}}(t,{\rm x})$ will be called the {\em subflow} from subcompartment $j_k$ to $i_k$ at time $t$.
It can be formulated as follows: 
\begin{equation}
\label{eq:cons}
f_{{i_k}{j_k}}(t,{\rm x}) = x_{j_k}(t) \, \frac{ f_{ij}(t,x) }{x_j(t)} = 
x_{j_k}(t) \, q^x_{ij}(t,x) =
d_{j_k}({\rm x}) \, f_{ij}(t,x) 
\end{equation}
where the coefficients $d_{j_k}({\rm x}) = {x_{j_k}(t)} / {x_j(t)}$ will be called the {\em decomposition factors}. Consequently, due to the exhaustiveness of the system partitioning, we have
\begin{equation}
\label{eq:flow_part}
f_{ij}(t,x) = \sum \limits_{k=0}^n  f_{i_k j_k}(t,{\rm x}) , \quad i,j=1,\ldots,n. 
\end{equation}

In summary, the dynamic system partitioning methodology explicitly generates mutually exclusive and exhaustive subsystems running within the original system. The $k^{th}$ subsystem is composed of all $k^{th}$ subcompartments of each compartment together with the corresponding subflows and substorages. These subsystems have the same structure and dynamics as the original system itself, except for their environmental inputs and initial conditions (see Figs.~\ref{fig:sd} and~\ref{fig:sc}). Each subsystem, except the initial one\textemdash which is driven by the initial stocks\textemdash is generated by a single environmental input. Therefore, the number of non-intersecting subcompartments in each compartment is equal to the number of inputs or compartments, plus one for the initial stocks. If an input or all initial conditions are zero, the corresponding subsystem becomes null. Consequently, for a system with $n$ compartments, each compartment has $n+1$ non-intersecting subcompartments, and therefore the system has $n+1$ mutually exclusive subsystems indexed by $k=0,\ldots,n$. The initial subsystem ($k=0$) represents the evolution of the initial stocks, receives no environmental input, and has the same initial conditions as the original system. The initial conditions for all the other subcompartments ($k \neq 0$) are zero, since they are initially assumed to be empty.

The governing equation for each subcompartment $i_k$ then becomes
\begin{equation}
\label{eq:model_sc}
\dot{x}_{i_k}(t) = \left( z_{i_k}(t,{\rm x}) + \sum_{\substack{j=1}}^n f_{{i_k}{j_k}}(t,{\rm x})  \right) - \left( y_{i_k}(t,{\rm x}) + \sum_{\substack{j=1}}^n f_{{j_k}{i_k}}(t,{\rm x}) \right)
\end{equation} 
for $i=1,\ldots,n$, $k=0,\ldots,n$. There are $n \times (n + 1)$ of such governing equations, one for each subcompartment. In order to track the evolution of environmental inputs within the system individually and separately, all except the initial subcompartments are assumed to be initially empty, as mentioned above. Therefore, the initial conditions become 
\begin{equation}
\label{eq:ic}
x_{i_k} (t_0) = \left \{
\begin{aligned}
x_{i,0}, \quad k=0 \\
0. \quad k \neq 0
\end{aligned}
\right. 
\end{equation}
The governing system of equations, Eq.~\ref{eq:model_sc}, is solved numerically with the initial conditions, Eq.~\ref{eq:ic}. The result yields the substorages at any time $t$, that is, $x_{i_k}(t)$.

The total subcompartmental inflows and outflows at compartment $i$ at time $t$ generated by the environmental input into compartment $k$, $z_k(t)$, during $[t_0,t]$ can then be defined, respectively, as
\begin{equation}
\label{eq:in_out_flows_sc}
\check{\tau}_{i_k}(t,{\rm x}) = z_{i_k}(t,{\rm x}) + \sum_{\substack{j=1 }}^n f_{{i_k}{j_k}}(t,{\rm x}) \quad \mbox{and} \quad \hat{\tau}_{i_k}(t,{\rm x}) = y_{i_k}(t,{\rm x}) + \sum_{\substack{j=1 }}^n f_{{j_k}{i_k}}(t,{\rm x})
\end{equation}
for $k=0,1,\ldots,n$. The functions $\check{\tau}_{i_k}(t,{\rm x})$ and $\hat{\tau}_{i_k}(t,{\rm x})$ will respectively be called {\em inward} and {\em outward subthroughflow} at subcompartment $i_k$ at time $t$ (see Fig.~\ref{fig:utilityfigs}). Therefore, the system partitioning enables dynamically partitioning composite compartmental flows and storages into subcompartmental segments based on their constituent sources from the initial stocks and environmental inputs.

We define the $n \times n$ {\em substorage} and associated {\em inward} and {\em outward subthroughflow matrix} functions, $X(t)$, $\check{T}(t,{\rm x})$, and $\hat{T}(t,{\rm x})$, respectively, as follows:
\begin{equation}
\label{eq:Tmatrix}
X_{ik}(t) = x_{i_k}(t), \quad \check{T}_{ik}(t,{\rm x}) = \check{\tau}_{i_k}(t,{\rm x}), \quad \mbox{and} \quad \hat{T}_{ik}(t,{\rm x}) = \hat{\tau}_{i_k}(t,{\rm x}) , 
\end{equation}
for $i,k=1,\ldots,n$. The {\em substorage} and associated {\em inward} and {\em outward subthroughflow vector} functions of size $n$ for the initial subsystem, ${x}_0(t)$, $\check{\tau}_0(t,{\rm x})$, and $\hat{\tau}_0(t,{\rm x})$, can also be defined, respectively, as 
\begin{equation}
\label{eq:Tvectors}
\begin{aligned}
{x}_0(t) & = \left[ x_{1_0}(t), \ldots, x_{n_0}(t) \right]^T,  \\
\check{\tau}_0(t,{\rm x}) & = \left[ \check{\tau}_{1_0}(t,{\rm x}), \ldots, \check{\tau}_{n_0}(t,{\rm x}) \right]^T , \, \, \,  \mbox{and} \, \, \, \hat{\tau}_0(t,{\rm x}) = \left[ \hat{\tau}_{1_0}(t,{\rm x}), \ldots, \hat{\tau}_{n_0}(t,{\rm x}) \right]^T .
\end{aligned}
\end{equation}
We use the constant vector notation $x_0$ for the initial conditions and the function notation $x_0(t)$ for the evolution of these initial stoks for $t > t_0$ with $x_0(t_0) = x_0$.

The notation $\diag({x(t)})$ will be used to represent the diagonal matrix whose diagonal elements are the elements of vector $x(t)$, and $\diag({X(t)})$ to represent the diagonal matrix whose diagonal elements are the same as the diagonal elements of matrix $X(t)$. The $n \times n$ diagonal {\em storage}, {\em output}, and {\em input} matrix functions, $\mathcal{X}(t)$, $\mathcal{Y}(t,x)$, and $\mathcal{Z}(t,x)$ will be defined, respectively, as
\[ \mathcal{X}(t) = \operatorname{diag}(x(t)), \quad \mathcal{Y}(t,x) = \operatorname{diag}(y(t,x)), \quad  \mbox{and} \quad \mathcal{Z}(t,x) = \operatorname{diag}(z(t,x)) . \]
Using Eq.~\ref{eq:cons}, the subthroughflow matrices can then be formulated as follows: 
\begin{equation}
\label{eq:Tmatrix2} 
\begin{aligned}
\check{T}(t,{\rm x}) &= \mathcal{Z}(t,x) + F(t,x) \, \mathcal{X}^{-1}(t) \, X(t) \\ 
\hat{T}(t,{\rm x}) &= \left( \mathcal{Y}(t,x) + \operatorname{diag} \left(F^T(t,x) \, \mathbf{1} \right) \right) \, \mathcal{X}^{-1}(t) \, X(t) \\
&= \mathcal{T}(t,x) \, \mathcal{X}^{-1}(t) \, X(t) \\
\end{aligned}
\end{equation}
where $\mathcal{T}(t,x) = \diag{(\hat{\tau}(t,x))} = \mathcal{Y}(t,x) + \operatorname{diag} \left(F^T(t,x) \, \mathbf{1} \right) $. Note that,
\begin{equation}
\label{eq:dyn_vec}
\begin{aligned}
x(t) & = {x}_0(t)+ X(t) \, \mathbf{1}, \\  \check{\tau}(t,x) & = \check{\tau}_0(t,{\rm x}) + \check{T}(t,{\rm x}) \, \mathbf{1}, \quad \mbox{and} \quad \hat{\tau}(t,x) = \hat{\tau}_0(t,{\rm x}) + \hat{T}(t,{\rm x}) \, \mathbf{1} .
\end{aligned}
\nonumber
\end{equation}

The governing equations for the decomposed system, Eq.~\ref{eq:model_sc}, can be expressed in terms of the vector and matrix functions introduced above as follows:
\begin{equation}
\label{eq:model_mat}
\begin{aligned}
{\dot X}(t) & = \check{T}(t,{\rm x}) - \hat{T}(t,{\rm x}) , \quad \, \, X (t_0) = \mathbf{0} , \\
\dot{{x}}_{0}(t) & = \check{\tau}_{0}(t,{\rm x}) - \hat{\tau}_{0}(t,{\rm x}) , \quad {x}_{0} (t_0) = x_{0} .
\end{aligned} 
\end{equation} 
We define an $n \times n$ matrix function $A(t,x)$ as
\begin{equation}
\label{eq:matrix_A}
\begin{aligned}
A(t,x) &= \left( F (t,x) - \mathcal{Y}(t,x) - \operatorname{diag} \left( F^T (t,x) \, \mathbf{1} \right) \right) \, \mathcal{X}^{-1} (t) \\
& = \left( F(t,x) - \mathcal{T}(t,x) \right) \, \mathcal{X}^{-1} (t) \\
& = Q^x (t,{x}) - \mathcal{R}^{-1}(t,{x}) 
\end{aligned}
\end{equation} 
where $Q^x (t,{x}) = F (t,{x}) \, \mathcal{X}^{-1}(t)$ and $\mathcal{R}^{-1}(t,{x}) = \mathcal{T}(t,{x}) \, \mathcal{X}^{-1}(t)$, assuming $\mathcal{R}(t,{x})$ is invertible. Note that the first term in the definition of $A(t,x)$, $Q^x (t,{x})$, represents the intercompartmental flow intensity defined in Eq.~\ref{eq:cons}, and the second term, $\mathcal{R}^{-1}(t,x)$, represents the outward throughflow intensity. Consequently, we will call $A(t,x)$ the {\em flow intensity matrix} per unit storage. It is sometimes called the {\em compartmental matrix} \cite{Jacquez1993}. The matrix measure $\mathcal{R}(t,{x})$ will be called the {\em residence time matrix}, and the matrix measure $Q^x (t,{x})$ will be called the {\em flow intensity matrix} per unit storage or the {\em flow distribution matrix} for system storages \cite{Coskun2017NDP,Coskun2017SCSA}. The governing equations, Eq.~\ref{eq:model_mat}, can be expressed using the flow intensity matrix in the following from:
\begin{equation}
\label{eq:model_M}
\begin{aligned}
\dot{X}(t) & = \mathcal{Z} (t,x) + A(t,x) \, X(t) ,
\quad X(t_0) = \mathbf{0} , \\
\dot{{x}}_{0}(t) & = A(t,x) \, x_0(t) ,
\quad \quad \quad \quad \quad  {x}_{0} (t_0) = x_{0} .
\end{aligned}
\end{equation} 

The dynamic system partitioning methodology that yields a decomposed system of $n^2+n$ governing equations for all subcompartments, Eq.~\ref{eq:model_M}, from the original system of $n$ governing equations for all compartments, Eq.~\ref{eq:model}, can algebraically be schematized as follows (see Figs.~\ref{fig:sd} and~\ref{fig:sc} for graphical illustrations):
\begin{center}
\begin{tikzpicture}
  \matrix (m) [matrix of math nodes,row sep=0em,column sep=7em,minimum width=1em]
  {
      \hspace{0cm} \dot{x}(t) = {\tau}(t,x) 
   & 
    {} 
   \\ 
  \hspace{0cm}  \veq & {} 
  \\
     {
 \begin{bmatrix}
  \dot{x}_{1} \\
  \dot{x}_{2} \\
  \vdots  \\
  \dot{x}_{n} 
 \end{bmatrix} 
=     
 \begin{bmatrix}
  {\tau}_{1} \\
  {\tau}_{2} \\
  \vdots  \\
  {\tau}_{n} 
 \end{bmatrix} 
 }
 & 
 { }
\\
{
  \begin{bmatrix}
  \dot{x}_{1_0} \\
  \dot{x}_{2_0} \\
  \vdots  \\
  \dot{x}_{n_0} 
 \end{bmatrix} 
= 
 \begin{bmatrix}
  {\tau}_{1_0} \\
  {\tau}_{2_0} \\
  \vdots  \\
  {\tau}_{n_0} 
 \end{bmatrix} 
}
&
\hspace{-2.5cm}
 {
\quad \mbox{and} \quad
 \begin{bmatrix}
  \dot{x}_{1_1} & \cdots & \dot{x}_{1_n} \\
  \dot{x}_{2_1} & \cdots & \dot{x}_{2_n} \\
  \vdots  & \ddots & \vdots  \\
  \dot{x}_{n_1} & \cdots & \dot{x}_{n_n} 
 \end{bmatrix} 
=
 \begin{bmatrix}
  {\tau}_{1_1} & \cdots & {\tau}_{1_n} \\
  {\tau}_{2_1} & \cdots & {\tau}_{2_n} \\
  \vdots & \ddots & \vdots  \\
  {\tau}_{n_1} & \cdots & {\tau}_{n_n} 
 \end{bmatrix} 
 }
 \\
  \veq &  \hspace{-1cm} \veq  \\
   {\dot{x}_0(t) = {\tau}_0(t,{\rm x})} & \hspace{-1cm} \dot{X}(t) = T(t,{\rm x}) \\
}; 
  \path[-stealth, decorate]            
    (m-3-1.east|-m-3-2) edge node [below] {} 
            node [above] {dynamic system partitioning} (m-3-2) ;            
\end{tikzpicture}
\end{center}
In the diagram above, the {\em net subthroughflow matrix}, $T(t,{\rm x})$, as well as the {\em net throughflow} and {\em initial throughflow vectors}, ${\tau}(t,{\rm x})$ and ${\tau}_0(t,{\rm x})$, are defined as the difference between the corresponding inward and outward throughflows. That is,
\begin{equation}
\label{eq:Ttautau0}
\begin{aligned}
T(t,{\rm x}) & = \check{T}(t,{\rm x}) - \hat{T}(t,{\rm x}) , \\
{\tau}(t,x) & = \check{\tau}(t,x) - \hat{\tau}(t,x) , \quad \text{and} \quad {\tau}_0(t,{\rm x}) = \check{\tau}_0(t,{\rm x}) - \hat{\tau}_0(t,{\rm x}) .
\end{aligned}
\end{equation}

The system partitioning introduced in this section is input-oriented. The governing system can be partitioned based on environmental outputs instead of inputs, by conceptually reversing all system flows. The following condition on the direct flows, instead of Eq.~\ref{eq:sf}, ensures the possibility of the system partitioning and analysis in both the input- and output-orientations:
\begin{equation}
\label{eq:sfO}
f_{ij}(t,{x}) = q^x_{ij}(t,{x}) \, {x}_i (t) \, {x}_j (t) , \quad i,j = 1. \ldots, n .
\end{equation}
This form of flow rates makes both the original and reversed decomposed systems well-defined.

\subsection{Subsystem flows and storages}
\label{appsec1:subsystem_part}

The system partitioning methodology dynamically decomposes a nonlinear system into mutually exclusive and exhaustive subsystems through the formulation of a set of governing equations derived from subcompartmentalization and flow partitioning components, as introduced above. This methodology enables the dynamic analysis of subsystems generated by the initial stocks and environmental inputs individually and separately. The subsystem flows and storages are formulated in matrix form in this section.

We define the $k^{th}$ {\em direct subflow matrix} function for the $k^{th}$ subsystem as $F_k(t,{\rm x})  = \left(f_{i_k j_k}(t,{\rm x}) \right)$, $k=0,\ldots,n$. Using the relationships formulated in Eq.~\ref{eq:cons}, this matrix can be expressed as follows:
\begin{equation}
\label{eq:directflow}
\begin{aligned}
F_k(t,{\rm x}) = F(t,x) \, \mathcal{X}^{-1}(t) \, \mathcal{X}_k(t) 
\end{aligned}
\end{equation}
where $\mathcal{X}_k(t) = \diag \left( [ x_{1_k}(t), \ldots, x_{n_k}(t) ] \right)$ is the diagonal matrix of the substorage functions in the $k^{th}$ subsystem. The matrix $\mathcal{X}_k(t)$ will accordingly be called the $k^{th}$ {\em substorage matrix} function. The $k^{th}$ output and input matrix functions then become
\[ \mathcal{Y}_k(t,{\rm x}) = \mathcal{Y}(t,x) \, \mathcal{X}^{-1}(t) \, \mathcal{X}_k(t) \quad \mbox{and} \quad \mathcal{Z}_k(t,{\rm x}) = \diag \left(z_k(t) \, \bm{e}_k \right) \]
where $\bm{e}_k$ is the elementary unit vector whose components are all zero except the $k^{th}$ element, which is $1$, and we set $\bm{e}_0 = {\bm 0}$. The $k^{th}$ direct subflow matrix, $F_k(t,{\rm x})$, and the $k^{th}$ {\em input} and {\em output vector functions}, defined as $\check{z}_k(t,{\rm x}) = \mathcal{Z}_k(t,{\rm x}) \, \bm{1}$ and $\hat{y}_k(t,{\rm x}) = \mathcal{Y}_k(t,{\rm x}) \, \bm{1}$, are the counterparts for the $k^{th}$ subsystem of the direct flow matrix, $F(t,x)$, and the input and output vectors, $z(t,x)$ and $y(t,x)$, for the original system. Altogether, they represent the subflow regime of the $k^{th}$ subsystem.

Using the notations and definitions of Eqs.~\ref{eq:Tmatrix2}  and~\ref{eq:directflow}, the $k^{th}$ {\em inward} and {\em outward subthroughflow matrices}, $\check{\mathcal{T}}_k(t,{\rm x}) = \diag{\left([\check{\tau}_{1_k}(t,{\rm x}),\ldots,\check{\tau}_{n_k}(t,{\rm x})] \right)}$ and $\hat{\mathcal{T}}_k(t,{\rm x}) = \diag{ \left( [\hat{\tau}_{1_k}(t,{\rm x}),\ldots,\hat{\tau}_{n_k}(t,{\rm x})] \right) }$, for the $k^{th}$ subsystem can be expressed as follows:
\begin{equation}
\label{eq:Tsubs}
\begin{aligned}
\check{\mathcal{T}}_k(t,{\rm x}) & = \mathcal{Z}_k(t,{\rm x}) + \diag{ \left( F(t,{x}) \, \mathcal{X}^{-1}(t) \, \mathcal{X}_k(t) \, \bm{1} \right) } , \\ 
\hat{\mathcal{T}}_k(t,{\rm x}) & =  \left ( \mathcal{Y}(t,{x}) + \diag { \left( F^T(t,{x}) \, \mathbf{1} \right) } \right ) \, \mathcal{X}^{-1}(t) \, \mathcal{X}_k(t)  \\
& = \mathcal{T}(t,{x}) \, \mathcal{X}^{-1}(t) \, \mathcal{X}_k(t) .
\end{aligned}
\end{equation} 

We also define the {\em decomposition} and $k^{th}$ {\em decomposition matrices}, ${D}({\rm x}) = (d_{i_k}({\rm x}))$ and $\mathcal{D}_k({\rm x}) = \diag{([d_{1_k}({\rm x}),\ldots,d_{n_k}({\rm x})])}$, as
\begin{equation}
\label{eq:DDk_dist}
\begin{aligned}
{D}({\rm x}) &= \mathcal{X}^{-1}(t) \, {X}(t) = \mathcal{T}^{-1}(t,{x}) \, \hat{T}(t,{\rm x}),  \\
\mathcal{D}_k({\rm x}) &= \mathcal{X}^{-1}(t) \, \mathcal{X}_k(t) = \mathcal{T}^{-1}(t,{x}) \, \hat{\mathcal{T}}_k(t,{\rm x}) .
\end{aligned}
\end{equation} 
The second equalities in the definitions of ${D}({\rm x})$ and $\mathcal{D}_k({\rm x})$ are due to Eq.~\ref{eq:Tmatrix2} and~\ref{eq:Tsubs}, respectively. Note that ${D}({\rm x})$ and $\mathcal{D}_k({\rm x})$ decompose the compartmental throughflow matrix, $\mathcal{T}(t,{x})$, into the subthroughflow and $k^{th}$ subthroughflow matrices as indicated in Eqs.~\ref{eq:Tmatrix2} and~\ref{eq:Tsubs}, similar to the decomposition of ${F}_k(t,{\rm x})$ as formulated below in Eq.~\ref{eq:F_k}. That is,
\begin{equation}
\label{eq:F_k2} 
\begin{aligned}
\hat{T}(t,{\rm x}) =  \mathcal{T}(t,{x}) \, D({\rm x})  \quad \mbox{and} \quad 
\hat{\mathcal{T}}_k(t,{\rm x}) =  \mathcal{T}(t,{x}) \, \mathcal{D}_k({\rm x}) .
\end{aligned}
\end{equation} 

The $k^{th}$ direct subflow and substorage matrices, $F_k(t,{\rm x})$ and $\mathcal{X}_k(t)$, can then be written in the following various forms:
\begin{equation}
\label{eq:F_k}
\begin{aligned}
{F}_k(t,{\rm x}) & = F(t,{x}) \, \mathcal{D}_k({\rm x}) =  Q^x(t,{x}) \, \mathcal{X}_k(t)  = Q^\tau(t,{x}) \, \hat{\mathcal{T}}_k(t,{\rm x})  \\
\mathcal{X}_k(t)  & = \mathcal{R}(t,{x}) \, \hat{\mathcal{T}}_k(t,{\rm x}) 
\end{aligned}
\end{equation} 
where $Q^\tau(t,{x}) = F(t,{x}) \, \mathcal{T}^{-1}(t,{x})$ will be called the {\em flow intensity matrix} per unit throughflow or the {\em flow distribution matrix} for system flows in the context of the proposed methodology \cite{Coskun2017SCSA}. Note that the elements of $Q^\tau (t,{x})$, $q^\tau_{ij}(t,{x})$, are sometimes called {\em transfer coefficients}, {\em technical coefficients} in economics, or {\em stoichiometric coefficients} in chemistry. The system level counterpart of Eq.~\ref{eq:F_k} can also be formulated as follows:
\begin{equation}
\label{eq:F_dec} 
\begin{aligned}
\tilde{T}(t,{\rm x}) & = F(t,{x}) \, {D}({\rm x}) =  Q^x(t,{x}) \, {X}(t)  = Q^\tau(t,{x}) \, \hat{T}(t,{\rm x}) , \\
{X}(t)  & = \mathcal{R}(t,{x}) \, \hat{T}(t,{\rm x}) ,
\end{aligned}
\end{equation}
using Eq.~\ref{eq:Tmatrix2}. The matrix function $\tilde{T} (t,{\rm x}) = \check{T} (t,{\rm x}) - \mathcal{Z} (t,{x})$ will be called the {\em intercompartmental subthroughflow matrix}. Componentwise, $\tilde{T} (t,{\rm x}) = (\tilde{\tau}_{i_k}(t,{\rm x}))$ can be expressed as $\tilde{\tau}_{i_k}(t,{\rm x}) = \check{\tau}_{i_k}(t,{\rm x}) - z_{i_k}(t)$. Consequently, for any given storage organization within the system, $X(t)$, the corresponding intercompartmental flow distributions\textemdash that is, intercompartmental inward and outward subthroughflow matrices, $\tilde{T} (t,{\rm x})$ and $\hat{T} (t,{\rm x})$\textemdash can be determined at any time $t$ as follows:
\begin{equation}
\label{eq:dist_works} 
\begin{aligned}
\tilde{T}(t,{\rm x}) =  Q^x(t,{x}) \, {X}(t)  
\quad \mbox{and} \quad
\hat{T}(t,{\rm x})  = \mathcal{R}^{-1}(t,{x}) \, {X}(t) .
\end{aligned}
\end{equation}

It is worth noting that the residence time matrix can be written in the following various forms:
\begin{equation}
\label{eq:res_matrix} 
\begin{aligned} 
\mathcal{R}(t,{x}) = \mathcal{X}(t) \, \mathcal{T}^{-1}(t,{x}) = \mathcal{X}_k(t) \, \hat{\mathcal{T}}^{-1}_k(t,{\rm x}) = {X}(t) \, \hat{T}^{-1}(t,{\rm x}) 
\end{aligned} 
\end{equation}
as formulated in Eqs.~\ref{eq:matrix_A},~\ref{eq:F_k}, and~\ref{eq:F_dec}. For later use, we also define three diagonal matrices as follows:
\begin{equation}
\label{eq:two_matx} 
\begin{aligned} 
\tilde{\mathsf{T}}(t,{\rm x}) = \diag{( \tilde{T}(t,{\rm x}) )} , \quad
\check{\mathsf{T}}(t,{\rm x}) = \diag{( \check{T}(t,{\rm x}) )} , \quad
\hat{\mathsf{T}}(t,{\rm x}) = \diag{( \hat{T}(t,{\rm x}) )} .
\end{aligned} 
\end{equation}  

\subsection{Analytic solution to linear systems}
\label{appsec:lin}

This section formulates analytic solutions to linear systems with time-dependent inputs. The system partitioning methodology yields a linear system, if the original system is linear. That is, if Eq.~\ref{eq:model_c} is linear, Eq.~\ref{eq:model_M} is also linear. Due to the cancellations in matrix $A(t,x)$ defined in Eq.~\ref{eq:matrix_A}, the decomposed linear system, Eq.~\ref{eq:model_M}, can be expressed in the following matrix form:
\begin{equation}
\label{eq:model_lin}
\begin{aligned}
\dot{X}(t) & = \mathcal{Z}(t) + A(t) \, X(t) ,
\quad X(t_0) = \mathbf{0} , \\
\dot{{x}}_{0}(t) & = A(t) \, x_0(t) ,
\quad \quad \quad \quad  {x}_{0} (t_0) = x_{0} .
\end{aligned} 
\end{equation} 
Let $V(t)$ be the fundamental matrix solution to the system Eq.~\ref{eq:model_lin}, as defined by \cite{Coskun2017NDP}. That is, let $V(t)$ be the unique solution of the system 
\begin{equation}
\label{eq:model_UV}
\begin{aligned}
{\dot {V}}(t) & = A(t) \, {V}(t) , \quad  {V}(t_0) = I .
\nonumber
\end{aligned} 
\end{equation} 
The solutions to Eq.~\ref{eq:model_lin} for substorage matrix, $X(t)$, and initial substorage vector, $x_0(t)$, in terms of $V(t)$, become
\begin{equation}
\label{eq:model_U}
\begin{aligned}
X(t) & =  \int_{t_0}^{t}  {V}(t) \, {V}^{-1}(s) \, \mathcal{Z}(s) \, ds  \quad \mbox{and} \quad x_0(t) = {V}(t) \, {x}_0 ,
\end{aligned} 
\end{equation}
as formulated by \cite{Coskun2017NDP}. Therefore, the solution to the original system, Eq.~\ref{eq:model} can be written as
\begin{equation}
\label{eq:model_Uv}
\begin{aligned}
x(t) & = {V}(t) \, {x}_0 + \int_{t_0}^{t}  {V}(t) \, {V}^{-1}(s) \, z(s) \, ds .
\end{aligned} 
\nonumber
\end{equation} 

For the special case of {\em constant} diagonalizable flow intensity matrix $A$, we have
\begin{equation}
\label{eq:model_linfund_sln}
\begin{aligned}
{V} (t)  & = {\exp} \left( \int_{t_0}^t A \, ds \right ) = {\mathrm{e}}^{ \left(t-t_0 \right) \,A} = \Omega \, {\mathrm{e}}^{(t-t_0) \, \Lambda} \, \Omega^{-1}
\end{aligned} 
\end{equation}
where $\Omega$ is the matrix whose columns are the eigenvectors of A, and $\Lambda$ is the diagonal matrix whose diagonal elements are the eigenvalues of $A$. For this particular case, Eq.~\ref{eq:model_U} takes the following form:
\begin{equation}
\label{eq:model_lin_sln2}
\begin{aligned}
X(t) & =  \int_{t_0}^{t} {\mathrm{e}}^{ \left(t-s \right) \, A} \, \mathcal{Z}(s) \, ds  \quad \mbox{and} \quad x_0(t) = {\mathrm{e}}^{ \left(t-t_0 \right) \, A} \, {x}_0 .
\end{aligned} 
\end{equation} 
Consequently,
\begin{equation}
\label{eq:model_lin_sln2V}
\begin{aligned}
x(t) & = {\mathrm{e}}^{ \left(t-t_0 \right) \, A} \, {x}_0 + \int_{t_0}^{t} {\mathrm{e}}^{ \left(t-s \right) \, A} \, z(s) \, ds  .
\end{aligned} 
\nonumber
\end{equation} 

A subsystem scaling argument is proposed to analyze static system behavior {\em per unit input} by \cite{Coskun2017SCSA}.  The scaled substorage matrix, $S(t) = X(t) \, \mathcal{Z}^{-1}$, can be expressed for constant invertible input matrix, $\mathcal{Z}(t) = \mathcal{Z} > {\bm 0}$, as follows:
\begin{equation}
\label{eq:model_lin_S}
\begin{aligned}
S(t) & = \int_{t_0}^{t}  {V}(t) \, {V}^{-1}(s) \, ds 
= \int_{t_0}^{t} {\mathrm{e}}^{ \left(t-s \right) \, A} \, ds 
= \Omega \, \left( \int_{t_0}^{t} {\mathrm{e}}^{ \left(t-s \right) \, \Lambda} \, ds \right) \, \Omega^{-1} ,
\end{aligned} 
\end{equation} 
using Eq.~\ref{eq:model_U} and~\ref{eq:model_lin_sln2}. The static version of this measure $S(t)$ is widely used in static ecological network analyses as outlined in the next section \cite{Coskun2017SCSA}.

An example of the analytic solution to a linear ecosystem model with time dependent environmental input is presented in Section~\ref{appsec:ex_hippe}.

\subsection{Static ecological system analysis}
\label{appsec:sstate}

At steady state, the time derivatives of the state variables are zero, and all system flows and storages are constant. That is,
\begin{equation}
\label{eq:model_ssTD}
\begin{aligned}
\dot{X}(t) = \mathbf{0}  \quad \mbox{and} \quad  \dot{x}_0(t) =\mathbf{0} .
\end{aligned}
\nonumber
\end{equation} 
The constant static quantities will be denoted by the same symbols without the time argument. The constant substorage matrix, for example, will be denoted by ${X}(t) = X$.
 
Summing up the equations in Eq.~\ref{eq:model_sc} side by side over index $k$ yields Eq.~\ref{eq:model_c} because of the relationship  
\[ {\dot x}_i(t) = \sum_{k=0}^{n} {\dot x}_{i_k}(t), \quad i = 1, \ldots, n, \]
deduced from Eq.~\ref{eq:partition} and the definition of the decomposition factors, $d_{i_k}({\rm x})$, given in Eq.~\ref{eq:cons}. Therefore, if the partitioned system, Eq.~\ref{eq:model_sc}, is at steady state, the original system, Eq.~\ref{eq:model_c}, is also at steady state. The static version of the proposed dynamic methodology is introduced by \cite{Coskun2017SCSA,Coskun2017SESM}, as summarized below in this section. 

Since $A$ is a strictly diagonally dominant constant matrix, it is invertible. It can be expressed as
\begin{equation}
\label{eq:matrix_A_apx}
\begin{aligned}
A = \left( F - \mathcal{T} \right) \, \mathcal{X}^{-1} = Q^x - \mathcal{R}^{-1} .
\end{aligned}
\end{equation} 
We then have the following solutions to Eq.~\ref{eq:model_M} for the substorage matrix, $X(t)$, and initial substorage vector, $x_0(t)$, at steady state:
\begin{equation}
\label{eq:model_ss}
\begin{aligned}
X = - A^{-1} \, \mathcal{Z} = \mathcal{X} \, \left( \mathcal{T} - F \right)^{-1} \, \mathcal{Z} \quad \mbox{and} \quad x_0 = \mathbf{0} .
\end{aligned}
\end{equation}  
From Eq.~\ref{eq:Tmatrix2} and the fact that ${\tau} = \hat{\tau} = \check{\tau}$ and $T = \check{T} = \hat{T}$ at steady state, the throughflow matrix can be written in terms of system flows only:
\begin{equation}
\label{eq:model_ss2}
\begin{aligned}
T & = \mathcal{Z} + F \, \mathcal{X}^{-1} \, X = \mathcal{Z} + F  \, \mathcal{T}^{-1} \, T  \quad
\Rightarrow  \quad T = \left ( I - F  \, \mathcal{T}^{-1} \right )^{-1} \, \mathcal{Z} .
\end{aligned}
\end{equation}  
The residence time matrix $\mathcal{R}$ can also be expressed as
\begin{equation}
\label{eq:DtR}
\begin{aligned}
\mathcal{R} & = \mathcal{X} \, \mathcal{T}^{-1}  = \mathcal{X}_k \, \mathcal{T}_k^{-1}  = X \, T^{-1}  
\end{aligned}
\end{equation}
similar to Eq.~\ref{eq:res_matrix}.

The scaled substorage and subthroughflow matrices are defined for system analysis per unit input by \cite{Coskun2017SCSA}. They are formulated as $S=X \, \mathcal{Z}^{-1}$ and $N= T \, \mathcal{Z}^{-1}$, where $\mathcal{Z}$ is invertible. Using Eqs.~\ref{eq:model_ss} and~\ref{eq:model_ss2}, these matrix measures can be expressed as follows:
\begin{equation}
\label{eq:SandN}
\begin{aligned} 
S = -A^{-1}  \quad  & \mbox{and} \quad 
N =  \left ( I - F  \, \mathcal{T}^{-1} \right )^{-1} .
\end{aligned}
\end{equation} 
Note that $S(t)$ formulated in Eq.~\ref{eq:model_lin_S} is equivalent to $S$ at steady state, that is $$\displaystyle \lim_{t \to \infty} S(t) = S .$$ 
The matrices $N$ and $S$ are called the {\em cumulative flow} and {\em storage distribution matrices} in the context of the proposed methodology \cite{Coskun2017SCSA}.  

Although the derivation rationales are different, the proposed matrix measures $S$ and $N$ are equivalent to the ones formulated in the current static ecological network analyses, as shown by \cite{Coskun2017SCSA}. These matrices are treated separately in the current static methodologies, although they are naturally related by a factor of the residence time matrix. Equation~\ref{eq:DtR} implies that
\[ T = \mathcal{R}^{-1} \, X  \, \, \Rightarrow  \, \, S = \mathcal{R} \, N . \]
This relationship enables the {\em holistic view} of static ecological networks:
\begin{equation}
\label{eq:holistic}
\begin{aligned} 
x  = S \, z = \mathcal{R} \, N \, z  = \mathcal{R} \, \tau  \quad  & \mbox{and} \quad 
X = S \, \mathcal{Z} = \mathcal{R} \, N \, \mathcal{Z}  = \mathcal{R} \, T 
\end{aligned}
\end{equation}
as introduced by \cite{Coskun2017SESM}.
 
The substorage and subthroughflow matrices can be scaled by output matrix instead, for the {\em output-oriented system analysis}. We use a bar notation over the output-oriented counterparts of the input oriented quantities. In the output-oriented analysis, we assume that all system flows are conceptually reversed. That is,
\[ \bar{F} = F^T, \quad \mathcal{\bar Y} = \mathcal{Z}, \quad \mbox{and} \quad \mathcal{\bar Z}= \mathcal{Y} .\] 
The counterparts of the Eq.~\ref{eq:holistic} for the output-oriented analysis then become
\begin{equation}
\label{eq:holisticO} 
\begin{aligned} 
\bar{x}  = \bar{S} \, y = \mathcal{R} \, \bar{N} \, y  = \mathcal{R} \, \bar{\tau}  \quad & \mbox{and} \quad 
\bar{X} = \bar{S} \, \mathcal{Y} = \mathcal{R} \, \bar{N} \, \mathcal{Y}  = \mathcal{R} \, \bar{T}
\end{aligned}
\end{equation}
where $\bar{S}=\bar{X} \, \mathcal{Y}^{-1}$, $\bar{N}= \bar{T} \, \mathcal{Y}^{-1}$, and the diagonal output matrix, $\mathcal{Y}$, is assumed to be invertible. Since $x=\bar{x}$ and $\tau = \bar{\tau}$ at steady state, $\mathcal{\bar R} = \mathcal{R}$.

It is worth noting that the input- and output-oriented cumulative flow and storage distribution matrices are similar. This duality can be expressed as
\begin{equation}
\label{eq:SSp}
\begin{aligned}
S \, \mathcal{X} = \mathcal{X} \, \bar{S}^T \quad \mbox{and} \quad N \mathcal{T} = \mathcal{T} \, \bar{N}^T .
\end{aligned}
\end{equation}
The holistic input- and output-oriented static ecological system analyses and their duality have recently been introduced by \cite{Coskun2017SCSA,Coskun2017SESM}.

\subsection{Subsystem partitioning methodology}
\label{sec:dsp}

In this section, we introduce the {\em dynamic subsystem partitioning} methodology for further partitioning or segmentation of subsystems along a given set of mutually exclusive and exhaustive subflow paths, as a simplified version of the \emph{subsystem decomposition methodology} recently proposed by \cite{Coskun2017NDP}. The subsystem partitioning methodology dynamically apportions arbitrary composite intercompartmental flows and the associated storages generated by these flows into transient subflow and substorage segments along given subflow paths. The subsystem partitioning, therefore, determines the distribution of arbitrary intercompartmental flows and the organization of the associated storages generated by these flows within the subsystems. In other words, this methodology enables tracking the evolution of arbitrary intercompartmental flows and associated storages within and monitoring their spread throughout the system.

The {\em natural subsystem decomposition} is defined as the set of mutually exclusive and exhaustive subflow paths whose local inputs and outputs, except for the closed paths, are environmental inputs and outputs, respectively \cite{Coskun2017NDP,Coskun2017SCSA}. By {\em mutually exclusive} subflow paths, we mean that no given subflow path is a {\em subpath}, that is, completely inside of another path in the same subsystem. {\em Exhaustiveness} in this context means that such mutually exclusive subflow paths\textemdash together with the corresponding transient subflows and substorages along the paths\textemdash all together sum to the entire subsystem so partitioned. The natural subsystem decomposition of each subsystem then results in a mutually exclusive and exhaustive decomposition of the entire system.

We will first introduce the transient flows and storages below. Thereafter, they will then be used for the formulation of the \texttt{diact} flows and storages in the subsequent section.

\hfill 
\begin{flushright}
{\em No man ever steps in the same river twice. -- Heraclitus (535-475 BC) }
\end{flushright}

\subsubsection{Transient flows and storages}
\label{apxsec:transient_flows}

As indicated in the famous {\em dictum} by Heraclitus that ``everything flows,'' flows are one of the most important physical phenomena of existence. In this section, we formulate the {\em transient flows} and the associated {\em storages} generated by these flows.

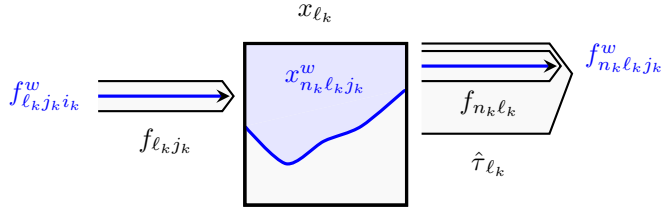
\begin{figure}[t]
\begin{center}
\begin{tikzpicture} 
   \draw[very thick, fill=gray!5, draw=black] (-.05,-.05) rectangle node(R1) [pos=.5] { } (2.1,2.1) ;
   \draw [very thick, fill=blue!10, draw=blue] plot [smooth] coordinates {(-0.05,1) (.5,.5) (1,.8) (1.5,1) (2.1,1.5)};
   \draw [fill=blue!10, draw=none]  (-0.05,1) -- (2.1,1.5) -- (2.1,2.1) -- (-.05,2.1) ;   
   \draw[very thick, draw=black, text=blue] (-.05,-.05) rectangle node(R1) [pos=.5, yshift=.6cm] { $x^w_{n_k \ell_k j_k}$ } (2.1,2.1) ;   
   \draw [thick, fill=gray!5, draw=black]  (-2,1.2) -- (-0.35,1.2) -- (-0.2,1.4) -- (-.35,1.6) -- (-2,1.6) ;      
   \draw [thick, fill=gray!5, draw=black]  (2.3,.9) -- (4,.9) -- (4.3,1.7) -- (4,2.1) -- (2.3,2.1) ;           
   \draw [thick, fill=gray!5, draw=black]  (2.3,1.6) -- (4,1.6) -- (4.15,1.8) -- (4,2) -- (2.3,2) ;
    \node (x) at (.9,2.5) {${x}_{\ell_k}$};  
    \node (x) at (-1.1,.8) {$f_{\ell_k j_k}$};
   \draw [very thick, -stealth, draw=blue]  (-2,1.4) -- (-.25,1.4) ;
    \node [text=blue] (x) at (-2.7,1.4) {$f^w_{\ell_k j_k i_k}$};
   \draw [very thick, -stealth, draw=blue]  (2.3,1.8) -- (4.1,1.8) ; 
    \node [text=blue] (x) at (5,1.9) {$f^w_{n_k \ell_k j_k}$}; 
    \node (x) at (3.2,1.3) {$f_{n_k \ell_k}$};   
    \node (x) at (3.2,.5) {${\hat{\tau}}_{\ell_k}$};    
\end{tikzpicture}
\end{center}
\caption{Schematic representation of the dynamic subsystem decomposition. The transient inflow and outflow rate functions, $f^w_{\ell_k j_k i_k}(t)$ and $f^w_{n_k \ell_k j_k}(t)$, at and associated transient substorage, $x^w_{n_k \ell_k j_k}(t)$, in subcompartment ${\ell_k}$ along subflow path $p^w_{n_k j_k}= i_k \mapsto j_k \to \ell_k \to n_k$. }
\label{fig:subsystemp}
\end{figure}

The {\em transient} and {\em cumulative transient subflows} along a subflow path are defined as follows: Along a given subflow path $p^w_{n_k j_k}= i_k \mapsto j_k \to \ell_k \to n_k$, the {\em transient inflow} at subcompartment ${\ell_k}$, $f^{w}_{\ell_k j_k i_k}(t)$, generated by the local input from $i_k$ to ${j_k}$ during $[t_1,t]$, $t_1 \geq t_0$, is the input segment that is transmitted from $j_k$ to ${\ell_k}$ at time $t$. Similarly, the {\em transient outflow} generated by the transient inflow at ${\ell_k}$ during $[t_1,t]$, $f^{w}_{n_k \ell_k j_k}(t)$, is the inflow segment that is transmitted from ${\ell_k}$ to the next subcompartment, ${n_k}$, along the path at time $t$. The associated {\em transient substorage} in subcompartment ${\ell_k}$ at time $t$, $x^{w}_{n_k \ell_k j_k}(t)$, is the substorage segment governed by the transient inflow and outflow balance during $[t_1,t]$  (see Fig.~\ref{fig:subsystemp}).

The transient outflow at subcompartment ${\ell_k}$ at time $t$ along subflow path $p^w_{n_k j_k}$ from ${j_k}$ to ${n_k}$, $f^{w}_{n_k \ell_k j_k}(t)$, can be formulated as follows:
\begin{equation}
\label{eq:out_in_fs}
\begin{aligned}
f^{w}_{n_k \ell_k j_k}(t) = \frac{ f_{n_k \ell_k}(t,\rm{x}) }{x_{\ell_k}(t) } \, x^{w}_{n_k \ell_k j_k}(t) ,
\end{aligned}
\end{equation}
similar to Eq.~\ref{eq:cons}, due to the equivalence of flow and subflow intensities, where the transient substorage, $x^{w}_{n_k \ell_k j_k}(t)$, is determined by the governing mass balance equation 
\begin{equation}
\label{eq:out_in_fs2}
\begin{aligned}
\dot{x}^{w}_{n_k \ell_k j_k}(t) & = f^{w}_{\ell_k j_k i_k}(t) - \frac{ {\hat{\tau}}_{\ell_k}(t,\rm{x}) }{ x_{\ell_k}(t) } \, {x}^{w}_{n_k \ell_k j_k}(t) , \quad {x}^{w}_{n_k \ell_k j_k}(t_1) = 0 .
\end{aligned}
\end{equation}
The equivalence of the throughflow and subthroughflow intensities, as well as the flow and subflow intensities in the same direction, that is
\begin{equation}
\label{eq:eqiv_intense}
\begin{aligned}
q^x_{n \ell}(t,x) = \frac{  f_{n \ell}(t,x) }{x_{\ell}(t) } = \frac{ f_{n_k \ell_k}(t,\rm{x})  }{x_{\ell_k}(t) } 
 \quad \mbox{and} \quad r^{-1}_\ell (t,x) = \frac{ {\hat{\tau}}_{\ell}(t,x) }{ x_{\ell}(t) } = \frac{ {\hat{\tau}}_{\ell_k}(t,\rm{x}) }{ x_{\ell_k}(t) }
 \end{aligned}
\end{equation}
are given by Eqs.~\ref{eq:cons} and~\ref{eq:Tmatrix2}, for $\ell,n=1,\ldots,n$, and $k=0,1,\ldots,n$ \cite{Coskun2017NDP}. Therefore, since the intensities in Eqs.~\ref{eq:out_in_fs} and~\ref{eq:out_in_fs2} can be expressed at both the compartmental and subcompartmental levels, the subsystem partitioning is actually independent from the system partitioning. That is, the same analysis can be done along flow paths within the system, instead of subflow paths within the subsystems. This allows the flexibility of tracking arbitrary intercompartmental flows and storages generated by all or individual environmental inputs within the system. The governing equations, Eqs.~\ref{eq:out_in_fs} and~\ref{eq:out_in_fs2}, establish the foundation of the {\em dynamic subsystem partitioning}. These equations for each subcompartment along a given flow path of interest will then be coupled with the partitioned system, Eq.~\ref{eq:model_sc}, or the original system, Eq.~\ref{eq:model_c}, and be solved simultaneously. The equations can alternatively be solved individually and separately once the original or partitioned system is solved.

The transient subflows and substorages are defined for linear subflow paths above. The sum of the transient inflows from subcompartment $j_k$ to $\ell_k$ and the outflows from $\ell_k$ to $n_k$ generated at subcompartment $\ell_k$ at time $t$ by the local input into the connection of a given non-self-intersecting closed subflow path $p^w_{n_k j_k}$ during $[t_1,t]$, $t_1 \geq t_0$, will respectively be called the {\em inward} and {\em outward cumulative transient subflow} at subcompartment ${\ell_k}$ at time $t$. The associated storage generated by the inward cumulative transient subflow will be called {\em cumulative transient substorage}. These inward and outward cumulative transient subflows will be denoted by $\check{\tau}^{w}_{\ell_k}(t)$ and $\hat{\tau}^{w}_{\ell_k}(t)$, respectively, and associated cumulative transient substorage by ${x}^{w}_{\ell_k}(t)$. They can be formulated as
\begin{equation}
\label{eq:apxout_in_fs10}
\begin{aligned}
{x}^{w}_{\ell_k}(t) = \sum_{m=1}^{m_w} {x}^{w,m}_{n_k \ell_k j_k}(t), \quad 
\check{\tau}^{w}_{\ell_k}(t) = \sum_{m=1}^{m_w} {f}^{w,m}_{\ell_k j_k i_k}(t),  \quad 
\hat{\tau}_{\ell_k}^{w}(t) = \sum_{m=1}^{m_w} {f}^{w,m}_{n_k \ell_k j_k}(t) 
\end{aligned}
\end{equation}
for $k=0,1,\ldots,n$, where the superscript $m$ represents the cycle number, and ${m_w}$ is the number of cycles, that is, the number of times the path $p^w_{n_k i_k}$ pass through subcompartment $\ell_k$. Large number of terms, ${m_w}$, in computation of these summations reduce truncation errors and, thus, improve the approximations. 

Using the equivalence of flow intensities as formulated in Eq.~\ref{eq:eqiv_intense}, it has been recently shown for compartmental systems that the parallel subflows and the corresponding subthroughflows and substorages are proportional \cite{Coskun2017NDP}. By {\em parallel subflows}, we mean the intercompartmental flows that transit through different subcompartments of the same compartment along the same flow path at the same time. This proportionality can be formulated as 
\begin{equation}
\label{eq:thr_dense3New}
\begin{aligned}
\frac{{\hat{\tau}}_{k_\ell}(t,{\rm x}) }{{\hat{\tau}}_{k_k}(t,{\rm x}) } = \frac{{x}_{k_\ell}(t) }{{x}_{k_k}(t)} = \frac{{f}_{i_\ell k_\ell}(t,{\rm x}) }{{f}_{i_k k_k}(t,{\rm x})} 
\end{aligned}
\end{equation} 
for $k=1,\ldots,n$ and $\ell=0,\ldots,n$, where the denominators are nonzero.

\subsubsection{The \texttt{diact} flows and storages}
\label{apxsec:flows}

In this section, we formulate five main transaction types for nonlinear systems at both the subcompartmental and compartmental levels using two approaches: the {\em direct} (\texttt{d}), {\em indirect} (\texttt{i}), {\em cycling} (\texttt{c}), {\em acyclic} (\texttt{a}), and {\em transfer} (\texttt{t}) {\em flows} and the associated {\em storages} generated by these \texttt{diact} flows. The first approach based on the subsystem partitioning methodology will be called the {\em path-based approach}, and the second approach based on the system partitioning methodology will be called the {\em dynamic approach}.

\begin{figure}[t]
\begin{center}
\begin{tikzpicture}
   \draw[very thick, fill=gray!10, draw=black] (-.05,-.05) rectangle node(R1) [pos=.5] { } (2.1,2.4) ;
   \draw[very thick, fill=blue!3, draw=blue, text=blue] (0.3,0.1) rectangle node(R1) [pos=.5] {$x_{i_i}$} (1.2,1.3) ;
    \node (x) at (1,-.5) {${x}_{i}$};
   \draw[very thick, fill=gray!10, draw=black] (8.95,-.05) rectangle node(R2) [pos=.5] { } (11.1,2.4) ;
   \draw[very thick, fill=gray!10, draw=black] (8.95,1.69) rectangle node(R2) [pos=.5] { } (10.5,2.3) ;   
   \draw[very thick, fill=blue!3, draw=blue, text=blue] (9.05,0.05) rectangle node(R1) [pos=.5] {$x_{j_i}$} (10.05,1.1) ;
    \node (x) at (10,-.5) {${x}_{j}$};
    \node (x) at (9.7,2) {${x}_{j_0}$};    
   \draw[fill=gray!10] (3.05,1.3) -- ++ (-0.6,0) -- ++ (-0.3,.3) -- ++ (.3,.3) -- ++ (.6,0);
   \draw[fill=gray!10] (8.1,1.3) -- ++ (0.84,0) -- ++ (0,.4) -- ++ (-.84,0) ;
   \draw[fill=gray!10] (8.1,1.7) -- ++ (0.84,0) -- ++ (0,.43) -- ++ (-.84,0) ;
    \node[] (t2) at (2.8,1) {${\tau}^\texttt{t}_{i j}$};
    \node[] (t3) at (8.5,2.4) {$\hat{\tau}_{j}$};
    \node[] (t3) at (8.5,1.92) {$\hat{\tau}_{j_0}$};    
    \node[] (t2) at (4,2.2) {${\tau}^\texttt{c}_{i j}$};
    \node[] (x) at (4.5,1.1) {${\tau}^\texttt{d}_{i j} $};
    \draw[very thick,-stealth]  (8.8,1.4) -- (2.4,1.4);
   \draw[fill=blue!10] (2.95,0.3) -- ++ (-0.84,0) -- ++ (0,.4) -- ++ (.84,0) ;
    \draw[very thick,-stealth,draw=blue]  (-.8,.5) -- (.25,.5) ;
    \node (z) [text=blue] at (-.45,.8) {$z_i$};
    \node[blue] (t3) at (2.7,0.05) {$\hat{\tau}_{i_i}$};
   \draw[fill=blue!10] (8,0.1) -- ++ (0.6,0) -- ++ (0.3,.3) -- ++ (-.3,.3) -- ++ (-.6,0);
    \node[blue] (t1) at (8.4,1) {$\tilde{\tau}_{j_i}$};
    \draw[very thick,draw=blue, -stealth]  (1.3,0.6) -- (8.7,0.6) ;
    \node[blue] (x) at (6.7,0.9) {${\tau}^\texttt{d}_{j_i} $};
   \node[blue,anchor=west] at (3.3,0.1) (a) {${\tau}^\texttt{i}_{j_i}$};
   \node[blue,anchor=east] at (7,0.4) (b) {};
   \node[blue,anchor=west] at (.9,0.4) (e) {};
   \node[blue,anchor=east] at (3.3,0.1) (f) {};
\draw [very thick,draw=blue, dashed, -stealth] plot [smooth, tension=1] coordinates { (1.3,0.4) (4,0.4) (5,-0.1) (4.5,-0.6) (4,-0.1) (5,0.4) (8.7,0.4) };
\draw [very thick,draw=blue, dashed, -stealth] plot [smooth, tension=1] coordinates { (9,0.4) (9.6,-0.1) (9,-0.5) (7,-0.5) (6.5,-0.1) (7,0.2) (8.7,0.2) };
    \node[blue] (t2) at (7,-0.2) {${\tau}^\texttt{c}_{j_i}$};
   \node[anchor=east] at (7.2,2.1) (c) {$\tau^\texttt{i}_{i j}$};
   \node[blue,anchor=west] at (2.1,1.7) (d) {};
\draw [very thick,dashed, -stealth] plot [smooth, tension=1] coordinates { (8.8,1.6) (6.5,1.6)  (5.5,2.1)  (6,2.6) (6.5,2.1) (5.5,1.6) (2.4,1.6) };
\draw [very thick, dashed, -stealth] plot [smooth, tension=1] coordinates { (2,1.6) (1.5,2.1) (2,2.6) (4,2.6) (4.5,2.1) (4,1.8) (2.4,1.8) };
   \node[blue,anchor=east] at (5.9,2.1) (h) {};
   \node[blue,anchor=west] at (8,1.7) (g) {};
\end{tikzpicture}
\end{center}
\caption{Schematic representation of the simple and composite \texttt{diact} flows. Solid arrows represent direct flows, and dashed arrows represent indirect flows through other compartments (not shown).
The composite \texttt{diact} flows (black) generated by outward throughflow $\hat{\tau}_{j}(t,{x}) - \hat{\tau}_{j_0}(t,{\rm x})$ (i.e. derived from all environmental inputs): direct flow, $\tau^\texttt{d}_{i j}(t)$, indirect flow, $\tau^\texttt{i}_{ij}(t)$, acyclic flow, $\tau^\texttt{a}_{ij}(t) = \tau^\texttt{t}_{ij}(t) - \tau^\texttt{c}_{ij}(t)$, cycling flow, $\tau^\texttt{c}_{ij}(t)$, and transfer flow, $\tau^\texttt{t}_{ij}(t)$.
The simple \texttt{diact} flows (blue) generated by outward subthroughflow $\hat{\tau}_{i_i}(t,{\rm x})$ (i.e.  derived from single environmental input $z_{i}(t)$): direct flow, ${\tau}^\texttt{d}_{j_i}(t) = {\tau}^\texttt{d}_{j_i i_i}(t) $, indirect flow, ${\tau}^\texttt{i}_{j_i}(t) = \tau^\texttt{i}_{j_i i_i}(t)$, acyclic flow, ${\tau}^\texttt{a}_{j_i}(t) = \tau^\texttt{a}_{j_i i_i}(t) = {\tau}^\texttt{t}_{j_i}(t) - {\tau}^\texttt{c}_{j_i}(t)$, cycling flow, ${\tau}^\texttt{c}_{j_i}(t) = \tau^\texttt{c}_{j_i i_i}(t)$, and transfer flow, $\tau^{\texttt{t}}_{j_i}(t) = \tilde{\tau}_{j_i}(t,{\rm x}) = \check{\tau}_{j_i}(t,{\rm x}) - z_{j_i}(t)$. Note that the cycling flows at the terminal (sub)compartment may include the segments of the direct and/or indirect flows at that (sub)compartment, if the cycling flows indirectly pass through the corresponding initial (sub)compartment (see Fig.~\ref{fig:cycindices}). Therefore, the acyclic flows are composed of the segments of the direct and/or indirect flows.
}
\label{fig:utilityfigs}
\end{figure}
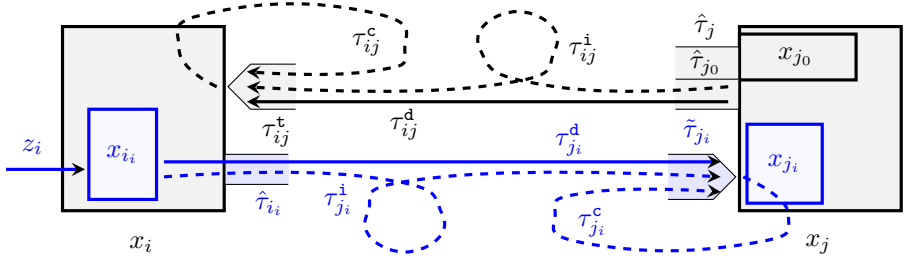

The {\em composite transfer flow} will be defined as the total intercompartmental transient flow that is generated by all environmental inputs from one compartment, {\em directly} or {\em indirectly} through other compartments, to another. The {\em composite direct}, {\em indirect}, {\em acyclic}, and {\em cycling flows} from the initial compartment to the terminal compartment are then defined as the direct, indirect, non-cycling, and cycling segments at the terminal compartment of the composite transfer flow (see Fig.~\ref{fig:utilityfigs}). The cycling and acyclic flows can, therefore, be interpreted as the flows that visit the terminal compartment {\em multiple times} and {\em only once}, respectively, after being transmitted from the initial compartment.

The {\em composite transfer subflow} within the initial subsystem can also be defined as the total intercompartmental transient subflow that is derived from all initial stocks from one initial subcompartment, {\em directly} or {\em indirectly} through other initial subcompartments, to another. The {\em composite direct}, {\em indirect}, {\em acyclic}, and {\em cycling subflows} within the initial subsystem from the initial subcompartment to the terminal subcompartment are then defined as the direct, indirect, non-cycling, and cycling segments at the terminal subcompartment of the composite transfer subflow.

The {\em simple transfer flow} will be defined as the total intercompartmental transient subflow that is generated by the single environmental input from an input-receiving subcompartment, {\em directly} or {\em indirectly} through other compartments, to another subcompartment. The {\em simple direct}, {\em indirect}, {\em acyclic}, and {\em cycling flows} from the initial input-receiving subcompartment to the terminal subcompartment are then defined as the direct, indirect, non-cycling, and cycling segments at the terminal subcompartment of the simple transfer flow (see Fig.~\ref{fig:utilityfigs}). The associated simple and composite \texttt{diact} storages are defined as the storages generated by the corresponding \texttt{diact} flows.

Let $P^\texttt{t}_{i_k j_k}$ be the set of mutually exclusive subflow paths $p^w_{i_k j_k}$ from subcompartment $j_k$ directly or indirectly to $i_k$ in subsystem $k$. The sets $P^\texttt{d}_{i_k j_k}$ and $P^\texttt{i}_{i_k j_k}$ are also defined as the sets of mutually exclusive {\em direct} and {\em indirect subflow paths} $p^w_{i_k j_k}$ from subcompartment $j_k$ {\em directly} and {\em indirectly} to $i_k$, respectively. Similarly, the sets $P^\texttt{c}_{i_k j_k}$ and $P^\texttt{a}_{i_k j_k}$ are defined as the sets of mutually exclusive {\em cyclic} and {\em acyclic subflow paths} $p^w_{i_k j_k}$ from $j_k$ to $i_k$ with a {\em closed} and {\em linear} subpath at terminal subcompartment $i_k$, respectively (see Fig.~\ref{fig:utilityfigs}). The {\em cyclic subflow set}, ${P}^{\texttt{c}}_{i_k}$, can alternatively be defined as the set of mutually exclusive subflow paths ${p}^w_{i_k}$ from subcompartment $i_k$ {\em indirectly} back to itself. The number of subflow paths in $P^\texttt{*}_{i_k j_k}$ will be denoted by $w_k$, where the superscript $(^\texttt{*})$ represent any of the \texttt{diact} symbols. 

The {\em composite} \texttt{diact} {\em subflow} from subcompartment $j_k$ to $i_k$, $\tau^\texttt{*}_{i_k j_k}(t)$, is defined as the sum of the cumulative transient subflows, $\check{\tau}_{i_k}^{w}(t)$, generated by the outward subthroughflow at subcompartment $j_k$, $\hat{\tau}_{j_k}(t,{\rm x})$, during $[t_1,t]$, $t_1 \geq t_0$, and transmitted into $i_k$ at time $t$ along all subflow paths $p^w_{i_k j_k} \in P^\texttt{*}_{i_k j_k}$. The associated composite \texttt{diact} {\em substorage}, $x^\texttt{*}_{i_k j_k}(t)$, at subcompartment $i_k$ at time $t$ is the sum of the cumulative transient substorages, $x^{w}_{i_k}(t)$, generated by the cumulative transient inflows, $\check{\tau}_{i_k}^{w}(t)$, during $[t_1,t]$. Alternatively, $x^\texttt{*}_{i_k j_k}(t)$ can be defined as the storage segment generated by the composite \texttt{diact} inflow $\tau^\texttt{*}_{i_k j_k}(t)$ in subcomaprtment $i_k$ during $[t_1,t]$. Note that, for the cycling case, the first entrance of the transient subflows and substorages into $i_k$ are not considered as cycling subflows and substorages.

The composite \texttt{diact} subflows and substorages can then be formulated as follows:
\begin{equation}
\label{eq:out_in_fsDT_diact}
\begin{aligned}
{\tau}^\texttt{*}_{i_k j_k}(t) = 
\sum_{w=1}^{w_k}  \check{\tau}_{i_k}^{w}(t) 
\quad \mbox{and} \quad 
x^\texttt{*}_{i_k j_k}(t) = \sum_{w=1}^{w_k} x^{w}_{i_k}(t) .
\end{aligned}
\end{equation}
The sum of all composite \texttt{diact} subflows and substorages from subcompartment $j_k$ to $i_k$ within each subsystem $k \neq 0$ will be called the {\em composite} \texttt{diact} {\em flow} and {\em storage} from compartment $j$ to $i$ at time $t$, ${\tau}^\texttt{*}_{i j}(t)$ and $x^\texttt{*}_{i j}(t)$, generated by all environmental inputs during $[t_1,t]$. They can be formulated as
\begin{equation} 
\label{eq:out_in_diact} 
\begin{aligned}
\tau^\texttt{*}_{i j}(t) = \sum_{k=1}^{n} \tau^\texttt{*}_{i_k j_k}(t) 
\quad & \mbox{and} \quad 
x^\texttt{*}_{i j}(t) = \sum_{k=1}^{n} x^\texttt{*}_{i_k j_k}(t) .
\end{aligned}
\end{equation}

For notational convenience, we define $n \times n$ matrix functions ${T}^{\texttt{*}}_k(t)$ and ${X}^{\texttt{*}}_k(t)$ whose $(i,j)-$elements are $\tau^\texttt{*}_{i_kj_k}(t)$ and $x^\texttt{*}_{i_kj_k}(t)$, respectively. That is,
\begin{equation}
\label{eq:fmT_diact} 
\begin{aligned}
{T}^{\texttt{*}}_k(t) = \left( \tau^\texttt{*}_{i_k j_k}(t) \right) 
\quad \mbox{and} \quad 
{X}^{\texttt{*}}_k(t) = \left( x^\texttt{*}_{i_k j_k}(t) \right) ,
\end{aligned}
\end{equation}
for $k=0,\ldots,n$. These matrix measures ${T}^{\texttt{*}}_k(t)$ and ${X}^{\texttt{*}}_k(t)$ are called the $k^{th}$ {\em composite} \texttt{diact} {\em subflow} and associated {\em substorage matrix} functions. The corresponding {\em composite} \texttt{diact} {\em flow} and associated {\em storage matrix} functions generated by environmental inputs are ${T}^{\texttt{*}}(t) = \left( \tau^\texttt{*}_{i j}(t) \right)$ and ${X}^{\texttt{*}}(t) = \left( x^\texttt{*}_{i j}(t) \right)$, respectively \cite{Coskun2017NDP}.

The {\em simple} \texttt{dicat} {\em flows} and {\em storages} generated by single environmental inputs can be formulated in terms of their composite counterparts as follows:
\begin{equation}
\label{eq:simple_diact_in}
\begin{aligned}
{\tau}^\texttt{*}_{i_k}(t) = {\tau}^\texttt{*}_{i_k k_k}(t) \quad \mbox{and} \quad {x}^\texttt{*}_{i_k}(t) = {x}^\texttt{*}_{i_k k_k}(t) .
\end{aligned}
\end{equation}
To distinguish the composite and simple \texttt{diact} flow and storage matrices, we use a tilde notation over the simple versions. That is, the simple \texttt{diact} flow and storage matrices, for example, will be denoted by $\tilde{T}^{\texttt{*}}(t) = \left( \tau^\texttt{*}_{i_k}(t) \right)$ and $\tilde{X}^{\texttt{*}}(t) = \left( x^\texttt{*}_{i_k}(t) \right) $.

The {\em simple} \texttt{diact} {\em throughflow} and {\em compartmental storage matrices} and {\em vectors} can then be formulated as
\begin{equation}
\label{eq:comp_diact} 
\begin{aligned}
\mathcal{\tilde T}^\texttt{*}(t) = \diag{(\tilde {T}^\texttt{*}(t) \, \bm{1} )} \, \, \, & \Rightarrow \, \, \, 
\tilde {\tau}^\texttt{*}(t) = \mathcal{\tilde T}^\texttt{*}(t) \, \bm{1} 
\quad \mbox{and} \\
\mathcal{\tilde X}^\texttt{*}(t) = \diag{(\tilde {X}^\texttt{*}(t) \, \bm{1} )} \, \, \, & \Rightarrow \, \, \, 
\tilde{x}^\texttt{*}(t) = \mathcal{\tilde X}^\texttt{*}(t) \, \bm{1} .
\end{aligned}
\end{equation}
The composite counterparts of these quantities can similarly be formulated in parallel.

The difference between the composite and simple \texttt{diact} flows, $\tau^\texttt{*}_{ik}(t)$ and $\tau^\texttt{*}_{i_k}(t)$, and storages, $x^\texttt{*}_{ik}(t)$ and $x^\texttt{*}_{i_k}(t)$, is that the composite flow and storage from compartment $k$ to $i$ are generated by outward throughflow $\hat{\tau}_k(t,{ x}) - \hat{\tau}_{k_0}(t,{\rm x})$ derived from all environmental inputs and their simple counterparts from input-receiving subcompartment $k_k$ to $i_k$ are generated by outward subthroughflow $\hat{\tau}_{k_k}(t,{\rm x})$ derived from single environmental input $z_k(t, {x})$ (see Fig.~\ref{fig:utilityfigs}). In that sense, the composite and simple \texttt{diact} flows and storages measure the influence of one compartment on another induced by all and a single environmental input, respectively. The composite \texttt{diact} subflows and substorages within the initial subsystem, $\tau^\texttt{*}_{i_0 k_0}(t)$ and $x^\texttt{*}_{i_0 k_0}(t)$, from compartment $k$ to $i$ are then generated by outward throughflow $\hat{\tau}_{k_0}(t,{\rm x})$ derived from all initial stocks. 
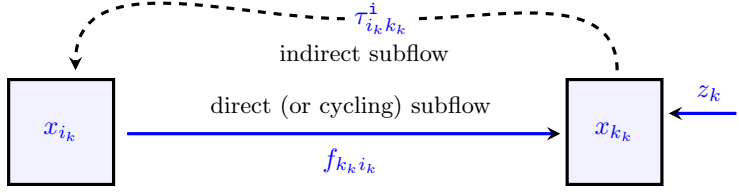
\begin{figure}[t]
\begin{center}
\begin{tikzpicture} [scale=.9]
   \draw[very thick, fill=blue!5, draw=black, text=blue] (-.05,-.05) rectangle node(R1) [pos=.5] {$ {x}_{i_k} $} (1.5,1.5) ;
   \draw[very thick, fill=blue!5, draw=black, text=blue] (8.2,-.05) rectangle node(R2) [pos=.5] { ${x}_{k_k}$ } (9.6,1.5) ;
    \draw[very thick,stealth-,draw=blue]  (9.7,1) -- (10.7,1) ;     
    \node (z) [text=blue] at (10.3,1.3) {$z_k$};       
    \draw[very thick,draw=blue,-stealth] (1.7,0.7) --  (8.1,0.7);  
    \node[blue] (x) at (5,.3) {${f}_{k_k i_k}$};  
   \node[text=black] at (5,1.1) { \small{direct (or cycling) subflow} };  
   \node[text=black] at (5.2,1.9) {\small{indirect subflow} };
   \node[blue,anchor=east] at (6,2.4) (c) {${\tau}^\texttt{i}_{i_k k_k}$};
   \node[blue,anchor=west] at (.8,1.5) (d) {};
   \draw[very thick,draw=blue, dashed, -stealth]  (c) edge[out=180,in=90] (d);       
   \node[blue,anchor=east] at (5.9,2.4) (h) {};
   \node[blue,anchor=west] at (8.8,1.5) (g) {};
   \draw[very thick,draw=blue, dashed]  (g) edge[out=90,in=0] (h); 
\end{tikzpicture}
\end{center}
\caption{Schematic representation for the complementary nature of the simple indirect and cycling flows within the $k^{th}$ subsystem. The composite direct subflow, $f_{k_k i_k}(t,{\rm x})$, is represented by solid arrow. This subflow also contributes to the simple cycling flow at subcompartment $k_k$. The simple indirect subflow, ${\tau}^\texttt{i}_{i_k k_k}(t)$, through other compartments (not shown) is represented by dashed arrow.}
\label{fig:cycindices}
\end{figure}

The simple and composite \texttt{diact} flows have been explicitly formulated for static systems through the system partitioning methodology in a recent study by \cite{Coskun2017SCSA}. This {\em static approach} will be extended to dynamic systems pointwise in time, that is, at each time step, in what follows. In addition to the {\em path-based approach} introduced above, this alternative approach to formulate the dynamic \texttt{diact} flows and storages will be called the {\em dynamic approach}.

The simple transfer flow from an input-receiving subcompartment $k_k$ to $i_k$ can be expressed as follows:
\begin{equation}
\label{eq:prop_ss1_tr}
\begin{aligned}
\tau^\texttt{t}_{i_k }(t) = \tau^\texttt{t}_{i_k k_k}(t) = \sum_{\substack{j=1}}^n  f_{i_k j_k}(t,{\rm x}) = {\check{\tau}}_{i_k} (t,{\rm x}) - z_{i_k}(t,{\rm x}) = {\tilde{\tau}}_{i_k} (t,{\rm x}) 
\end{aligned}
\end{equation} 
for $i,k=1,\ldots,n$. Note that the simple transfer flow at subcompartment $i_k$ is equal to the intercompartmental subthroughflow at that subcompartment. The simple direct flow from $k_k$ to $i_k$ is
\begin{equation}
\label{eq:propdf}
\begin{aligned}
\tau^\texttt{d}_{i_k}(t) = f_{i_k k_k}(t,{\rm x}) .
\end{aligned}
\end{equation}
The simple indirect flow from $k_k$ to $i_k$ at time $t$ can then be formulated as the simple transfer flow, $\tau^\texttt{t}_{i_k}(t)$, diminished by the simple direct flow, $\tau^\texttt{d}_{i_k}(t)$. That is,
\begin{equation}
\label{eq:propif}
\begin{aligned}
\tau^\texttt{i}_{i_k}(t) = \tau^\texttt{i}_{i_k k_k}(t) = \sum_{\substack{j=1 \\ j \neq k}}^n  f_{i_k j_k}(t,{\rm x}) = {\check{\tau}}_{i_k} (t,{\rm x}) - z_{i_k}(t,{\rm x}) - f_{i_k k_k}(t,{\rm x}) .
\end{aligned}
\end{equation}

Due to the reflexivity of the cycling flow, the simple cycling flow from an input-receiving subcompartment $k_k$ back into itself can be formulated in terms of the simple indirect or transfer flows as follows:
\begin{equation}
\label{eq:propif2}
\begin{aligned}
\tau^{\texttt{c}}_{k_k}(t) = \tau^{\texttt{c}}_{k_k k_k}(t) = \tau^\texttt{t}_{k_k}(t) = \tau^\texttt{i}_{k_k }(t) = \sum_{\substack{j=1}}^n  f_{k_k j_k}(t,{\rm x}) = {\check{\tau}}_{k_k}(t,{\rm x}) - z_{k_k}(t,{\rm x}) .
\end{aligned}
\end{equation}
That is, the simple cycling flow is the simple transfer or indirect flow from a subcompartment back into itself. The complementary nature of the cycling and indirect flows are schematized in Fig.~\ref{fig:cycindices}. The proportionality given in Eq.~\ref{eq:thr_dense3} implies that the simple (composite) cycling (sub)flow from an input-receiving (arbitrary) subcompartment $k_k$ ($i_k$) into $i_k$ is
\begin{equation}
\label{eq:thr_dense4}
\begin{aligned}
\tau^{\texttt{c}}_{i_k}(t) = \tau^{\texttt{c}}_{i_k k_k}(t) = \tau^{\texttt{c}}_{i_k i_k}(t) = 
\frac{ {\check{\tau}}_{i_i}(t,{\rm x}) - z_{i_i}(t,{\rm x}) } {{\hat{\tau}}_{i_i}(t,{\rm x}) } \, {\hat{\tau}}_{i_k}(t,{\rm x}) . 
\end{aligned}
\end{equation} 
The simple acyclic flow can, therefore, be formulated as
\begin{equation}
\label{eq:propif2_acyc1}
\begin{aligned}
\tau^{\texttt{a}}_{i_k}(t) & = \tau^{\texttt{a}}_{i_k k_k}(t) = {\tau}^{\texttt{t}}_{i_k}(t) - \tau^{\texttt{c}}_{i_k}(t) \\
& =  {\check{\tau}}_{i_k} (t,{\rm x}) - z_{i_k}(t,{\rm x})  - \frac{ {\check{\tau}}_{i_i}(t,{\rm x}) - z_{i_i}(t,{\rm x}) } {{\hat{\tau}}_{i_i}(t,{\rm x}) } \, {\hat{\tau}}_{i_k}(t,{\rm x}) .
\end{aligned}
\end{equation}
Note that the simple acyclic flow from subcompartment $k_k$ back into itself is zero:
\begin{equation}
\label{eq:propif2_acyc}
\begin{aligned}
\tau^{\texttt{a}}_{k_k}(t) = \tilde{\tau}_{k_k}(t) - \tau^{\texttt{c}}_{k_k}(t) = 0 .
\end{aligned}
\end{equation}
This means that there is, obviously, no acyclic flow from a subcompartment back into itself.

The proportionality of the parallel subflows and the corresponding subthroughflows and substorages formulated in Eq.~\ref{eq:thr_dense3New} can be expressed for the \texttt{diact} subflows as follows:
\begin{equation}
\label{eq:thr_dense3}
\begin{aligned}
 \tau^{\texttt{*}}_{i_\ell k_\ell}(t)  = \tau^{\texttt{*}}_{i_k k_k}(t)  \, \frac{{\hat{\tau}}_{k_\ell}(t,{\rm x}) }{{\hat{\tau}}_{k_k}(t,{\rm x}) } 
\end{aligned}
\end{equation} 
for $i,k=1,\ldots,n$ and $\ell=0,\ldots,n$, where the denominator is nonzero, ${\hat{\tau}}_{k_k}(t,{\rm x}) \neq 0$. Using this proportionality and the simple \texttt{diact} flows formulated above, the composite \texttt{diact} subflows from subcompartment $k_\ell$ to $i_\ell$ can also be formulated as follows:
\begin{equation}
\label{eq:comp_diact_subs}
\begin{aligned}
 \tau^{\texttt{d}}_{i_\ell k_\ell}(t) & =  \frac{ f_{i_k k_k}(t,{\rm x}) } {{\hat{\tau}}_{k_k}(t,{\rm x}) } \, {{\hat{\tau}}_{k_\ell}(t,{\rm x}) } = \frac{ f_{i k}(t,{x}) } {{\hat{\tau}}_{k}(t,{x}) } \, {{\hat{\tau}}_{k_\ell}(t,{\rm x}) } \\ 
 \tau^{\texttt{i}}_{i_\ell k_\ell}(t) & = \frac { {\check{\tau}}_{i_k} (t,{\rm x}) - z_{i_k}(t,{\rm x}) - f_{i_k k_k}(t,{\rm x}) } { {\hat{\tau}}_{k_k}(t,{\rm x}) } \, { {\hat{\tau}}_{k_\ell}(t,{\rm x}) } \\
 \tau^{\texttt{a}}_{i_\ell k_\ell}(t) & = \left[ \frac{ {\check{\tau}}_{i_k} (t,{\rm x}) - z_{i_k}(t,{\rm x}) } {{\hat{\tau}}_{k_k}(t,{\rm x}) } - \frac{ {\check{\tau}}_{i_i}(t,{\rm x}) - z_{i_i}(t,{\rm x}) } {{\hat{\tau}}_{i_i}(t,{\rm x}) } \, \frac{ {\hat{\tau}}_{i_k}(t,{\rm x}) } {{\hat{\tau}}_{k_k}(t,{\rm x}) }  \right] {{\hat{\tau}}_{k_\ell}(t,{\rm x}) } \\
 \tau^{\texttt{c}}_{i_\ell k_\ell}(t) & = \frac{ {\check{\tau}}_{i_i}(t,{\rm x}) - z_{i_i}(t,{\rm x}) } {{\hat{\tau}}_{i_i}(t,{\rm x}) } \, \frac{ {\hat{\tau}}_{i_k}(t,{\rm x}) } {{\hat{\tau}}_{k_k}(t,{\rm x}) }  \, {{\hat{\tau}}_{k_\ell}(t,{\rm x}) } \\
 \tau^{\texttt{t}}_{i_\ell k_\ell}(t) & = \frac{ {\check{\tau}}_{i_k} (t,{\rm x}) - z_{i_k}(t,{\rm x}) }  {{\hat{\tau}}_{k_k}(t,{\rm x}) } \, { {\hat{\tau}}_{k_\ell}(t,{\rm x}) }
\end{aligned}
\end{equation}
for $t > t_0$. Note that $\hat{\tau}_{k_k}(t_0,{\rm x})=0$ and we assume that $\hat{\tau}_{k_k}(t,{\rm x})$ is nonzero for all $t > t_0$. The second equality of the first equation in Eq.~\ref{eq:comp_diact_subs} for the composite direct subflow is due to the equivalence of flow and subflow intensities in the same direction, as formulated in Eq.~\ref{eq:eqiv_intense} \cite{Coskun2017NDP,Coskun2017SCSA}. 

The composite \texttt{diact} flows generated by environmental inputs then become 
\begin{equation}
\label{eq:props}
\begin{aligned}
{\tau}^{\texttt{*}}_{ik}(t) & = \sum_{\ell=1}^{n} {\tau}^\texttt{*}_{i_\ell k_\ell} (t) 
= \frac{ \tau^{\texttt{*}}_{i_k}(t) } {{\hat{\tau}}_{k_k}(t,{\rm x}) } 
\, \sum_{\ell=1}^{n} \hat{\tau}_{k_\ell} (t,{\rm x})
= n^\texttt{*}_{ik} (t) \, \left(\hat{\tau}_{k} (t,{x}) - \hat{\tau}_{k_0} (t,{\rm x}) \right)
\end{aligned}
\end{equation}
where the \texttt{diact} {\em flow distribution factor}, $n^{\texttt{*}}_{ik}(t)$, is $n^{\texttt{*}}_{ik}(t)  = { \tau^{\texttt{*}}_{i_k}(t) } / {{\hat{\tau}}_{k_k}(t,{\rm x}) } $. The dynamic \texttt{diact} {\em flow distribution matrix} function can be defined as $N^{\texttt{*}}(t) = ( n^{\texttt{*}}_{ik}(t) )$. The dynamic \texttt{diact} flow distribution matrices, as well as the dynamic simple and composite \texttt{diact} flow matrices are explicitly formulated in Table~\ref{tab:flow_stor1} based on their componentwise definitions in Eq.~\ref{eq:comp_diact_subs}, similar to their static counterparts introduced by \cite{Coskun2017SCSA}. The inverted matrices in the table are assumed to be invertible. For the sake of readability, the function arguments are dropped in the table.
\begin{table}
     \centering
     \caption{The dynamic \texttt{diact} flow distribution and the simple and composite \texttt{diact} (sub)flow matrices. The superscript ($^\texttt{*}$) in each equation represents any of the \texttt{diact} symbols. For the sake of readability, the function arguments are dropped.}
     \label{tab:flow_stor1}
     \begin{tabular}{c p{6cm} l }
     \hline
\texttt{diact} & {flow distribution matrix} & {flows} \\
     \hline
     \noalign{\vskip 2pt}
\texttt{d} & $ N^\texttt{d} =  F \, \mathcal{T}^{-1}  $  &
\multirowcell{5}{ 
$
\begin{aligned}
\hfill
{T}^\texttt{*} &= {N}^\texttt{*} \, (\mathcal{T} -  \mathcal{\hat T}_0) \\ 
{T}^\texttt{*}_\ell &= {N}^\texttt{*} \, \mathcal{\hat T}_\ell \\ 
\tilde{T}^\texttt{*} &= {N}^\texttt{*} \, \mathsf{\hat T}  
\end{aligned}
$
} 
\\
\texttt{i} & $ N^\texttt{i} = \displaystyle \tilde{T} \, \hat{\mathsf{T}}^{-1} - F \,\mathcal{T}^{-1}  $ &  \\
\texttt{a} & $ {N}^\texttt{a} = \displaystyle \tilde{T} \, \hat{\mathsf{T}}^{-1} - \tilde{\mathsf{T}} \, \hat{\mathsf{T}}^{-1} \, \hat{T} \, \hat{\mathsf{T}}^{-1} $ & \\
\texttt{c} & $ {N}^\texttt{c} = \displaystyle \tilde{\mathsf{T}} \, \hat{\mathsf{T}}^{-1} \, \hat{T} \, \hat{\mathsf{T}}^{-1} $
& \\
\texttt{t} & $ N^\texttt{t} = \displaystyle  \tilde{T} \, \hat{\mathsf{T}}^{-1}   $ & \\
\noalign{\vskip 1pt}
\hline
     \end{tabular}
\end{table}

The composite \texttt{diact} substorages can also be formulated using the corresponding \texttt{diact} subflows as transient inflows in Eq.~\ref{eq:out_in_fs2} as follows:
\begin{equation}
\label{eq:out_in_diact2}
\begin{aligned}
\dot{x}^{\texttt{*}}_{i_\ell k_\ell}(t) & = \tau^{\texttt{*}}_{i_\ell k_\ell}(t) - \frac{ {\hat{\tau}}_{i}(t,{x}) }{ x_{i}(t) } \, {x}^{\texttt{*}}_{i_\ell k_\ell}(t) , \quad {x}^{\texttt{*}}_{i_\ell k_\ell}(t_1) = 0 
\end{aligned}
\end{equation}
for $t_1 > t_0$, $i,k=1,\ldots,n$ and $\ell=0,\ldots,n$. The solution to this governing equation, ${x}^{\texttt{*}}_{i_\ell k_\ell}(t)$, represents the \texttt{diact} substorage in subcompartment $i_\ell$ at time $t \geq t_1$ generated by the corresponding \texttt{diact} subflow, $\tau^{\texttt{*}}_{i_\ell k_\ell}(t)$, during $[t_1,t]$ (see Fig.~\ref{fig:subsystemp}).

The analytic solutions for linear systems are introduced in Section~\ref{appsec:lin}. In this case, the governing equation for the \texttt{diact} substorages, Eq.~\ref{eq:out_in_diact2}, can be solved explicitly for $x^\texttt{*}_{i_\ell k_\ell}(t)$ as well. The solution becomes
\begin{equation}
\label{eq:sln_transient}
\begin{aligned}
x^\texttt{*}_{i_\ell k_\ell}(t) = \int_{t_1}^t {\rm e}^{-\int_s^t r_i^{-1}(s',{x}) \, ds'} \, \tau^\texttt{*}_{i_\ell k_\ell} (s) \, ds 
\end{aligned}
\end{equation}
where $r_i^{-1}(t,{x}) = { {\hat{\tau}}_{i}(t,{x}) } / { x_{i}(t) }$ is the outward throughflow intensity function.
  
The supplementary relationship between the simple direct and indirect subflows given in Eq.~\ref{eq:propif} can be expressed in matrix form, in terms of both flows and storages, as follows:
\begin{equation}
\label{eq:tr_fs}
\begin{aligned}
\tilde{T}^{\texttt{t}}(t) = \tilde{T}^\texttt{d}(t) + \tilde{T}^\texttt{i}(t) \quad \mbox{and} \quad  \tilde{X}^{\texttt{t}}(t)  = \tilde{X}^\texttt{d}(t) + \tilde{X}^\texttt{i}(t) .
\end{aligned}
\end{equation}
A similar supplementary relationship can be formulated between the simple cycling and acyclic flows and storages:
\begin{equation}
\label{eq:TtTad} 
\begin{aligned}
\tilde{T}^\texttt{t}(t) = \tilde{T}^\texttt{c}(t) + \tilde{T}^\texttt{a}(t) \quad \mbox{and} \quad \tilde{X}^\texttt{t}(t) = \tilde{X}^\texttt{c}(t) + \tilde{X}^\texttt{a}(t) ,
\end{aligned}
\end{equation}
due to Eq.~\ref{eq:propif2_acyc1}. The reflexivity of the simple cycling flows and storages, formulated in Eq.~\ref{eq:propif2}, can also be written in matrix form, as follows:
\begin{equation}
\label{eq:dyn_cyc} 
\begin{aligned}
\diag{( \tilde{T}^\texttt{c}(t) )} &= \diag{( \tilde{T}^{\texttt{t}}(t) )}= \diag{( \tilde{T}^\texttt{i}(t) )} , \\
\diag{( \tilde{X}^\texttt{c}(t) )} & = \diag{( \tilde{X}^{\texttt{t}}(t) )} = \diag{( \tilde{X}^\texttt{i}(t) )} .
\end{aligned}
\end{equation}
The matrix relationships formulated in Eqs.~\ref{eq:tr_fs},~\ref{eq:TtTad}, and~\ref{eq:dyn_cyc} can similarly be expressed for the composite \texttt{diact} flows and storages.

\subsection{Static subsystem partitioning and \texttt{diact} transactions}
\label{apxsec:ssd}

The static version of the dynamic subsystem partitioning methodology given in Eqs.~\ref{eq:out_in_fs} and~\ref{eq:out_in_fs2} has recently been formulated by \cite{Coskun2017SCSA}.  This static partitioning is summarized in this section. 

Since time derivatives are zero at steady state, we set $\dot{x}^{w}_{n_k \ell_k j_k}(t) =0$ in Eq.~\ref{eq:out_in_fs2}. The static transient outflow at subcompartment $\ell_k$ along subflow path $p^w_{n_k j_k}= i_k \mapsto j_k \to \ell_k \to n_k$ from $j_k$ to $n_k$, $f^{w}_{n_k \ell_k j_k}$, and the associated transient substorage generated in $\ell_k$, $x^{w}_{n_k \ell_k j_k}$, by the transient inflow, $f^{w}_{\ell_k j_k i_k}$, are then formulated as follows:
\begin{equation}
\label{eq:out_in_fs2_SS}
\begin{aligned}
{x}^w_{n_k \ell_k j_k} =
 \frac{ x_{\ell} }{ \tau_{\ell} }  \,  f^w_{\ell_k j_k i_k} 
\quad \mbox{and}
\quad f^w_{n_k \ell_k j_k} 
= \frac{ f_{n \ell} }{x_{\ell} }  \, {x}^w_{n_k \ell_k j_k}
= \frac{ f_{n \ell}  }{ \tau_{\ell} }  \, f^w_{\ell_k j_k i_k}  .
\end{aligned}
\end{equation}
The second equality for $f^w_{n_k \ell_k j_k}$ is obtained by using the first equation for ${x}^w_{n_k \ell_k j_k}$ in the first equality. The relationships in Eq.~\ref{eq:out_in_fs2_SS} establish the foundation of the {\em static subsystem partitioning}. 

Through the system partitioning methodology, the static $\texttt{diact}$ flows and storages are formulated in matrix form by \cite{Coskun2017SCSA}, as presented in Table~\ref{tab:flow_stor}. Note that the matrix $\mathcal{N}$ used in the table is defined as $\mathcal{N} =\operatorname{diag}(N )$.
\begin{table}
     \centering
     \caption{The input-oriented, flow-based \texttt{diact} flow and storage distribution and the simple and composite \texttt{diact} (sub)flow and (sub)storage matrices. The superscript ($^\texttt{*}$) in each equation represents any of the \texttt{diact} symbols.}
     \label{tab:flow_stor}
\resizebox{\linewidth}{!}{%
     \begin{tabular}{c l l l l }
     \hline
\texttt{diact} & \multicolumn{2}{c} {flow and storage distribution matrices} & \multicolumn{1}{c} {flows} & \multicolumn{1}{c} {storages} \\ 
     \hline
     \noalign{\vskip 2pt} 
\texttt{d} & $ N^\texttt{d} =  F \, \mathcal{T}^{-1}  $ & \multirow{5}{*}{ ${S}^\texttt{*} = \mathcal{R} \, {N}^\texttt{*}$ }  & 
\multirowcell{5}{ $ {T}^\texttt{*} = {N}^\texttt{*} \, \mathcal{T} $ \\ $ {T}^\texttt{*}_\ell = {N}^\texttt{*} \, \mathcal{T}_\ell $ \\ $ \tilde{T}^\texttt{*} = {N}^\texttt{*} \, \mathsf{T} $ } 
& \multirowcell{5}{ $ {X}^\texttt{*} = {S}^\texttt{*} \, \mathcal{T} $ \\ $ {X}^\texttt{*}_\ell = {S}^\texttt{*} \, \mathcal{T}_\ell $ \\ $ \tilde{X}^\texttt{*} = {S}^\texttt{*} \, \mathsf{T} $ } \\
\texttt{i} & $ N^\texttt{i} = \displaystyle (N-I) \, \mathcal{N}^{-1} - F \,\mathcal{T}^{-1}  $ & & & \\
\texttt{a} & $ {N}^\texttt{a} = \displaystyle (\mathcal{N}^{-1} \, N - I) \, \mathcal{N}^{-1} $ & & & \\
\texttt{c} & $ {N}^\texttt{c} = \displaystyle (N - \mathcal{N}^{-1} N) \, \mathcal{N}^{-1} $ & & & \\
\texttt{t} & $ N^\texttt{t} = \displaystyle (N-I) \, \mathcal{N}^{-1}   $ & & & \\
\noalign{\vskip 1pt} 
\hline     
     \end{tabular}
     }
\end{table}

\subsection{System measures and indices}
\label{sec:dsami}

The dynamic system partitioning methodology yields the subthroughflow and substorage vectors and matrices that measure the influence of the initial stocks and environmental inputs on system compartments in terms of the flow and storage generation. These vector and matrix measures enable tracking the evolution of the initial stocks, environmental inputs, as well as the associated storages sourced from these stocks and inputs individually and separately within the system. For the analysis of intercompartmental flow and storage dynamics, the dynamic subsystem partitioning methodology then formulates the transient and dynamic $\texttt{diact}$ flows and associated storages. The transient and \texttt{diact} transactions enable tracking of arbitrary intercompartmental flows and storages along a particular and all possible flow paths within the system and determining the influence of system compartments directly and indirectly on one another.

The proposed methodology constructs a foundation for the development of new mathematical system analysis tools as quantitative ecosystem indicators. In addition to the measures summarized above, multiple novel dynamic and static system analysis tools of matrix, vector, and scalar types have recently been introduced in separate works, based on the proposed methodology \cite{Coskun2017DESM,Coskun2017SESM}. More specifically, these manuscripts introduce measures and indices for the \texttt{diact} effects, utilities, exposures, and residence times, as well as the system efficiencies, stress, and resilience. 

\subsection{Quantitative classification of interspecific interactions} 
\label{sec:qdii}

An immediate ecological application of the proposed methodology is the quantitative analysis of food webs and chains. The system compartments of a food web ecosystem represent species, the conserved quantity in question becomes nutrient or energy, and flow paths correspond to food chains in the web. In this setting, direct flow rate $f_{ik}(t,x)$ represents an interspecific interaction, such as predation, between species $i$ and $k$ and measures the rate of nutrient or energy flow from species $k$ in a lower trophic level to $i$ in the next level at time $t$. Nutrient or energy stored in species $i$ through all trophic interactions is represented by $x_i(t)$. 

Community ecology classifies interspecific interactions qualitatively using network topology without regard for system flows \cite{Menge1995,Menge1997}. This structural determination, however, gets more complicated, if at all possible, with the increasing complexity of intricate food webs \cite{Holt1997,Wootton1994}. Multiple food chains of potentially different lengths between two species, for example, disallow the classification based on the length of the chains \cite{Patten2007}. A mathematical characterization and classification technique for the analysis of the nature and strength of food chains has recently been proposed by \cite{Coskun2017SCSA} for static systems. This section introduces a dynamic version of this technique with slight modifications.

The proposed methodology can quantitatively determine the net benefit in terms of flow and storage transfers received by the involved species from each other. The {\em sign analysis} of the \texttt{diact} {\em interspecific interactions} determines the neutral and antagonistic nature of the interactions\textemdash whether the interaction is beneficial or harmful to the species involved. The {\em strength analysis} then quantifies the strength of these \texttt{diact} interactions. The {\em sign} and {\em strength} of the \texttt{diact} {\em interactions} induced by environmental inputs between species $i$ and $j$ will be defined respectively as follows:
\begin{equation}
\label{eq:iidm}
\delta^\texttt{*}_{ij}(t) = \sgn{(\tau^\texttt{*}_{ij}(t) - \tau^\texttt{*}_{ji}(t) )} \quad \mbox{and} \quad \mu^\texttt{*}_{ij}(t) = 
\frac{| \tau^\texttt{*}_{ij}(t) - \tau^\texttt{*}_{ji}(t) | } { \check{\tau}_{i}(t,x) + \check{\tau}_{j}(t,x) } 
\end{equation} 
where $\sgn(\cdot)$ is the sign function, and the superscript ($^\texttt{*}$) represents any of the \texttt{diact} symbols. Following the convention of community ecology, instead of $(+1)$ and $(-1)$, $(+)$ and $(-)$ notations will be used for the sign of the \texttt{diact} interactions. The strength, $0 \leq \mu^\texttt{*}_{ij}(t) \leq 1$, is defined to be zero, if both terms in its denominator are zero. 

For the analysis of \texttt{diact} interactions ranging from the individual and local to the system-wide and global scale, the strength of the interactions can be formulated with the normalization by $\tau^\texttt{*}_{ij}(t) + \tau^\texttt{*}_{ji}(t)$, $\tau^\texttt{t}_{ij}(t) + \tau^\texttt{t}_{ji}(t)$, $\check{\tau}_{i}(t,x) + \check{\tau}_{j}(t,x)$, as in Eq.~\ref{eq:iidm}, or $\check{\sigma}^{\tau}(t) = \bm{1}^T \, \check{\tau}(t,x)$ in the given order, where $\check{\sigma}^{\tau}(t)$ will be called the {\em total inward system throughflow}. As an example, for the analysis of local \texttt{diact} interactions at the global scale, $\mu^\texttt{*}_{ij}(t)$ can be defined as $\mu^\texttt{*}_{ij}(t) = {| \tau^\texttt{*}_{ij}(t) - \tau^\texttt{*}_{ji}(t) | } / { \check{\sigma}^{\tau}(t) }$.

The \texttt{diact} {\em neutral relationship} between species $i$ and $j$ and ``{\em predation}'' of species $i$ on $j$ can quantitatively be characterized, respectively, as follows:
\begin{equation}
\label{eq:ii}
\begin{aligned}
\tau^\texttt{*}_{ij}(t) - \tau^\texttt{*}_{ji}(t) =0 \quad & \Rightarrow \quad \delta^\texttt{*}_{ij}(t) = (0) , \\
\tau^\texttt{*}_{ij}(t) - \tau^\texttt{*}_{ji}(t) > 0 \quad &  \Rightarrow \quad \delta^\texttt{*}_{ij}(t) = (+) .
\end{aligned}
\end{equation} 
This classification can also be extended to the {\em transient interactions} between two species along a given food chain using the transient inflows. We will use the notations $\delta^w_{ij}(t)$ and $\mu^w_{ij}(t)$ for the sign and strength of the net flow from compartment $j$ indirectly to $i$ through trophic interactions along a given food chain, $p^w_{i j}$.

The classification of the \texttt{diact} interactions induced by the initial stocks can be determined by using the composite \texttt{diact} subflows for the initial subsystem, $\tau^\texttt{*}_{i_0j_0}(t)$, in Eqs.~\ref{eq:iidm} and~\ref{eq:ii}. The storage-based quantitative definition of the \texttt{diact} interactions can be formulated in parallel by substituting the \texttt{diact} storages for the corresponding \texttt{diact} flows in the definitions above as well. The storage-based formulations represent the history of the \texttt{diact} interactions during $[t_1,t]$ while the flow-based formulations represent simultaneous interactions at time $t$. We will use the superscript $x$ to distinguish the storage-based measures, $\delta^{\texttt{*},x}_{ij}(t)$ and $\mu^{\texttt{*},x}_{ij}(t)$. For the classification of the \texttt{diact} interactions induced by individual environmental inputs, the simple \texttt{diact} flows and storages can be used instead of their composite counterparts in Eqs.~\ref{eq:iidm} and~\ref{eq:ii}. A tilde notation will be used over the measures for simple \texttt{diact} interspecific interactions, $\tilde{\delta}^\texttt{*}_{ij}(t)$ and $\tilde{\mu}^\texttt{*}_{ij}(t)$. 

A mathematical technique for the dynamic characterization and classification of the main interspecific interaction types, such as neutralism, mutualism, commensalism, competition, and exploitation, has also been developed recently in a separate paper \cite{Coskun2017DESM}.

\section{Results}
\label{sec:results}

The proposed dynamic methodology is applied to a linear and nonlinear dynamic ecosystem models. Numerical results for the system analysis tools developed in this manuscript, such as the substorage and subthroughflow matrix measures, as well as the transient and dynamic \texttt{diact} flows and storages, are presented in this section.

The results indicate that the proposed methodology precisely quantifies system functions, properties, and behaviors, enables tracking the evolution of the initial stocks, environmental inputs, and intercompartmental flows, as well as associated storages individually and separately within the system, is sensitive to perturbations due to even a brief unit impulse, and, thus, can be used for rigorous dynamic analysis of nonlinear ecological systems. It is worth noting, however, that this present work proposes a mathematical {\em method}\textemdash a systematic technique designed for solving and analyzing any nonlinear dynamic compartmental model\textemdash and it is not itself a {\em model}. Therefore, we focus more on demonstrating the efficiency and wide applicability of the method. It is expected that once the method is accessible to a broader community of environmental ecologists, it can be used for ecological inferences and the holistic analysis of specific models of interest.

\subsection{Case study}
\label{appsec:ex_hippe}

A linear dynamic ecosystem model introduced by \cite{Hippe1983} is analyzed through the proposed methodology in this case study to demonstrate the capability of the method to analytically solve linear systems with time-dependent inputs. The graphical representations of the results are presented.

\begin{figure}[t]
\begin{center}
\begin{tikzpicture}
\centering
   \draw[very thick,  fill=blue!5, draw=black] (-.05,-.05) rectangle node(R1) {$x_1(t)$} (1.5,1.5) ;
   \draw[very thick,  fill=blue!5, draw=black] (3.95,-.05) rectangle node(R2) {$x_2(t)$} (5.5,1.5) ;   
       \draw[very thick,-stealth,draw=red, line width=5pt, opacity=.3]  (1.6,1) -- (3.8,1) -- (4.5,1) -- (4.5,.5) -- (1.6,.5) ;
       \node (p) [text=red] at (3,1.5) {$p^1_{1_1}$};            
       \draw[very thick,stealth-,draw=black]  (1.6,.5) -- (3.8,.5) ;
       \draw[very thick,-stealth,draw=black]  (1.6,1) -- (3.8,1) ;  
       \draw[very thick,stealth-,draw=black]  (5.6,.5) -- (6.6,.5) ;
       \draw[very thick,-stealth,draw=black]  (5.6,1) -- (6.6,1) ;  
       \draw[very thick,stealth-,draw=black]  (-1.2,.5) -- (-0.15,.5) ;
       \draw[very thick,-stealth,draw=black]  (-1.2,1) -- (-0.15,1) ;  
\end{tikzpicture}
\end{center}
\caption{Schematic representation of the model network. Subflow path $p^1_{1_1}$, along which the cycling flow and storage functions are computed, is red (subsystems are not shown) (Case study~\ref{appsec:ex_hippe}).} 
\label{fig:hippe_diag}
\end{figure}
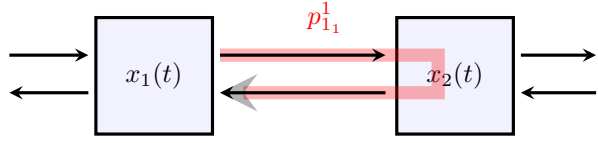

The model has two compartments, $x_1(t)$ and $x_2(t)$ (see Fig.~\ref{fig:hippe_diag}). The system flows are described as
\begin{equation}
\label{eq:hippe_flows}
\begin{aligned} 
F(t,x) = 
\begin{bmatrix}
    0  & \frac{2}{3} x_2(t)  \\
    \frac{4}{3} x_1(t)  & 0 \\
\end{bmatrix},
\quad
z(t,x) = 
\begin{bmatrix}
    z_1(t) \\
    z_2(t) 
\end{bmatrix},
\quad 
y(t,x) = 
\begin{bmatrix}
    \frac{1}{3} x_1(t) \\
    \frac{5}{3} x_2(t)
\end{bmatrix} .
\nonumber
\end{aligned}
\end{equation}
The governing equations take the following form:
\begin{equation}
\label{eq:hippe_model}
\begin{aligned} 
\dot x_1(t) & = z_1(t) + \frac{2}{3} x_2(t) - \left(\frac{4}{3} + \frac{1}{3} \right) \, x_1(t) \\
\dot x_2(t) & = z_2(t) + \frac{4}{3} x_1(t) - \left(\frac{2}{3} + \frac{5}{3} \right) \, x_2(t)
\end{aligned} 
\end{equation}
with the initial conditions $[x_{1,0}, x_{2,0}]^T = [3,3]^T$.

The subcompartmentalization step yields the substate variables that represent the substorage values as follows: 
\[ x_{1_k}(t) \quad \mbox{and} \quad x_{2_k}(t)
\quad \mbox{with} \quad 
x_i(t) = \sum_{k=0}^2 x_{i_k} (t) . \]
The flow partitioning then yields the subflows for the subsystems:
\begin{equation}
\label{eq:hippe_flows_SC}
\begin{aligned} 
F_k(t,{\rm x}) = 
\begin{bmatrix}
    0  & \frac{2}{3} \, d_{2_k} \, x_2 \\
    \frac{4}{3} d_{1_k} \, x_1  & 0 \\
\end{bmatrix},
\quad
\check{z}_k(t,{\rm x}) = 
\begin{bmatrix}
    \delta_{1k} \, z_1 \\
    \delta_{2k} \, z_2
\end{bmatrix},
\quad 
\hat{y}_k(t,{\rm x}) = 
\begin{bmatrix}
    \frac{1}{3} d_{1_k} \, x_1 \\
    \frac{5}{3} \, d_{2_k} \, x_2
\end{bmatrix} ,
\end{aligned}
\nonumber
\end{equation}
where the decomposition factors $d_{i_k}({\rm x})$ are defined by Eq.~\ref{eq:cons}. The dynamic system partitioning methodology then yields the following governing equations for the decomposed system: 
\begin{equation}
\label{eq:hippe_sc}
\begin{aligned} 
\dot x_{1_k}(t) & = z_{1_k}(t) + \frac{2}{3} x_{2_k}(t) - \left(\frac{4}{3} + \frac{1}{3} \right) \, x_{1_k}(t) \\
\dot x_{2_k}(t) & = z_{2_k}(t) + \frac{4}{3} x_{1_k}(t) - \left(\frac{2}{3} + \frac{5}{3} \right) \, x_{2_k}(t)
\end{aligned}
\nonumber
\end{equation}
with the initial conditions 
\begin{equation}
\label{eq:hippe_ic}
x_{i_k} (t_0) = \left \{
\begin{aligned}
3, \quad k=0 \\
0, \quad k \neq 0
\end{aligned}
\right.
\nonumber
\end{equation}
for $i = 1,2$. There are $n \times (n+1) = 2 \times 3 = 6$ equations in the system. 

The system can be written in matrix form as formulated in Eq.~\ref{eq:model_lin}:
\begin{equation}
\label{eq:model_Mlin}
\begin{aligned}
\dot{X}(t) & = \mathcal{Z} (t) + A \, X(t) ,
\quad X(t_0) = \mathbf{0} \\
\dot{{x}}_{0}(t) & = A \, x_0(t) ,
\quad \quad \quad \quad  {x}_{0} (t_0) = x_{0}
\end{aligned} 
\end{equation} 
where the constant flow intensity matrix $A$, as defined in Eq.~\ref{eq:matrix_A_apx}, becomes
\begin{equation}
\label{eq:hippe_matAS}
A = 
\begin{bmatrix}
  -({4}/{3} + {1}/{3} ) & {2}/{3} \\ 
  {4}/{3} &  -({2}/{3} + {5}/{3} )
\end{bmatrix} 
=
\begin{bmatrix}
   -{5}/{3}  & {2}/{3}  \\ 
  {4}/{3} &  -{7}/{3} 
\end{bmatrix} .
\nonumber
\end{equation}

The governing decomposed system, Eq.~\ref{eq:model_Mlin}, is linear. We can, therefore, solve it analytically, as formulated in Section~\ref{appsec:lin}. Since the flow intensity matrix $A$ is constant, we have the following fundamental matrix solution as given in Eq.~\ref{eq:model_linfund_sln}:
\begin{equation}
\label{eq:fund_sln}
\begin{aligned} 
V(t) = \left[
\begin{array}{cc} 
\frac{2\,{\mathrm{e}}^{-t}}{3}+\frac{{\mathrm{e}}^{-3\,t}}{3} & 
\frac{{\mathrm{e}}^{-t}}{3}-\frac{{\mathrm{e}}^{-3\,t}}{3} \\ 
\frac{2\,{\mathrm{e}}^{-t}}{3}-\frac{2\,{\mathrm{e}}^{-3\,t}}{3} & 
\frac{{\mathrm{e}}^{-t}}{3}+\frac{2\,{\mathrm{e}}^{-3\,t}}{3} 
\end{array}
\right] .
\end{aligned} 
\nonumber
\end{equation}
For $z = [1,1]^T$, the solutions for the matrix equation, Eq.~\ref{eq:model_Mlin}, then become
\begin{equation}
\label{eq:XXb_sln}
\begin{aligned} 
X(t) & =
\left[
\begin{array}{cc} 
\frac{7}{9} - \frac{{\mathrm{e}}^{-3\,t}}{9}-\frac{2\,{\mathrm{e}}^{-t}}{3} & 
\frac{2}{9} + \frac{{\mathrm{e}}^{-3\,t}}{9}-\frac{{\mathrm{e}}^{-t}}{3} \\ 
\frac{4}{9} + \frac{2\,{\mathrm{e}}^{-3\,t}}{9}-\frac{2\,{\mathrm{e}}^{-t}}{3} & 
\frac{5}{9} - \frac{2\,{\mathrm{e}}^{-3\,t}}{9}-\frac{{\mathrm{e}}^{-t}}{3} 
\end{array}
\right] ,  \quad 
x_0(t) = 
\left[
\begin{array}{c} 
3 \, {\mathrm{e}}^{-t} \\ 
3 \, {\mathrm{e}}^{-t} 
\end{array}
\right] ,
\end{aligned} 
\end{equation}
as given in Eq.~\ref{eq:model_lin_sln2}. Therefore, the solution to the original system, Eq.~\ref{eq:hippe_model}, in vector form, is
\begin{equation}
\label{eq:vec_sln}
\begin{aligned} 
x(t) & = x_0(t) + X(t) \, \bm{1} =
\left[
\begin{array}{c} 
x_1(t) \\ 
x_2(t) 
\end{array}
\right]
= 
\left[
\begin{array}{c} 
2\,{\mathrm{e}}^{-t}+1 \\ 
2\,{\mathrm{e}}^{-t}+1 
\end{array}
\right] .
\end{aligned} 
\nonumber
\end{equation}
The subthroughflow matrices, $\check{T}(t)$ and $\hat{T}(t)$, can also be expressed as formulated in Eq.~\ref{eq:Tmatrix2}, using the solution for the substorage matrix, $X(t)$.

The steady state solutions can also be computed as formulated in Eq.~\ref{eq:model_ss}:
\begin{equation}
\label{eq:hippe_sc_ex3}
\begin{aligned} 
X & = - A^{-1} = \mathcal{X} \, \left ( \mathcal{T} - F \right )^{-1} = 
\left[
\begin{array}{cc} 
{7}/{9} & {2}/{9} \\ 
{4}/{9} & {5}/{9} 
\end{array}
\right]
\quad \mbox{and} \quad x_0 = \bm{0} .
\end{aligned} 
\end{equation}
It can easily be seen that, this steady-state solution is the same as the limit of the dynamic solution, Eq.~\ref{eq:XXb_sln}, as $t$ tends to infinity. That is, $\lim_{t \rightarrow \infty} X(t) = X$.

We also analyze the system with a time dependent input $z(t) = [3+\sin(t),3+\sin(2 t)]^T$. Similar computations with the same fundamental matrix, $V(t)$, lead us to the following initial substorage vector, $x_{i_0}(t)$, and substorage matrix components, $x_{i_k}(t)$:
\begin{equation}
\label{eq:hippe_pedX}
\begin{aligned} 
x_{1_0}(t) & = x_{2_0}(t) =  3\,{\mathrm{e}}^{-t} , \\ 
x_{1_1}(t) & = \frac{7}{3} - \frac{11\,\cos\left(t\right)}{30} + \frac{13\,\sin\left(t\right)}{30} -\frac{5\,{\mathrm{e}}^{-t}}{3} - \frac{3\,{\mathrm{e}}^{-3\,t}}{10} , \\
x_{1_2}(t) & = \frac{2}{3} - \frac{16\,\cos\left(2\,t\right)}{195} - \frac{2\,\sin\left(2\,t\right)}{195} - \frac{13\,{\mathrm{e}}^{-t}}{15} + \frac{11\,{\mathrm{e}}^{-3\,t}}{39} , \\ 
x_{2_1}(t) & = \frac{4}{3} -\frac{4\,\cos\left(t\right)}{15} + \frac{2\,\sin\left(t\right)}{15} -\frac{5\,{\mathrm{e}}^{-t}}{3} + \frac{3\,{\mathrm{e}}^{-3\,t}}{5} , \\
x_{2_2}(t) & = \frac{5}{3} -\frac{46\,\cos\left(2\,t\right)}{195} + \frac{43\,\sin\left(2\,t\right)}{195} - \frac{13\,{\mathrm{e}}^{-t}}{15} - \frac{22\,{\mathrm{e}}^{-3\,t}}{39} .
\end{aligned} 
\end{equation}  
Using these solutions, we can express the solutions to the original system, Eq.~\ref{eq:hippe_model}, as
\[ x_1(t) = \sum_{k=0}^{2} x_{1_k} (t) \quad \mbox{and} \quad x_2(t) = \sum_{k=0}^{2} x_{2_k}(t) . \]
The proposed dynamic method solves linear systems analytically. Explicit solutions can be used to compute the quantities in question at any time $t$.

\begin{figure}[t]
    \centering
    \begin{subfigure}[b]{0.46\textwidth}
        \includegraphics[width=\textwidth]{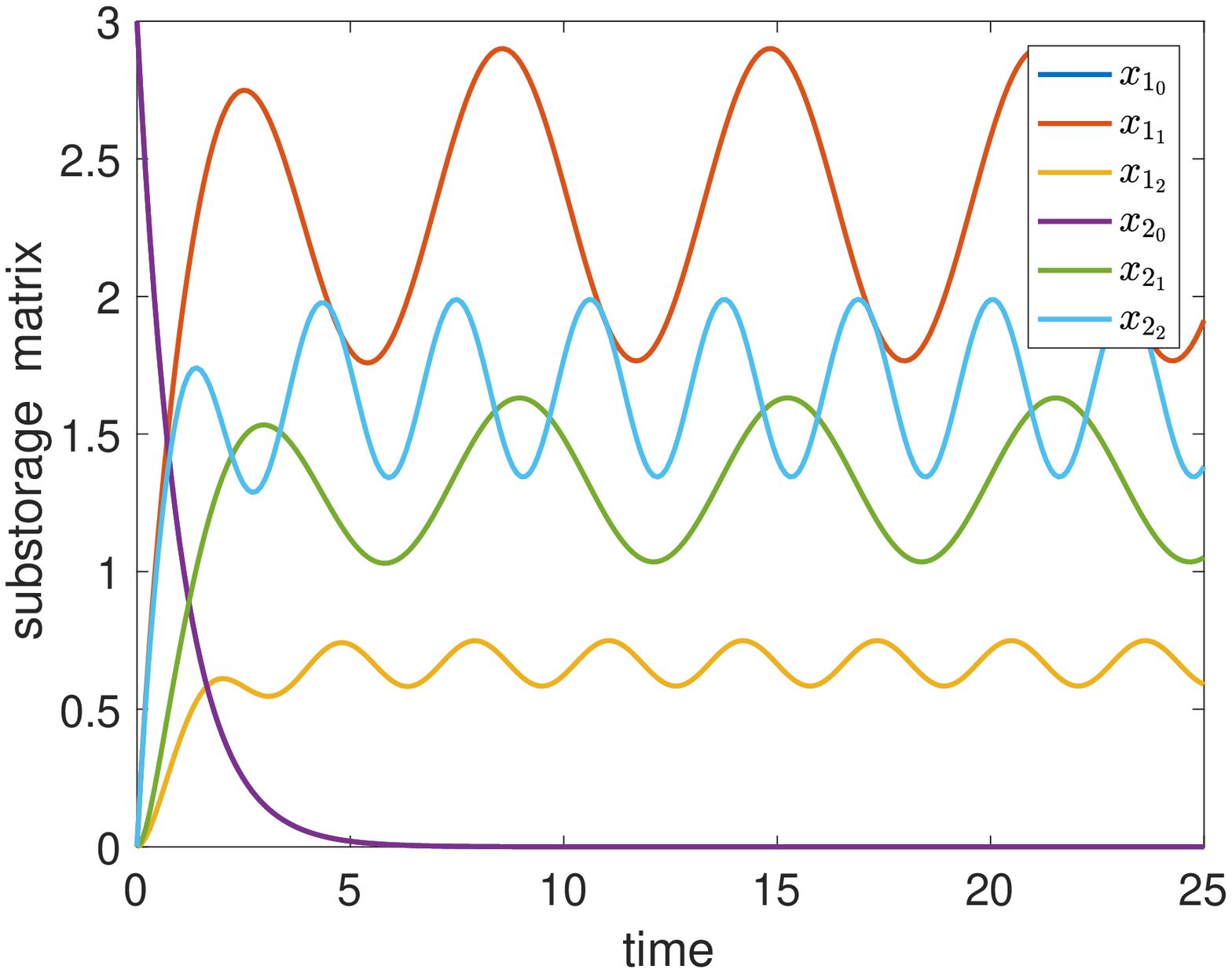}
        \caption{$X(t)$}
        \label{fig:X}
    \end{subfigure}
    \begin{subfigure}[b]{0.45\textwidth}
        \includegraphics[width=\textwidth]{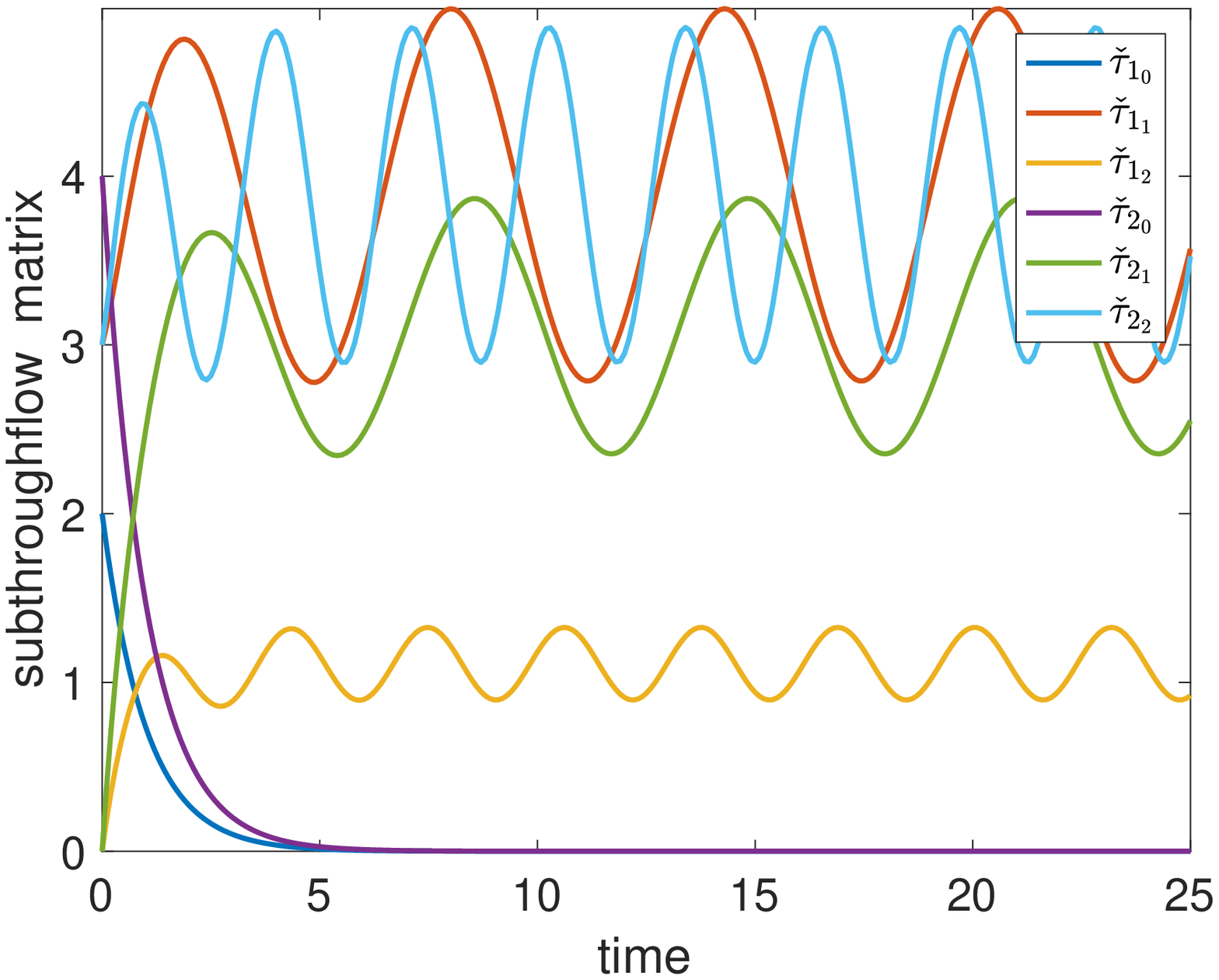}
        \caption{$\check{T}(t,{\rm x})$}
        \label{fig:T}
    \end{subfigure}
    \caption{The graphical representation of the substorage and inward subthroughflow matrices, $X(t)$ and $\check{T}(t,{\rm x})$, and the initial substorage and inward subthroughflow vectors, $x_0(t)$ and $\check{\tau}_{0}(t,{\rm x})$, for time-dependent input $z(t) = \left[ 3+\operatorname{sin}(t), 3+\operatorname{sin}(2 \, t) \right]^T$ (Case study~\ref{appsec:ex_hippe}).}
    \label{fig:hippe}
\end{figure}

The elements of the inward initial throughflow vector, $\check{\tau}_{0}(t)$, and subthroughflow matrix, $\check{T}(t)$, can be computed using Eq.~\ref{eq:Tmatrix2} as follows:
\begin{equation}
\label{eq:hippe_thr}
\begin{aligned} 
\check{\tau}_{1_0}(t) & = 2\,{\mathrm{e}}^{-t}, \quad \check{\tau}_{2_0}(t) = 4\,{\mathrm{e}}^{-t} , \\
\check{\tau}_{1_1}(t) & = \frac{35}{9} -\frac{8\,\cos\left(t\right)}{45}+\frac{49\,\sin\left(t\right)}{45} -\frac{10\,{\mathrm{e}}^{-t}}{9} + \frac{2\,{\mathrm{e}}^{-3\,t}}{5} , \\
\check{\tau}_{1_2}(t) & = \frac{742}{585} - \frac{184\, \cos^2\left(t\right)}{585} +\frac{86\,\sin\left(2\,t\right)}{585} -\frac{26\,{\mathrm{e}}^{-t}}{45}-\frac{44\,{\mathrm{e}}^{-3\,t}}{117} , \\ 
\check{\tau}_{2_1}(t) & = \frac{28}{9} -\frac{22\,\cos\left(t\right)}{45} + \frac{26\,\sin\left(t\right)}{45}  -\frac{20\,{\mathrm{e}}^{-t}}{9} - \frac{2\,{\mathrm{e}}^{-3\,t}}{5} , \\
\check{\tau}_{2_2}(t) & = \frac{2339}{585} -\frac{128\, \cos^2\left(t\right)}{585} +\frac{577\,\sin\left(2\,t\right)}{585} -\frac{52\,{\mathrm{e}}^{-t}}{45}+\frac{44\,{\mathrm{e}}^{-3\,t}}{117} .
\end{aligned} 
\end{equation}
The outward initial throughflows and throughflows can also be obtained similarly, using Eq.~\ref{eq:Tmatrix2}. The substorage and inward subthroughflow matrix functions, $X(t)$ and $\check{T}(t)$, given in Eqs.~\ref{eq:hippe_pedX} and~\ref{eq:hippe_thr}, determine the dynamic distribution of the environmental inputs and the organization of the associated storages generated by these inputs individually and separately within the system. In other words, using these functions, the evolution of the environmental inputs and associated storages can be tracked individually and separately throughout the system. The graphical representation of $X(t)$ and $\check{T}(t)$ for time dependent input $z(t) = \left[ 3+\operatorname{sin}(t), 3+\operatorname{sin}(2 \, t) \right]^T$ are depicted in Fig.~\ref{fig:hippe}. The evolution of initial stocks and inward throughflows, $x_{i_0}(t)$ and $\check{\tau}_{i_0}(t)$, are also presented in Fig.~\ref{fig:hippe}.

The residence time matrix for this model, defined in Eq.~\ref{eq:res_matrix}, becomes
\begin{equation}
\label{eq:hippe_res} 
\begin{aligned} 
\mathcal{R}(t,{x}) = \diag{( [0.6, 0.43] )} .
\end{aligned} 
\nonumber
\end{equation} 
The residence time of compartment $2$ is constantly smaller than that of compartment 1. That is, $r_2(t,x) = 0.43 < 0.6 = r_1(t,x)$. This result ecologically indicates that compartment $2$ is more active than $1$.

The subsystem partitioning methodology allows for the further analysis of the system and brings out additional insights that are not available through the state-of-the-art techniques. The composite transfer flow and storage from compartment $2$ to $1$, $\tau^\texttt{t}_{12}(t)$ and $x^\texttt{t}_{12}(t)$, and composite transfer subflow and substorage from initial subcompartment $2_0$ to $1_0$, $\tau^\texttt{t}_{1_02_0}(t)$ and $x^\texttt{t}_{1_02_0}(t)$, are computed below as an application of the proposed subsystem partitioning methodology. They can be expressed using Eq.~\ref{eq:out_in_diact} as
\begin{equation}
\label{eq:hippe_ss2}
\begin{aligned}
\tau^\texttt{t}_{12}(t) = \sum_{k=1}^2 \tau^\texttt{t}_{1_k 2_k}(t) 
\quad \mbox{and} \quad 
{x}^\texttt{t}_{12}(t) = \sum_{k=1}^2 {x}^\texttt{t}_{1_k 2_k}(t) .
\end{aligned}
\end{equation}
The sets of mutually exclusive subflow paths from $2_k$ to $1_k$, $P_{1_k 2_k}$, for $k=0,1,2$, can be formulated as follows: $P_{1_0 2_0} = \{p^1_{1_0 2_0}, p^2_{1_0 2_0} \}$, $P_{1_1 2_1} = \{ p^1_{1_1 2_1} \}$, $P_{1_2 2_2} = \{ p^1_{1_2 2_2}\}$ where $p_{1_0 2_0}^1 = 0_0 \mapsto 1_0 \rightsquigarrow 2_0 \to 1_0$, $p_{1_0 2_0}^2= 0_0 \mapsto 2_0 \to 1_0 \rightsquigarrow 2_0 $, $p_{1_1 2_1}^1= 0_1 \mapsto 1_1 \rightsquigarrow 2_1 \to 1_1$, and $p_{1_2 2_2}^1= 0_2 \mapsto 2_2 \to 1_2 \rightsquigarrow 2_2$. 

There are two subflow paths in the initial subsystem, $p_{1_0 2_0}^1$ and $p_{1_0 2_0}^2$, and therefore, $w_0 = 2$. The corresponding transfer subflow and associated substorage functions, as formulated in Eq.~\ref{eq:out_in_fsDT_diact}, become
\begin{equation}
\label{eq:hippe_ss4}
\begin{aligned}
\tau^\texttt{t}_{1_0 2_0}(t) & = \sum_{w=1}^2 \check{\tau}^{w}_{1_0}(t) = \check{\tau}^{1}_{1_0}(t) + \check{\tau}^{2}_{1_0}(t)  , \\
{x}^\texttt{t}_{1_0 2_0}(t) & = \sum_{w=1}^2 {x}^{w}_{1_0}(t) = {x}^1_{1_0}(t) + {x}^2_{1_0}(t)  .
\end{aligned}
\end{equation}
Similarly, we have 
\begin{equation}
\label{eq:hippe_ss14}
\begin{aligned}
\tau^\texttt{t}_{1_k 2_k}(t) & = \sum_{w=1}^1 \check{\tau}^{w}_{1_k}(t) = \check{\tau}^{1}_{1_k}(t)  \quad \mbox{and} \quad 
{x}^\texttt{t}_{1_k 2_k}(t) = \sum_{w=1}^1 {x}^{w}_{1_k}(t) = {x}^1_{1_k}(t) 
\end{aligned}
\end{equation}
for $k=1,2$, since there is only one subflow path in these subsystems ($w_k=1$).

\begin{figure}[t]
    \centering
    \begin{subfigure}[b]{0.32\textwidth}
        \includegraphics[width=\textwidth]{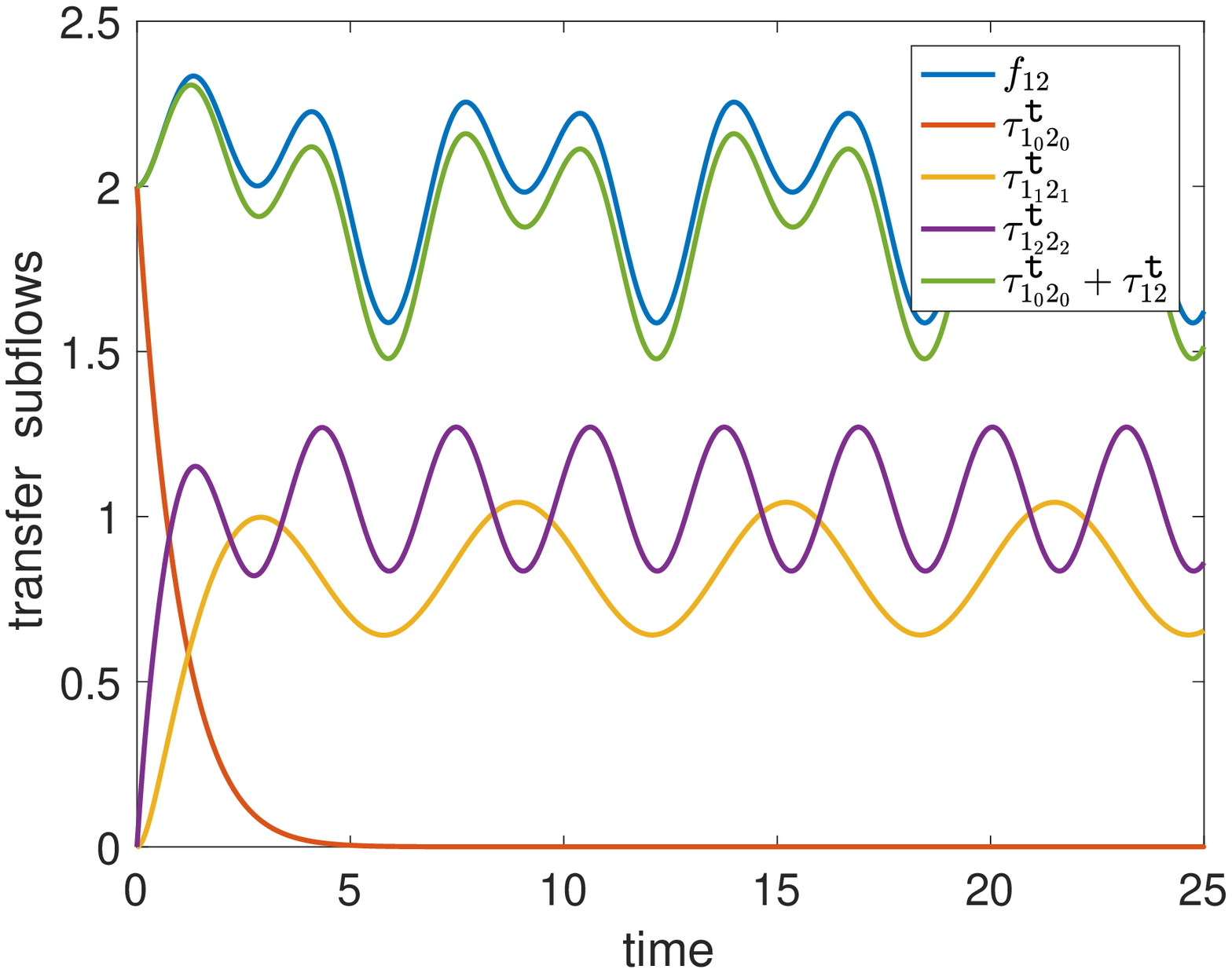}
        \caption{}
        \label{fig:tr_f}
    \end{subfigure}
    \begin{subfigure}[b]{0.32\textwidth}
        \includegraphics[width=\textwidth]{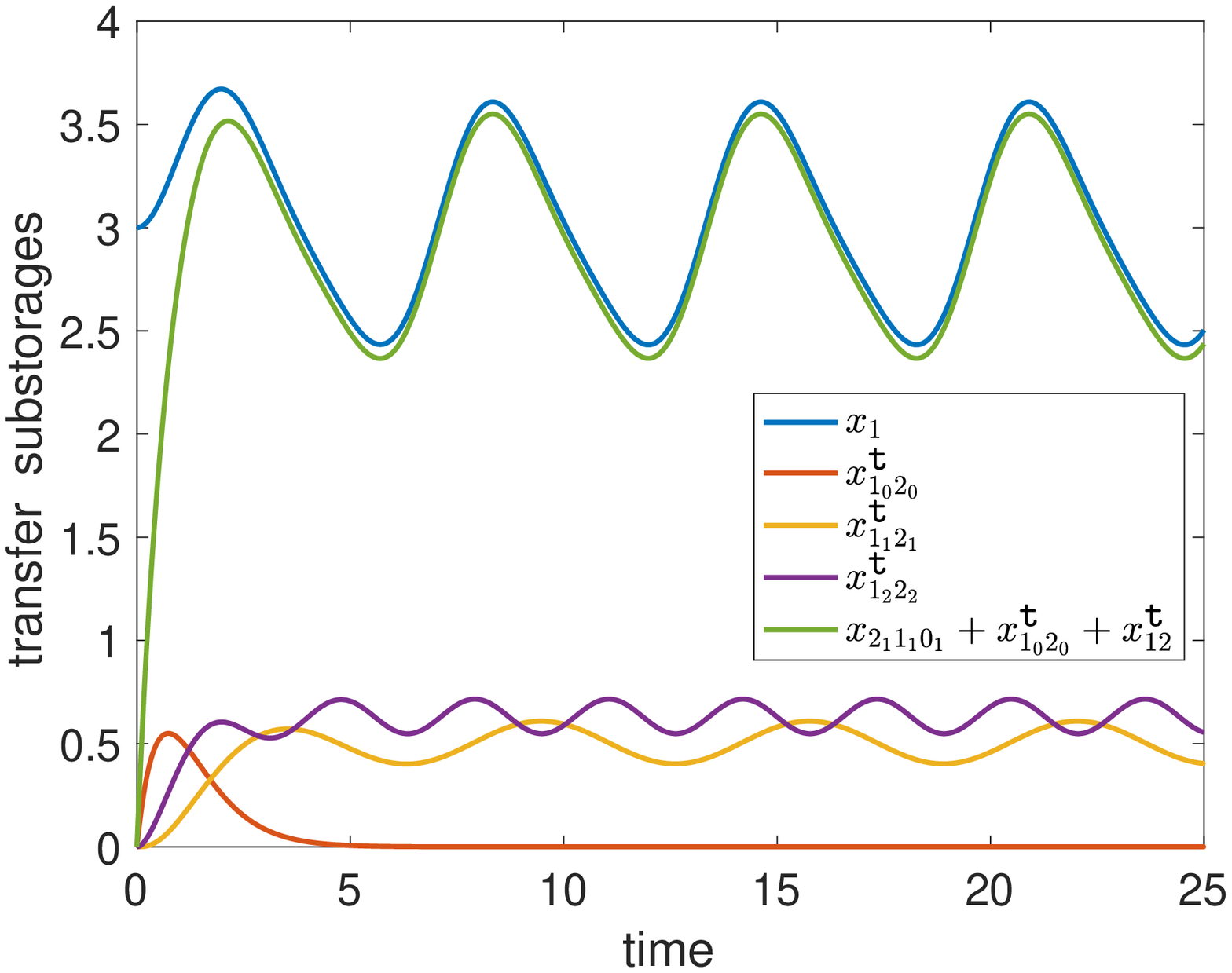}
        \caption{}
        \label{fig:tr_x}
    \end{subfigure} 
    \begin{subfigure}[b]{0.32\textwidth}
        \includegraphics[width=\textwidth]{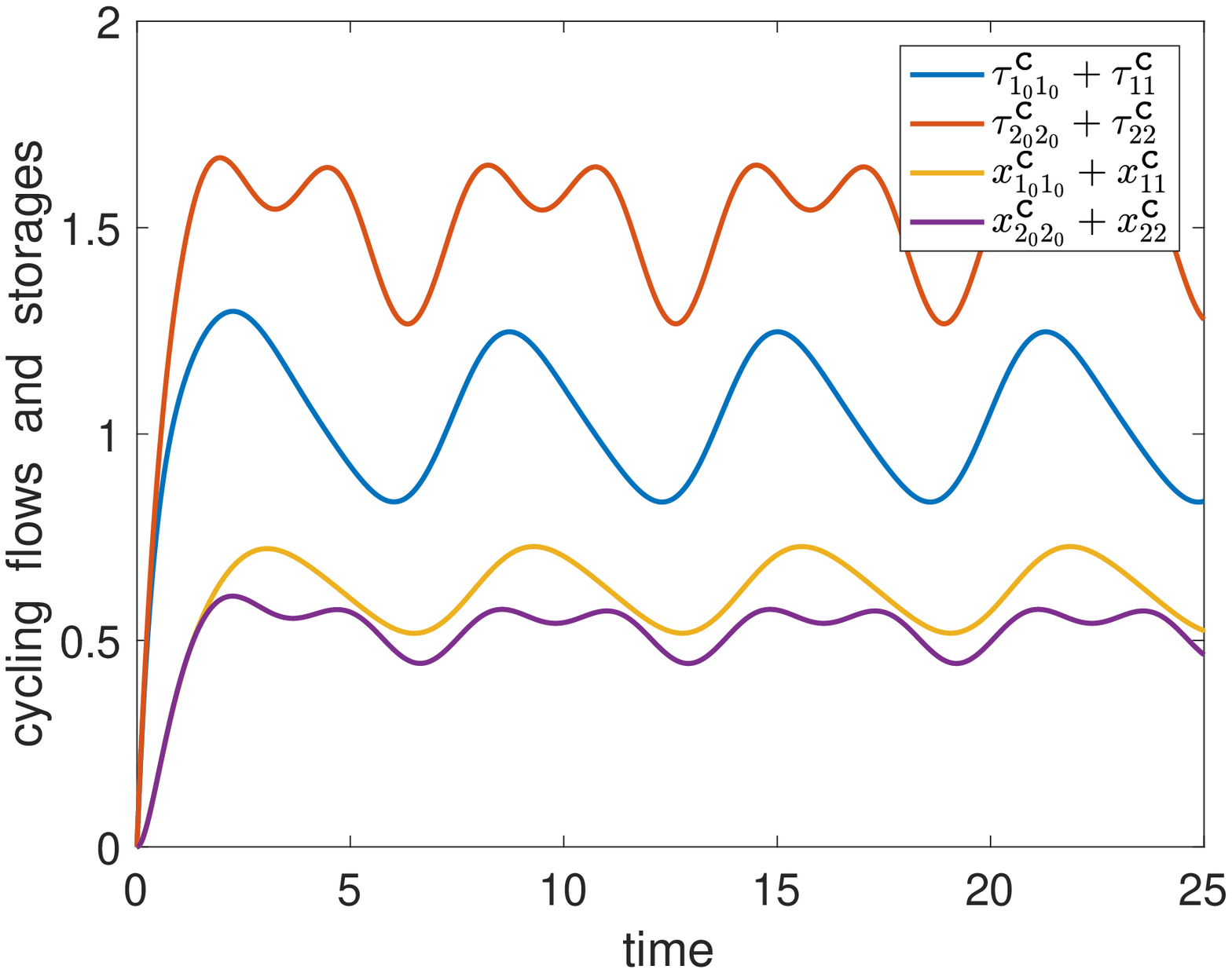}
        \caption{}
        \label{fig:cyc}
    \end{subfigure}
    \caption{The graphical representation of (a) the composite transfer flow and (b) storage, $\tau^\texttt{t}_{1 2}(t)$ and $x^\texttt{t}_{1 2}(t)$, together with the contributing composite transfer subflows and substorages, $\tau^\texttt{t}_{1_k 2_k}(t)$ and $x^\texttt{t}_{1_k 2_k}(t)$, as well as (c) the composite cycling flows and storages, $\tau^\texttt{c}_{i_0 i_0}(t)+\tau^\texttt{c}_{ii}(t)$ and $x^\texttt{c}_{i_0 i_0}(t)+x^\texttt{c}_{ii}(t)$ (Case study~\ref{appsec:ex_hippe}).}
    \label{fig:hippe_cyc}
\end{figure}

The links that directly contribute to the cumulative transient inflow and substorage, $\check{\tau}^{1}_{1_1}(t)$ and $x^1_{1_1}(t)$, at subcompartment $1_1$ along $p^1_{1_1 2_1}$ are marked with cycle numbers, $m$, in the extended subflow path diagram below:
\begin{equation}
\begin{aligned}
p_{1_1 2_1}^1= 0_1 \mapsto 1_1 \,  \rightsquigarrow \, 2_1 \xrightarrow{ 1 } 1_1 \, \rightsquigarrow \,   2_1 \xrightarrow{ 2 } 1_1 \, \rightsquigarrow \, 2_1 \xrightarrow{ 3 } 1_1  \rightsquigarrow   \cdots 
\end{aligned}
\nonumber
\end{equation}
The cumulative transient inflow and substorage will be approximated by two terms ($m_1=2$) using Eq.~\ref{eq:apxout_in_fs10}:
\begin{equation}
\label{eq:hippe_ss1}
\begin{aligned}
{x}^{1}_{1_1}(t) & \approx \sum_{m=1}^2 {x}^{1,m}_{2_1 1_1 2_1}(t) = {x}^{1,1}_{2_1 1_1 2_1}(t) + {x}^{1,2}_{2_1 1_1 2_1}(t) , \\ 
\check{\tau}^{1}_{1_1}(t) &  \approx  \sum_{m=1}^2 {f}^{1,m}_{1_1 2_1 1_1}(t) =  f^{1,1}_{1_1 2_1 1_1}(t) + f^{1,2}_{1_1 2_1 1_1}(t) .
\end{aligned}
\end{equation}
The governing equations, Eqs.~\ref{eq:out_in_fs} and~\ref{eq:out_in_fs2}, for the transient subflows and associated substorages, $f^{1,m}_{1_1 2_1 1_1}(t)$ and $x^{1,m}_{2_1 1_1 2_1}(t) $, and the other transient subflows and substorages involved in Eqs.~\ref{eq:hippe_ss4} and~\ref{eq:hippe_ss14}, are solved simultaneously, together with the decomposed system, Eq.~\ref{eq:model_sc}. Numerical results for the transfer subflows and associated substorages are presented in Fig.~\ref{fig:hippe_cyc}.

The subflow paths in $P_{1_k 2_k}$ for each subsystem $k$ are mutually exclusive and exhaustive. Therefore, $x_{1}(t) $ and $x_{1_0}(t) + x_{2_1 1_1 0_1}(t) + x^\texttt{t}_{1_02_0}(t) + x^\texttt{t}_{12}(t) $ must be the same, as well as $f_{12}(t)$ and $\tau^\texttt{t}_{1_0 2_0}(t)+\tau^\texttt{t}_{1 2}(t)$. The terms added to $x^\texttt{t}_{1_02_0}(t) + x^\texttt{t}_{12}(t)$ for a comparison, $x_{2_1 1_1 0_1}(t)$ and $x_{1_0}(t)$, are the transient substorage generated by environmental input in $1_1$ ($0_1 \mapsto 1_1$) and the initial substorage in compartment 1 (see Fig.~\ref{fig:X}). Therefore, they are not included in the transfer storage and initial substorage, $x^\texttt{t}_{12}(t)$ and $x^\texttt{t}_{1_02_0}(t)$. These quantities, however, are approximately equal as presented in Fig.~\ref{fig:hippe_cyc}: 
\[ x_{2_1 1_1 0_1}(t) + x^\texttt{t}_{1_0 2_0}(t) + x^\texttt{t}_{12}(t) \approx x_{1}(t)- x_{1_0}(t) \quad \mbox{and} \quad \tau^\texttt{t}_{1_0 2_0}(t)+\tau^\texttt{t}_{1 2}(t) \approx f_{12}(t).\]
The small differences are caused by the truncation errors in the computation of cumulative transient subflows, and larger ${m_w}$ values further improve the approximations. These close approximations demonstrate the accuracy and consistency of both the system and subsystem partitioning methodologies.

Instead of the path-based approach used in the numerical computations above, the \texttt{diact} flows can also be obtained analytically and explicitly using the dynamic approach as introduced in Section~\ref{apxsec:flows}. The composite transfer subflows $\tau^{\texttt{t}}_{1_k 2_k}(t)$ for $k=0,1,2$, and transfer flow $\tau^{\texttt{t}}_{12}(t)$ become:
\begin{equation}
\label{eq:trT_analytic}
\begin{aligned}
\tau^{\texttt{t}}_{1_0 2_0}(t) & = 2\,{\mathrm{e}}^{-t} , \\
\tau^{\texttt{t}}_{1_1 2_1}(t) & = \frac{8}{9} -\frac{8\,\cos\left(t\right)}{45}+\frac{4\,\sin\left(t\right)}{45} -\frac{10\,{\mathrm{e}}^{-t}}{9} + \frac{2\,{\mathrm{e}}^{-3\,t}}{5} , \\
\tau^{\texttt{t}}_{1_2 2_2}(t) & = -\frac{1013}{585} -\frac{184\,{\cos^2\left(t\right)}}{585} -\frac{499\,\sin\left(2\,t\right)}{585} -\frac{26\,{\mathrm{e}}^{-t}}{45}-\frac{44\,{\mathrm{e}}^{-3\,t}}{117} , \\
\tau^{\texttt{t}}_{12}(t) & = -\frac{493}{585} -\frac{8\,\cos\left(t\right)}{45}+\frac{4\,\sin\left(t\right)}{45} -\frac{998\,\cos\left(t\right)\,\sin\left(t\right)}{585} \\ & - \frac{184\,{\cos^2 \left(t\right)} }{585} + \frac{14\,{\mathrm{e}}^{-t}}{45}+\frac{14\,{\mathrm{e}}^{-3\,t}}{585} ,
\end{aligned}
\end{equation}
as formulated in Eq.~\ref{eq:comp_diact_subs}. The composite transfer substorages can then be obtained by coupling Eq.~\ref{eq:out_in_diact2} for the transfer subflows with the decomposed system, Eq.~\ref{eq:model_sc}, and solving them simultaneously. Alternatively, the corresponding transfer substorages can be obtained analytically as formulated in Eq.~\ref{eq:sln_transient}:
\begin{equation}
\label{eq:trX_analytic}
\begin{aligned}
x^{\texttt{t}}_{1_0 2_0}(t) & = 3\,{\mathrm{e}}^{-t}-3\,{\mathrm{e}}^{-\frac{5\,t}{3}} \\
x^{\texttt{t}}_{1_1 2_1}(t) & =\frac{8}{15} -\frac{26\,\cos\left(t\right)}{255}-\frac{2\,\sin\left(t\right)}{255} -\frac{5\,{\mathrm{e}}^{-t}}{3} + \frac{261\,{\mathrm{e}}^{-\frac{5\,t}{3}}}{170}-\frac{3\,{\mathrm{e}}^{-3\,t}}{10} \\
x^{\texttt{t}}_{1_2 2_2}(t) & = -\frac{17}{15} + \frac{2534\,\cos\left(2\,t\right)}{11895}-\frac{3047\,\sin\left(2\,t\right)}{11895} -\frac{13\,{\mathrm{e}}^{-t}}{15} \\ &+ \frac{459\,{\mathrm{e}}^{-\frac{5\,t}{3}}}{305} +\frac{11\,{\mathrm{e}}^{-3\,t}}{39} \\
x^{\texttt{t}}_{12}(t) & = -\frac{9671}{11895}  -\frac{26\,\cos\left(t\right)}{255}-\frac{2\,\sin\left(t\right)}{255} -\frac{6094\,\cos\left(t\right)\,\sin\left(t\right)}{11895} \\ & + \frac{5068\,{\cos^2 \left(t\right)}}{11895} + \frac{7\,{\mathrm{e}}^{-t}}{15} +\frac{417\,{\mathrm{e}}^{-\frac{5\,t}{3}}}{10370} -\frac{7\,{\mathrm{e}}^{-3\,t}}{390} .
\end{aligned}
\end{equation}
The graphs of these explicit transfer subflow and substorage functions in Eqs.~\ref{eq:trT_analytic} and ~\ref{eq:trX_analytic} obtained through the dynamic approach are exactly the same as the ones obtained by numerical computation through the path-based approach, Eq.~\ref{eq:hippe_ss2}, as depicted in Fig.~\ref{fig:hippe_cyc}.

The cycling flows and the associated storages generated by these flows are also calculated below for both compartments. The sets of mutually exclusive subflow paths from subcompartment $k_k$ to $1_k$ with a closed subpath at $1_k$, $P^\texttt{c}_{1_k k_k}$, are given as $P^\texttt{c}_{1_0 0_0} = \{p^1_{1_0 1_0} , p^2_{1_0 2_0}\}$, $P^\texttt{c}_{1_1 1_1} = \{ p^1_{1_1 1_1} \}$, $P^\texttt{c}_{1_2 2_2} = \{ p^1_{1_2 2_2}\}$, where $p_{1_0 1_0}^1 = 0_0 \mapsto 1_0 \rightsquigarrow 2_0 \to 1_0$, $p_{1_0 2_0}^2 = 0_0 \mapsto 2_0 \rightsquigarrow 1_0 \rightsquigarrow 2_0 \to 1_0 $, $p_{1_1 1_1}^1 =  0_1 \mapsto 1_1 \rightsquigarrow 2_1 \to 1_1$, and $p_{1_2 2_2}^1 = 0_2 \mapsto 2_2 \rightsquigarrow 1_2 \rightsquigarrow 2_2 \to 1_2$. For the subflow paths in $P^\texttt{c}_{1_0 0_0}$, the composite cycling subflows are derived from the initial stocks, and for the ones in $P^\texttt{c}_{1_1 1_1}$ and $P^\texttt{c}_{1_2 2_2}$, the simple cycling flows are generated by the respective environmental inputs of $z_1(t)$ and $z_2(t)$. The sets of subflow paths $P^\texttt{c}_{2_k k_k}$ for $k=0,1,2$ can similarly be defined.

The simple cycling subflow at subcompartment $1_2$ along the only subflow path ($w_2 = 1$) in subsystem $2$, $p^1_{1_2 2_2} \in P^\texttt{c}_{1_2 2_2}$, and associated substorage are
\begin{equation}
\label{eq:cyc1}
\begin{aligned}
{\tau}^\texttt{c}_{1_2}(t) = \sum_{w=1}^1 \check{\tau}^{w}_{1_2}(t) = \check{\tau}^{1}_{1_2}(t) \quad & \mbox{and} \quad x^\texttt{c}_{1_2}(t) = \sum_{w=1}^1 x^{w}_{1_2}(t) = x^{1}_{1_2}(t) ,
\end{aligned}
\nonumber
\end{equation}
as formulated in Eq.~\ref{eq:out_in_fsDT_diact}. The links contributing to the cycling subflow along the path are marked with cycle numbers in the extended subflow diagram below:
\begin{equation}
\begin{aligned}
p^1_{1_2 2_2} &= 0_2 \mapsto 2_2 \, \rightsquigarrow \, 1_2 \rightsquigarrow 2_2  \, \xrightarrow{ 1 } \,  1_2 \rightsquigarrow 2_2  \, \xrightarrow{ 2 } \,  1_2 \rightsquigarrow   \cdots 
\end{aligned}
\nonumber
\end{equation}
Note that the first flow entrance into $1_2$ is not considered as cycling flow. The cumulative transient inflow $\check{\tau}^1_{1_2}(t)$ and substorage $x^1_{1_2}(t)$ can be approximated by two terms ($m_1=2$) along the closed subpath as formulated in Eq.~\ref{eq:apxout_in_fs10}:
\begin{equation}
\label{eq:hippe_ss10}
\begin{aligned}
{x}^{1}_{1_2}(t) & = \sum_{m=1}^2 {x}^{1,m}_{2_2 1_2 2_2}(t) \approx {x}^{1,1}_{2_2 1_2 2_2}(t) + {x}^{1,2}_{2_2 1_2 2_2}(t) , \\ 
\check{\tau}^{1}_{1_2}(t) & =  \sum_{m=1}^2 {f}^{1,m}_{1_2 2_2 1_2}(t) \approx  f^{1,1}_{1_2 2_2 1_2}(t) + f^{1,2}_{1_2 2_2 1_2}(t) .
\end{aligned}
\nonumber
\end{equation}
The governing equations for the transient subflows and associated substorage functions, $f^{w,m}_{1_2 2_2 1_2}(t)$ and $x^{w,m}_{2_2 1_2 2_2}(t) $, as well as the other transient subflows and substorages involved in Eq.~\ref{eq:cyc2_sln}, as formulated in Eqs.~\ref{eq:out_in_fs} and~\ref{eq:out_in_fs2}, are coupled and solved simultaneously together with the decomposed system, Eq.~\ref{eq:model_sc}. The numerical results for the composite cycling flows and associated storages induced both by the environmental inputs and initial stocks, 
\begin{equation}
\label{eq:cyc2_sln}
\begin{aligned}
{\tau}^\texttt{c}_{i_0 i_0}(t) + {\tau}^\texttt{c}_{ii}(t) = \sum_{k=0}^2 {\tau}^\texttt{c}_{i_k i_k}(t)  \quad & \mbox{and} \quad x^\texttt{c}_{i_0 i_0}(t) + x^\texttt{c}_{ii}(t) = \sum_{k=0}^2 x^\texttt{c}_{i_k i_k}(t) 
\end{aligned}
\end{equation}
for $i=1,2$, are presented in Fig.~\ref{fig:cyc}. 

Note that, due to the reflexivity of cycling flows, the same computations can be done more practically in only two steps using the sets of closed subflow paths, $P^\texttt{c}_{i_k}$, instead, with the local inputs being the corresponding outwards subthroughflows. The subflow path $p_{1_1}^1 =  1_1 \mapsto 1_1 \rightsquigarrow 2_1 \to 1_1$ in subcompartment $1_1$ with local input $\hat{\tau}_{1_1}(t,{\mathbf x})$, for example, is depicted in Fig.~\ref{fig:hippe_diag}. The cycling flows can also be computed along closed paths at the compartmental level, where the local inputs are the outwards throughflows. 

The composite cycling subflows can also be computed analytically through the dynamic approach as formulated in Eq.~\ref{eq:thr_dense4}. As examples, $\tau^{\texttt{c}}_{1_0 1_0}(t)$ and $\tau^{\texttt{c}}_{2_2}(t) = \tau^{\texttt{c}}_{2_2 2_2}(t) $ become
\begin{equation}
\label{eq:cyc_analytic}
\begin{aligned}
\tau^{\texttt{c}}_{1_0 1_0}(t) & = -\frac{ 36 \, {\mathrm{e}}^{-t} + 80\,{\mathrm{e}}^{2\,t}-100\,{\mathrm{e}}^{1\,t}-16\,{\mathrm{e}}^{2\,t}\,\cos\left(t\right)+8\,{\mathrm{e}}^{2\,t}\,\sin\left(t\right) }
{9 + 50\,{\mathrm{e}}^{2\,t}-70\,{\mathrm{e}}^{3\,t}+11\,{\mathrm{e}}^{3\,t}\,\cos\left(t\right)-13\,{\mathrm{e}}^{3\,t}\,\sin\left(t\right) } \\
\tau^{\texttt{c}}_{2_2 2_2}(t) & = \frac{584}{585} -\frac{52\,{\mathrm{e}}^{-t}}{45}+\frac{44\,{\mathrm{e}}^{-3\,t}}{117}-\frac{8\,\sin\left(2\,t\right)}{585} - \frac{128\, \cos^2(t)}{585}.
\end{aligned}
\end{equation}
The composite cycling storages can then be obtained by coupling Eq.~\ref{eq:out_in_diact2} for the cycling flows and storages with the decomposed system, Eq.~\ref{eq:model_sc}, and solving them simultaneously. Alternatively, they can be obtained analytically by using Eq.~\ref{eq:sln_transient}, similar to the transfer storages presented above in this example. Because of the lengthy analytical formulations of the other cycling subflows and substorages, only $\tau^{\texttt{c}}_{1_0 1_0}(t)$ and $\tau^{\texttt{c}}_{2_2}(t)$ are presented in Eq.~\ref{eq:cyc_analytic} as examples.

\subsection{Case study}
\label{appsec:ex_hallam}

In this section, a nonlinear resource-producer-consumer ecosystem model introduced by \cite{Hallam1985} is analyzed through the proposed methodology. A comparison of the results is not possible, as the authors did not provide any computational or explicit results in the article. They only provided some results at steady state. Besides a constant environmental input, the system is also examined for a time dependent, symmetric Gaussian impulse to illustrate the efficiency of the proposed method in capturing the system response to disturbances. Such analysis can be used to quantify the system resistance and resilience in the face of disturbances and perturbations.
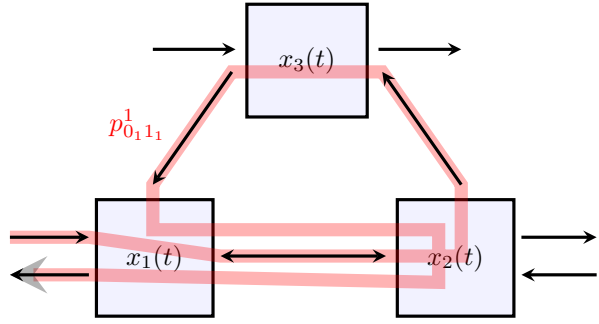
\begin{figure}[t]
\begin{center}
\begin{tikzpicture}
\centering
   \draw[very thick,  fill=blue!5, draw=black] (-.05,-.05) rectangle node(R1) {$x_1(t)$} (1.5,1.5) ;
   \draw[very thick,  fill=blue!5, draw=black] (3.95,-.05) rectangle node(R2) {$x_2(t)$} (5.5,1.5) ;   
   \draw[very thick,  fill=blue!5, draw=black] (1.95,2.6) rectangle node(R3) {$x_3(t)$} (3.55,4.1) ;         
       \draw[very thick,-stealth,draw=red, line width=5pt, opacity=.3]  (-1.2,1) -- (-0.15,1) -- (1.6,.75) -- (3.8,.75) -- (4.8,0.75) -- (4.8,1.7) -- (3.75,3.2) -- (1.75,3.2) -- (0.7,1.7) -- (0.7,1.1) -- (4.5,1.1) -- (4.5,.4) -- (-.15,.5) -- (-1.2,.5);   
       \draw[very thick,stealth-stealth,draw=black]  (1.6,.75) -- (3.8,.75) ;  
       \draw[very thick,stealth-,draw=black] (0.7,1.7) -- (1.75,3.2) ;     
       \draw[very thick,-stealth,draw=black]  (4.8,1.7) -- (3.75,3.2) ;  
       \draw[very thick,stealth-,draw=black]  (5.6,.5) -- (6.6,.5) ;      
       \draw[very thick,-stealth,draw=black]  (5.6,1) -- (6.6,1) ;         
       \draw[very thick,stealth-,draw=black]  (-1.2,.5) -- (-0.15,.5) ;  
       \draw[very thick,-stealth,draw=black]  (-1.2,1) -- (-0.15,1) ;    
       \draw[very thick,-stealth,draw=black]  (.7,3.5) -- (1.8,3.5) ;    
       \draw[very thick,-stealth,draw=black]  (3.7,3.5) -- (4.8,3.5) ;   
       \node (p) [text=red] at (.5,2.5) {$p^1_{0_1 1_1}$};           
\end{tikzpicture}
\end{center}
\caption{Schematic representation of the model network. Subflow path $p^1_{0_1 1_1}$ along which the transient subflows and substorages are computed is red (subsystems are not shown) (Case study~\ref{appsec:ex_hallam}).}
\label{fig:hallam_diag}
\end{figure}

The resource-producer-consumer model by \cite{Hallam1985} consists of the dynamics for three components: $x_1(t) = r(t)$ is the nutrient storage (such as phosphorus or nitrogen) present at time t; $x_2(t) = s(t)$ represents the nutrient storage in the producer (such as phytoplankton) population; and $x_3(t) = c(t)$ denotes the nutrient storage in the consumer (such as zooplankton) population (see Fig.~\ref{fig:hallam_diag}). The conservation of nutrient is the basic model assumption. The system flows are described as follows:
\begin{equation}
\label{eq:hallam_flows}
\begin{aligned} 
F(t,x) = 
\begin{bmatrix}
    0  & d_1 \, s(t) & d_2 \, c(t)  \\
    \frac{\alpha_1 \, s(t) \, r(t) }{\alpha_2 + r(t) }  & 0 & 0 \\
    0 & \frac{\beta_1 \, s(t) \, c(t) }{\beta_2 + s(t) } & 0 
\end{bmatrix},
\, \, \,
z(t) = 
\begin{bmatrix}
    z_1(t) \\
    z_2(t) \\
    z_3(t) 
\end{bmatrix},
\, \, \,
y(t) = 
\begin{bmatrix}
    r(t) \\
    s(t) \\
    c(t) 
\end{bmatrix}
\end{aligned} ,
\nonumber
\end{equation}
where the constant input is $z(t) = [1, 1, 1]^T$, and the parameters are given as
\[d_1=2.7, \, d_2=2.025, \, \alpha_2=0.098, \, \beta_1=2, \, \beta_2=20, \, \mbox{and } \, \alpha_1=1.\] 
The value for $\alpha_1$ was not provided in \cite{Hallam1985} and was chosen arbitrarily for this example. The governing equations take the following form:
\begin{equation}
\label{eq:hallam_ex}
\begin{aligned} 
\dot r(t) &= -r(t) + d_1 \, s(t) + d_2 \, c(t) - \frac{\alpha_1 \, s(t) \, r(t)}{\alpha_2 + r(t)} + z_1(t) \\
\dot s(t) &= -(1+d_1) \, s(t) + \frac{\alpha_1 \, s(t) \, r(t)}{\alpha_2 + r(t)} - \frac{\beta_1 \, c(t) \, s(t)}{\beta_2 + s(t)} + z_2(t) \\
\dot c(t) &= -(1+d_2) \, c(t) + \frac{\beta_1 \, c(t) \, s(t)}{\beta_2 + s(t)} + z_3(t)
\end{aligned}
\end{equation}
with the initial conditions of $[r_0,s_0,c_0] = [1, 1, 1]$. 
\begin{figure}[t]
    \centering
    \begin{subfigure}[b]{0.46\textwidth}
        \includegraphics[width=\textwidth]{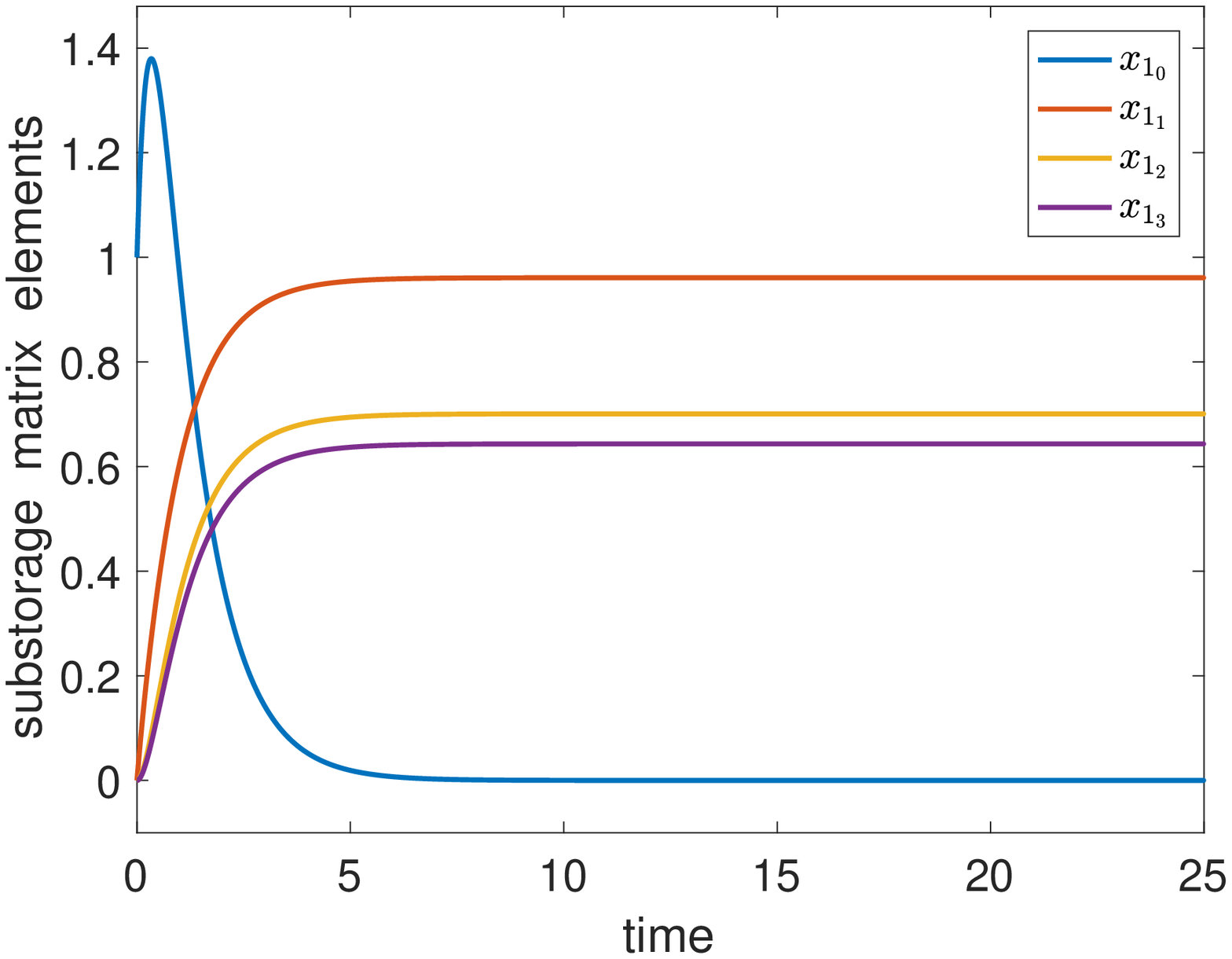}
        \caption{substorages of $x_1(t)$}
        \label{fig:x_1}
    \end{subfigure}
    \begin{subfigure}[b]{0.45\textwidth}
        \includegraphics[width=\textwidth]{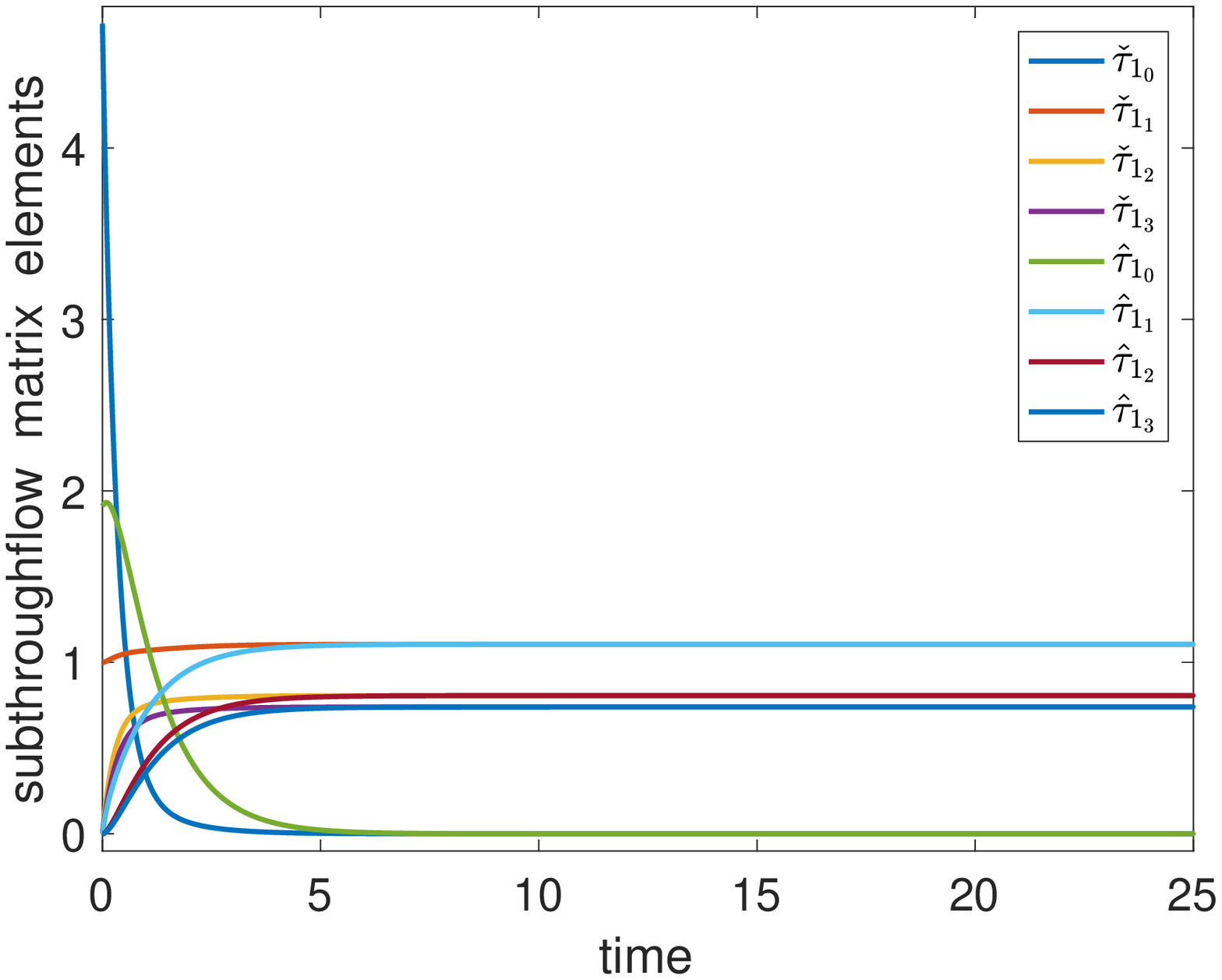}
        \caption{subthroughflows of $\check{\tau}_1(t,{\rm x})$ and $\hat{\tau}_1(t,{\rm x})$}
        \label{fig:tau_1}
    \end{subfigure}
    \caption{The numerical results for selected elements (first rows) of the substorage, $X(t)$, and subthroughflow matrices, $\check{T}(t,{\rm x})$ and $\hat{T}(t,{\rm x})$, and the initial substorage, $x_{0}(t)$, and subthroughflow vectors, $\check{\tau}_{0}(t,{\mathrm x})$ and $\hat{\tau}_{0}(t,{\mathrm x})$, for the system with constant input $z(t) = [1,1,1]^T$ (Case study~\ref{appsec:ex_hallam}).}
    \label{fig:hallam_s}
\end{figure}

The system partitioning methodology is composed of the subcompartmentalization and flow partitioning components. The subcompartmantalization yields 
\[ x_{1_k}(t) = r_k(t), \quad x_{2_k}(t) = s_k(t), \, \, \,  \mbox{and} \, \, \, x_{3_k}(t) = c_k(t)  \, \, \,  \mbox{with} \, \, \,  
x_i(t) = \sum_{k=0}^3 x_{i_k} (t) . \]
The flow partitioning then gives the flow regime for each subsystem as follows: 
\begin{equation}
\label{eq:hallam_flows_SC}
\begin{aligned} 
F_k(t,{\rm x}) = 
\begin{bmatrix}
    0  & d_{2_k} \, d_1 \, s & d_{3_k} \, d_2 \, c  \\
    d_{1_k} \, \frac{\alpha_1 \, s \, r}{\alpha_2 + r}  & 0 & 0 \\
    0 & d_{2_k} \, \frac{\beta_1 \, s \, c}{\beta_2 + s} & 0 
\end{bmatrix},
\, \, \,
\check{z}_k(t,{\rm x}) =  
\begin{bmatrix}
    \delta_{1k} \, z_1 \\
    \delta_{2k} \, z_2 \\
    \delta_{3k} \, z_3
\end{bmatrix},
\, \, \,
\hat{y}_k(t,{\rm x}) = 
\begin{bmatrix}
    d_{1_k} \, r \\
    d_{2_k} \, s \\
    d_{3_k} \, c
\end{bmatrix} ,
\end{aligned}
\nonumber
\end{equation}
where $F_k$, $\check{z}_k$, and $\hat{y}_k$ describe the $k^{th}$ direct flow matrix, input, and output vectors for the $k^{th}$ subsystem, and the decomposition factors $d_{i_k}({\rm x})$ are defined by Eq.~\ref{eq:cons}. Therefore, the dynamic system partitioning methodology yields the following governing equations for the decomposed system:
\begin{equation}
\label{eq:hallam_exsc}
\begin{aligned} 
{\dot r}_{k}(t) &=  \delta_{1k} \, z_1(t) + d_1 \, s_{k}(t) + d_2 \, c_{k}(t) - r_{k}(t) -  \frac{\alpha_1 \, s(t) \, r_{k}(t) }{\alpha_2 + r(t)}  \\
{\dot s}_{k}(t) &= \delta_{2k} \, z_2(t) + \frac{\alpha_1 \, s(t) \, r_{k}(t)}{\alpha_2 + r(t)} - s_{k}(t) - d_1 \, s_{k}(t) - \frac{\beta_1 \, c(t) \, s_{k}(t) }{\beta_2 + s(t)}  \\
{\dot c}_{k}(t) &= \delta_{3k} \, z_3(t) +  \frac{\beta_1 \, c(t) \, s_{k}(t) }{\beta_2 + s(t)} - c_{k}(t) - d_2 \, c_{k}(t) 
\end{aligned}
\end{equation}
with the initial conditions 
\begin{equation}
\label{eq:hallam_ic}
x_{i_k} (t_0) = \left \{
\begin{aligned}
1, \quad k=0 \\
0, \quad k \neq 0
\end{aligned}
\right.
\nonumber
\end{equation}
for $i = 1,\ldots,3$. There are $n \times (n+1) = 3 \times 4 = 12$ equations in this system. 
\begin{figure}[t]
    \centering
    \begin{subfigure}[b]{0.46\textwidth}
        \includegraphics[width=\textwidth]{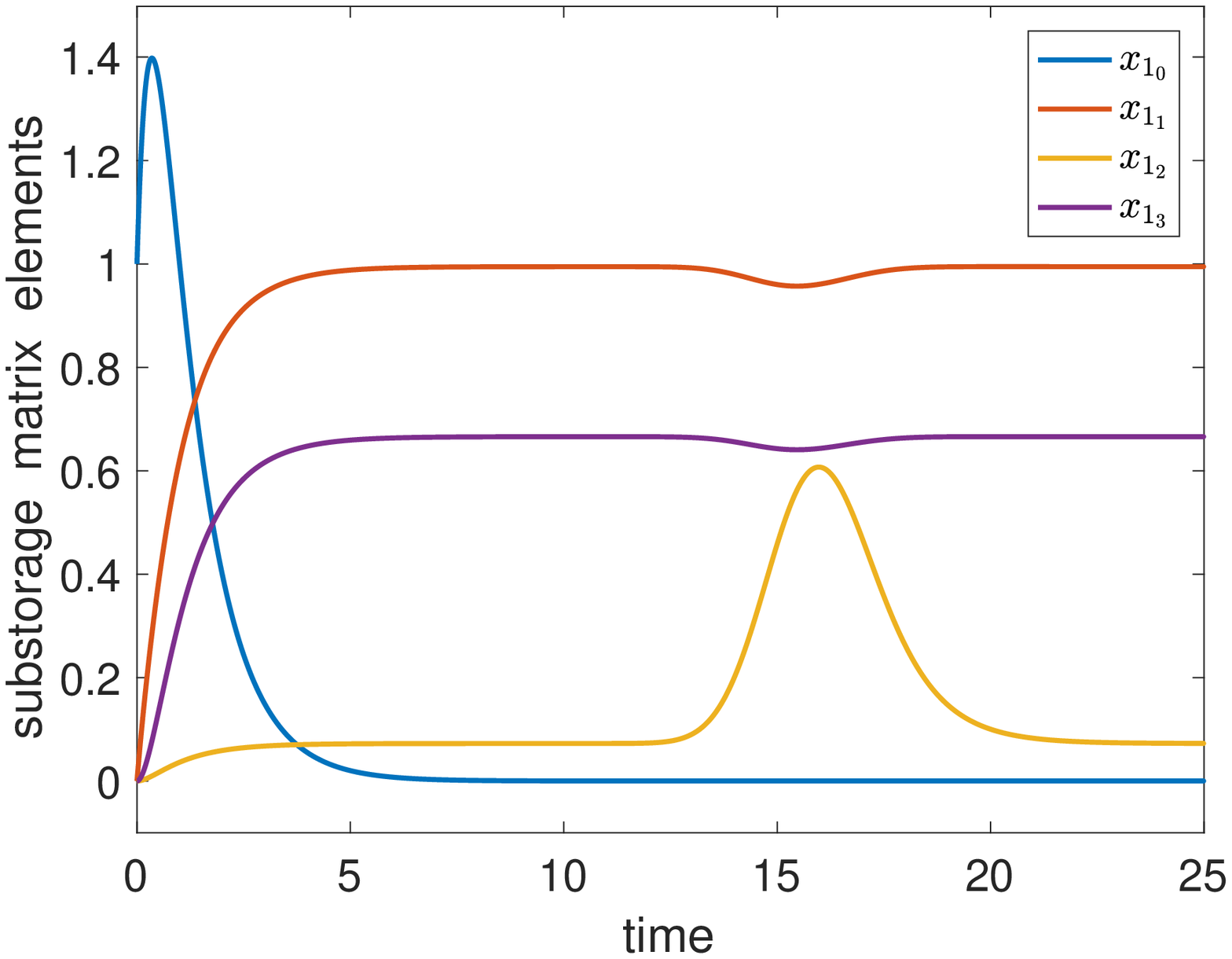}
        \caption{substorages of $x_1(t)$}
        \label{fig:x_1g}
    \end{subfigure}
    \begin{subfigure}[b]{0.45\textwidth}
        \includegraphics[width=\textwidth]{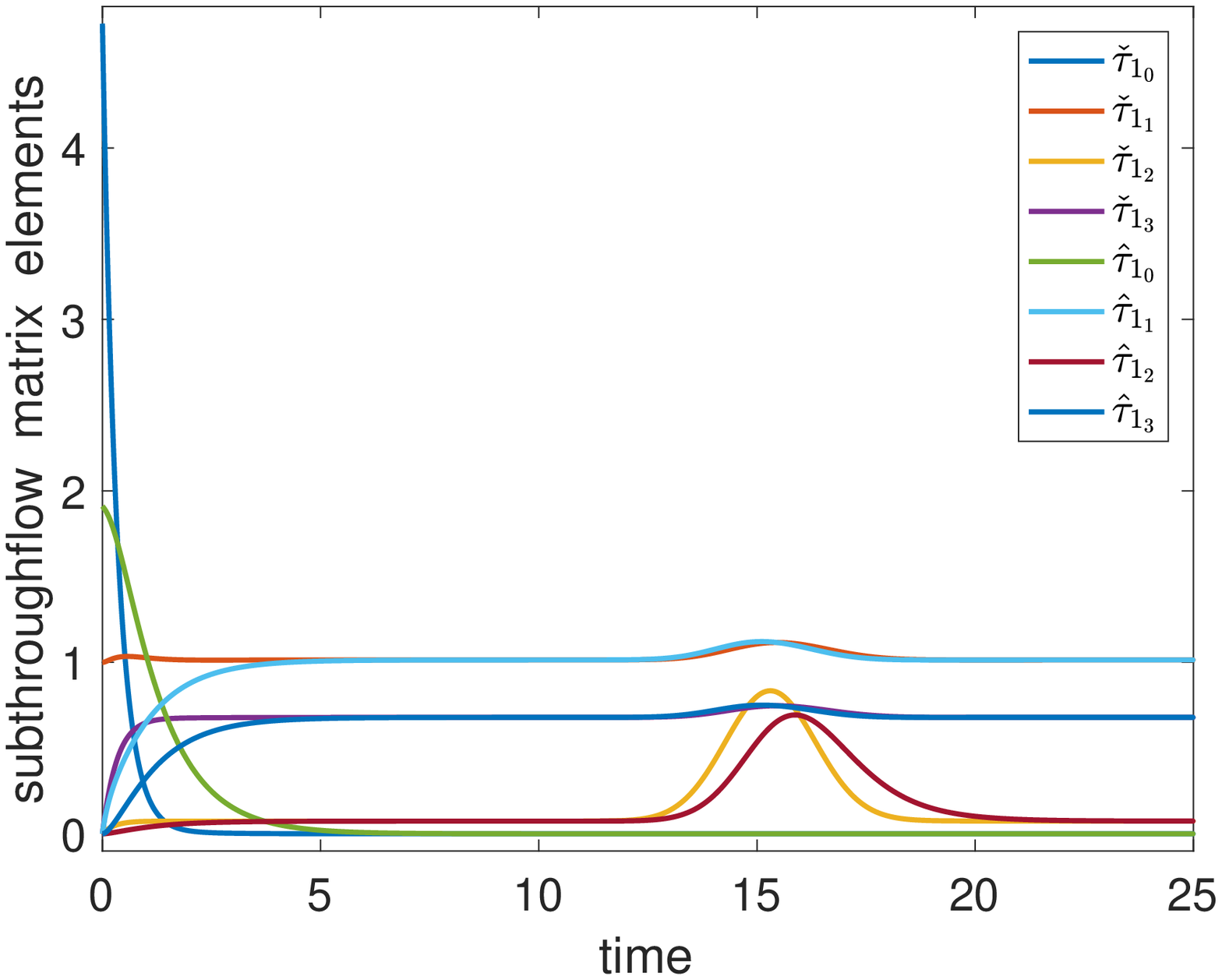}
        \caption{subthroughflows of $\check{\tau}_1(t,{\rm x})$ and $\hat{\tau}_1(t,{\rm x})$}
        \label{fig:tau_1g}
    \end{subfigure} 
    \caption{The numerical results for the selected elements (first rows) of the substorage, $X(t)$, and subthroughflow matrix functions, $\check{T}(t,{\rm x})$ and $\hat{T}(t,{\rm x})$, and the initial substorage, $x_{0}(t)$, and subthroughflow vectors, $\check{\tau}_{0}(t,{\mathrm x})$ and $\hat{\tau}_{0}(t,{\mathrm x})$, for the system with time-dependent environmental input (Gaussian impulse function) $z_{2}(t) = {\mathrm{e}}^{\frac{-(t-15)^2}{2}}+0.1$, and constant inputs $z_1(t)=1$ and $z_3(t)= 1$ (Case study~\ref{appsec:ex_hallam}).}
    \label{fig:hallam_g1}
\end{figure}

The system is solved numerically and the graphs for selected elements of the substorage and subthroughflow matrices are depicted in Fig.~\ref{fig:hallam_s}. As seen from the graphs, the system converges to a steady-state at about $t \approx 6$. The results show, for example, that the nutrient storage in the resource compartment ($i=1$) derived from nutrient input into the consumer compartment ($i=3$), $x_{1_3}(t)$, increases from $0$ to $0.64$ units until the system reaches the steady state, while the initial nutrient storage, $x_{1_0}$, first increases from $1$ to $1.38$ units and then vanishes. The throughflow into the resource compartment generated by nutrient input into the producer compartment ($i=2$), $\check{\tau}_{1_2}(t,{\rm x})$, increases until about $t \approx 2$. The outward throughflow at the same subcompartment, $\hat{\tau}_{1_2}(t,{\rm x})$, is slightly smaller than inward throughflow, $\check{\tau}_{1_2}(t,{\rm x})$, but has a similar behavior. As seen from these results, the distribution of environmental nutrient inputs and the organization of the associated nutrient storages generated by the inputs can be analyzed individually and separately within the system. 
\begin{figure}[t]
    \centering
    \begin{subfigure}[b]{0.45\textwidth}
        \includegraphics[width=\textwidth]{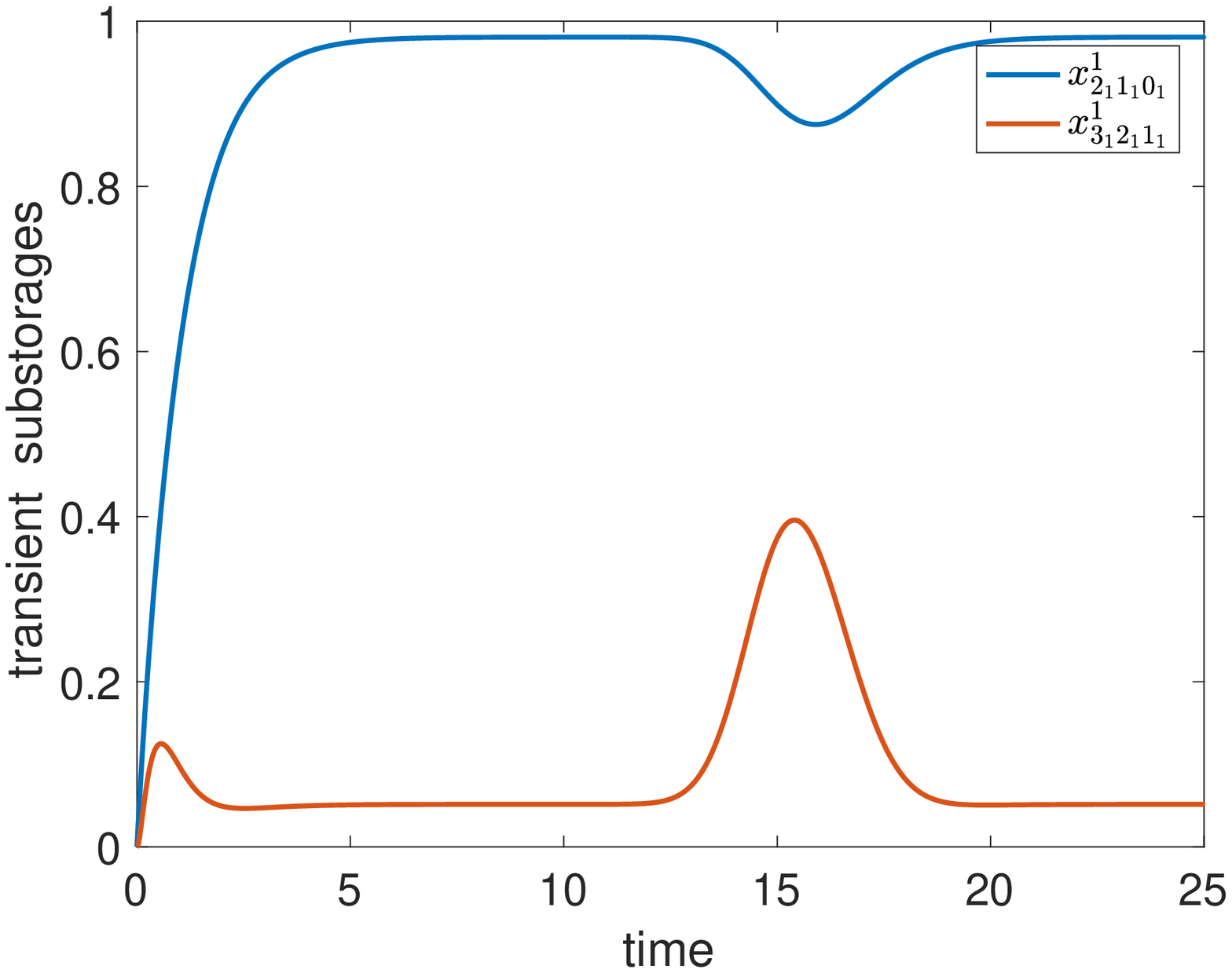}
        \caption{}
        \label{fig:tr_x_1ind}
    \end{subfigure}
    \begin{subfigure}[b]{0.46\textwidth}
        \includegraphics[width=\textwidth]{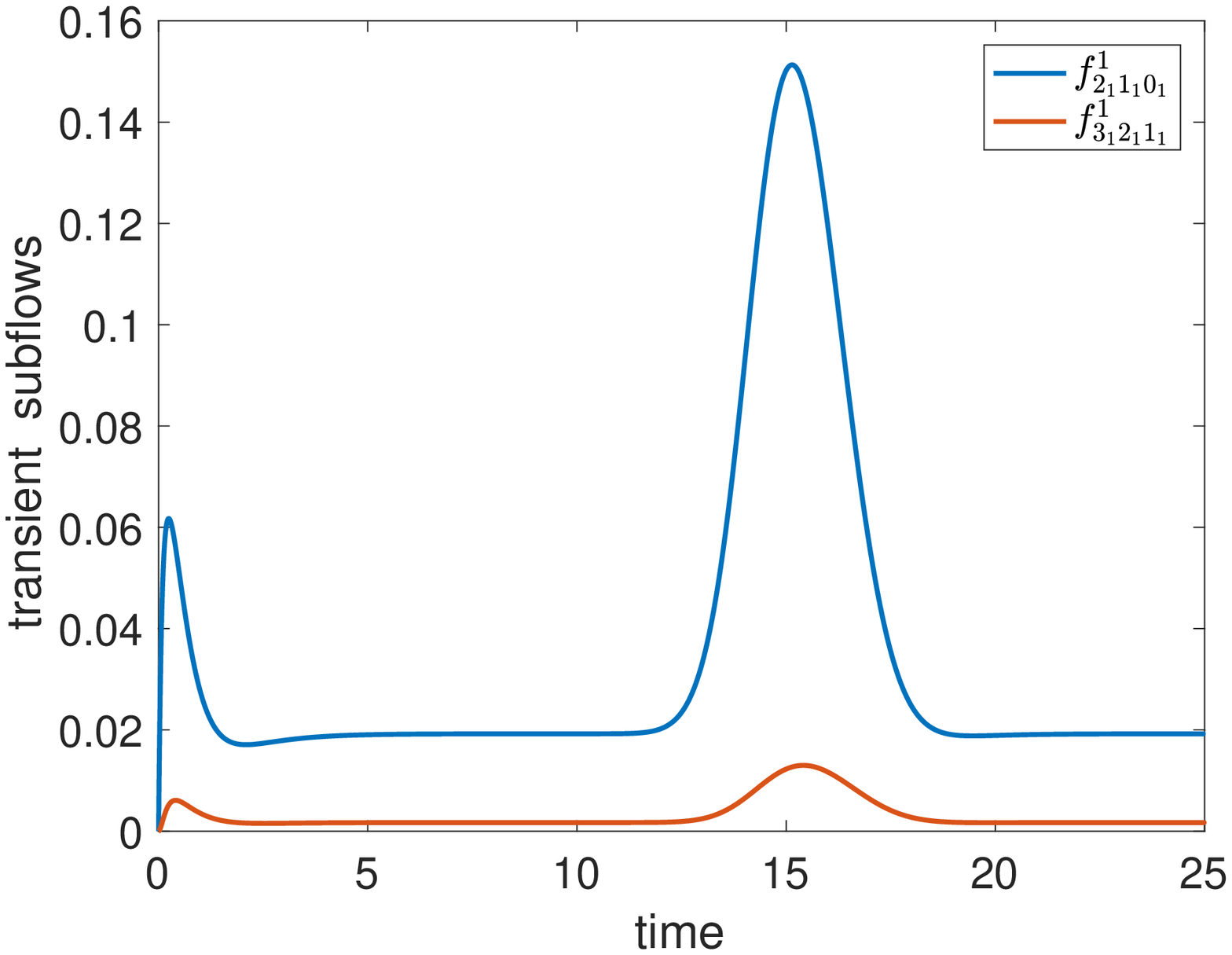}
        \caption{}
        \label{fig:tr_f_1ind}
    \end{subfigure} \\
    \begin{subfigure}[b]{0.45\textwidth}
        \includegraphics[width=\textwidth]{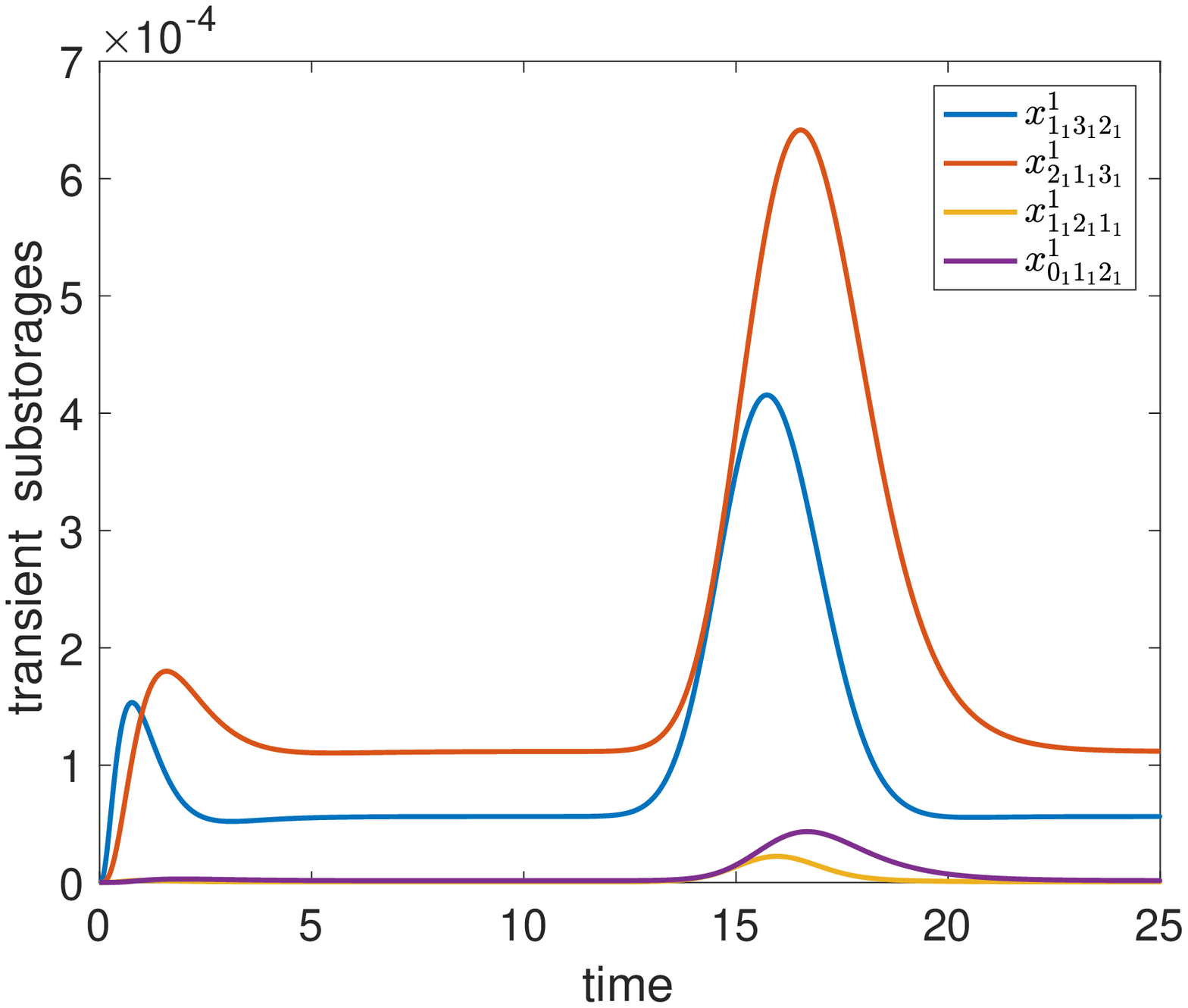}
        \caption{}
        \label{fig:tr_x_2ind}
    \end{subfigure} 
    \begin{subfigure}[b]{0.46\textwidth}
        \includegraphics[width=\textwidth]{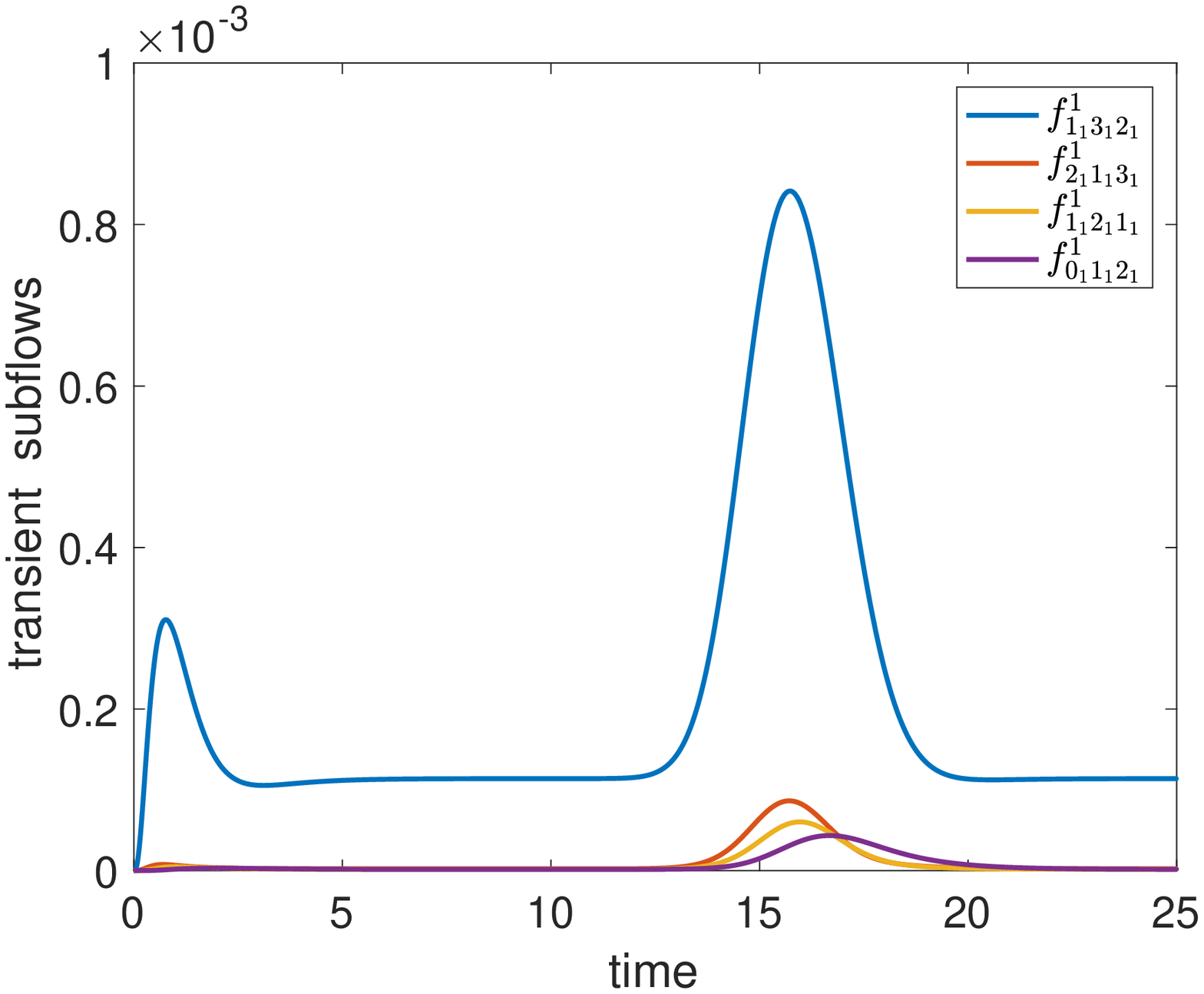}
        \caption{}
        \label{fig:tr_f_2ind}
    \end{subfigure} 
    \caption{The numerical results for the transient subflows and substorages at each step along subflow paths $p^1_{3_1 1_1}$ (a,b) and $p^1_{0_1 1_1}$ (a.b,c,d). In figures (a) and (b), functions $f^1_{3_1 2_1 1_1}(t)$ and $x^1_{3_1 2_1 1_1}(t)$ are scaled up by a factor of $ 10$ for clarity of the presentation (Case study~\ref{appsec:ex_hallam}).}
    \label{fig:hallam_ind}
\end{figure}

In general terms, the state variable $x_i(t)$ of the original system for the resource-producer-consumer dynamics, Eq.~\ref{eq:hallam_ex}, gives the nutrient storage in compartment $i$ at time $t$ based on its initial stock, $x_i(t_0)$. It cannot be used to distinguish the nutrient storage derived from individual environmental nutrient inputs. On the other hand, the state variable $x_{i_k}(t)$ of the decomposed system, Eq.~\ref{eq:hallam_exsc}, represents the nutrient storage in compartment $i$ that is derived from the specific environmental nutrient input into compartment $k$, $z_k(t)$. Similarly, the state variable ${x}_{i_0}(t)$ of the decomposed system represents the dynamics of the initial nutrient stocks in compartment $i$. Parallel interpretations are possible for the inward and outward throughflow functions of the original system, $\check{\tau}_i(t,x)$ and $\hat{\tau}_i(t,x)$, and the inward and outward subthroughflow functions of the decomposed system, $\check{\tau}_{i_k}(t,{\rm x})$ and $\hat{\tau}_{i_k}(t,{\rm x})$, as well.

The proposed dynamic system partitioning methodology, consequently, enables partitioning the compartmental composite nutrient flows and storages into subcompartmental segments based on their constituent sources from the initial stocks and environmental inputs. In other words, the system partitioning enables tracking the evolution of the initial nutrient stocks and environmental nutrient inputs a well as the associated storages generated by the stocks and inputs individually and separately within the system. This partitioning also allows for compiling a history of compartments visited by individual nutrient inputs separately.

The system is also perturbed with a Gaussian input $z_{2}(t) = {\mathrm{e}}^{\frac{-(t-15)^2}{2}} +0.1$, which represents a brief, unit local impulse at about $t=15$ to demonstrate the capability of the proposed method to analyze the influence of time dependent inputs on the system. The other two environmental nutrient inputs are kept constant as before for a comparison, that is, $z_1(t) = z_3(t) = 1$. The graphical representations for the selected elements of the substorage and subthroughflow matrices are given in Fig.~\ref{fig:hallam_g1}. It is clear from the graphs that the dynamic substorage and subthroughflow matrix measures reflect the impact of the unit impulse at about $t=15$. Note that, the system completely recovers after the disturbance in about $10$ time units. This time interval can be taken as a quantitative measure for the {\em restoration time} or {\em system resilience}. Therefore, the proposed measures can be used as quantitative ecological indicators for various ecosystem characteristics and behaviors.

The subsystem partitioning methodology is also applied to this model to track the fate of arbitrary intercompartmental flows and the associated storages generated by these flows within the subsystems. Along the subflow path $p_{3_1 1_1}^1= 0_1 \mapsto 1_1 \to 2_1 \rightsquigarrow 3_1$ from subcompartment $1_1$ to $3_1$ in subsystem $1$, the transient subflows and associated substorages are computed as formulated in Eqs.~\ref{eq:out_in_fs} and~\ref{eq:out_in_fs2}. The numerical results for the transient subflows, $f^1_{2_1 1_1 0_1}(t)$ and $f^1_{3_1 2_1 1_1}(t)$, and associated substorage functions, $x^1_{2_1 1_1 0_1}(t)$ and $x^1_{3_1 2_1 1_1}(t)$, are presented in Fig.~\ref{fig:tr_f_1ind} and~\ref{fig:tr_x_1ind}.
\begin{figure}[t]
    \centering
    \begin{subfigure}[b]{0.31\textwidth}
        \includegraphics[width=\textwidth]{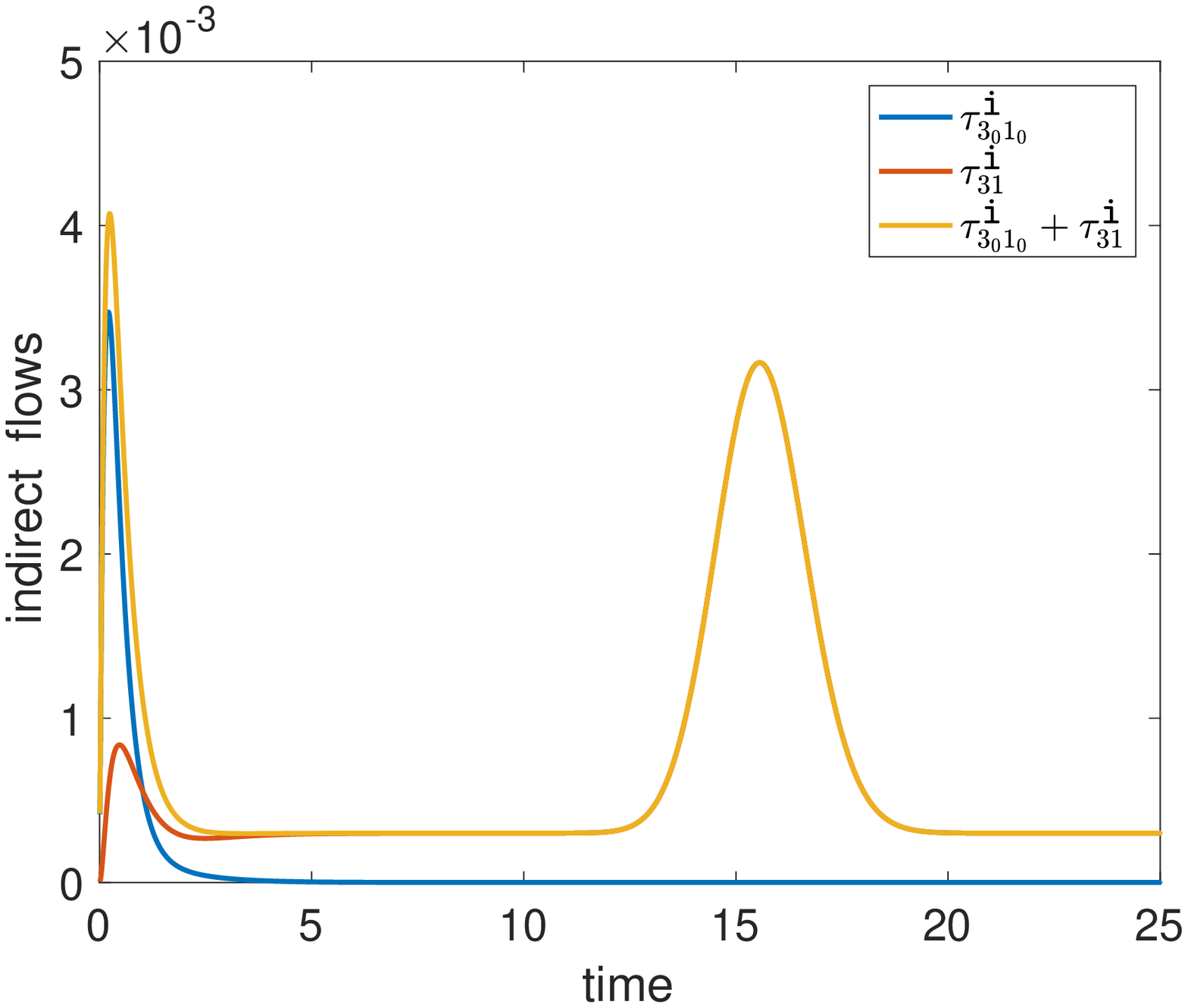}
        \caption{indirect (sub)flows}
        \label{fig:hallam_g_iT_31}
    \end{subfigure}
    \begin{subfigure}[b]{0.32\textwidth}
        \includegraphics[width=\textwidth]{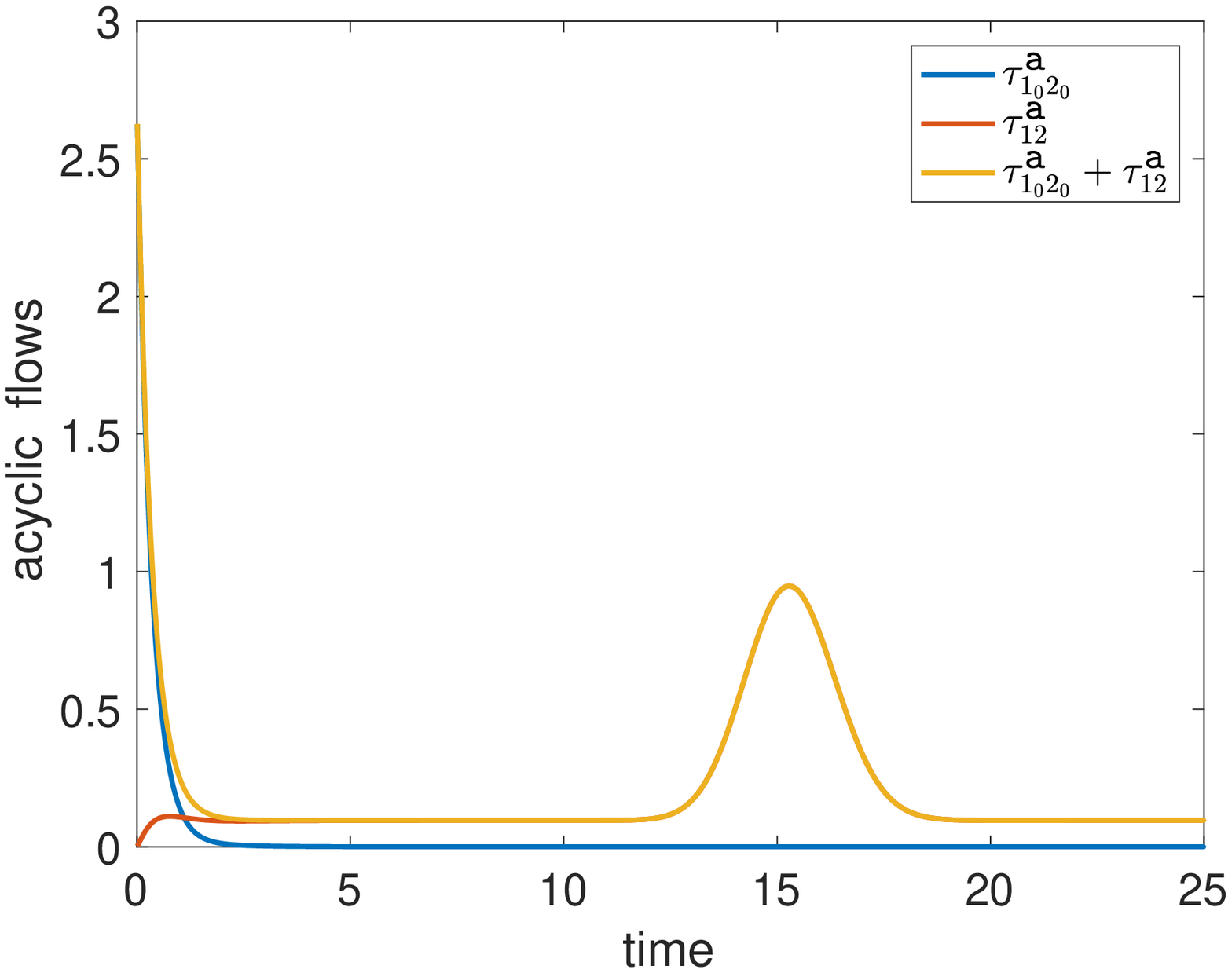}
        \caption{acyclic (sub)flows}
        \label{fig:hallam_g_aT_12}
    \end{subfigure} 
    \begin{subfigure}[b]{0.32\textwidth}
        \includegraphics[width=\textwidth]{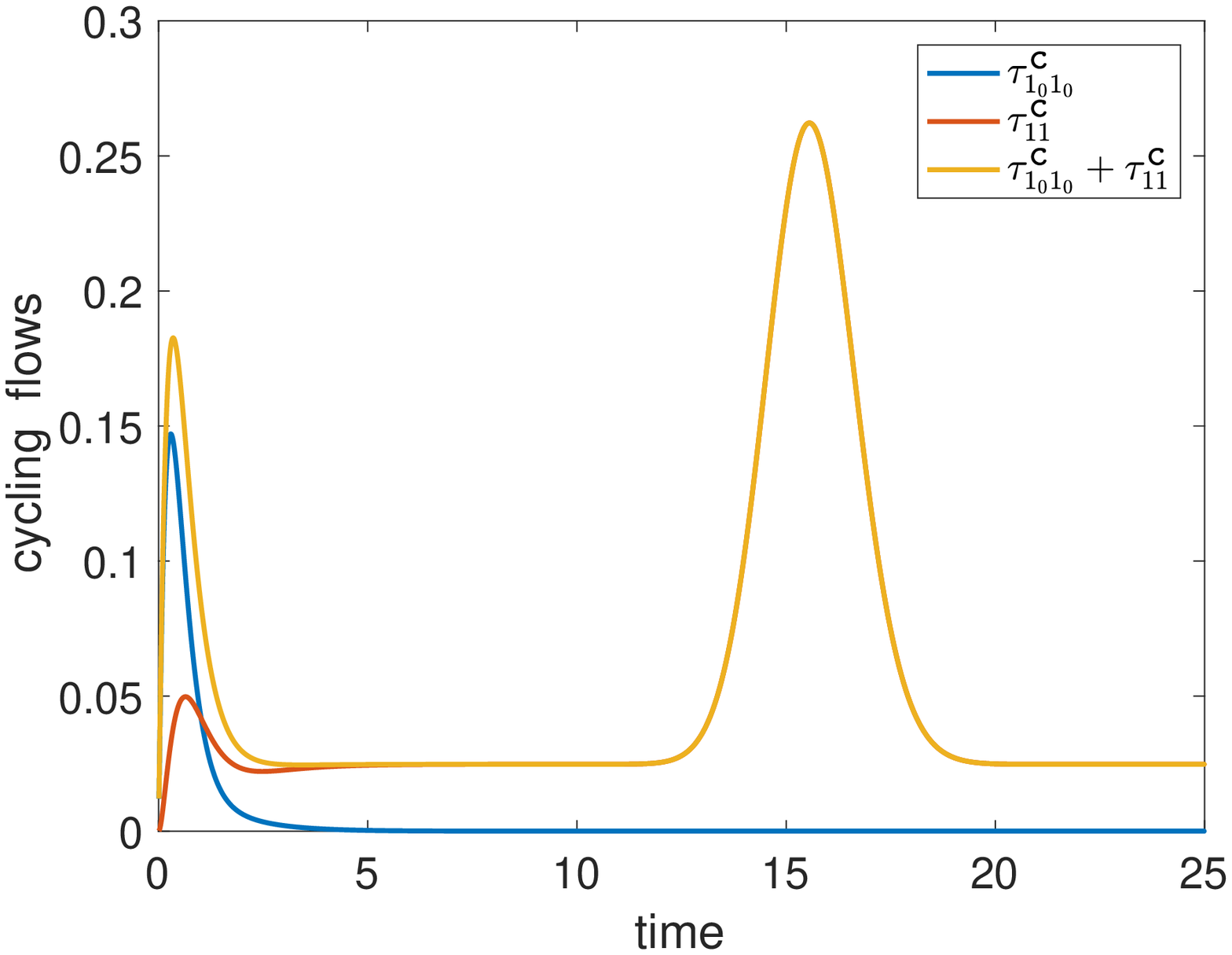}
        \caption{cycling (sub)flows}
        \label{fig:hallam_g_cT_11}
    \end{subfigure} \\
    \begin{subfigure}[b]{0.31\textwidth}
        \includegraphics[width=\textwidth]{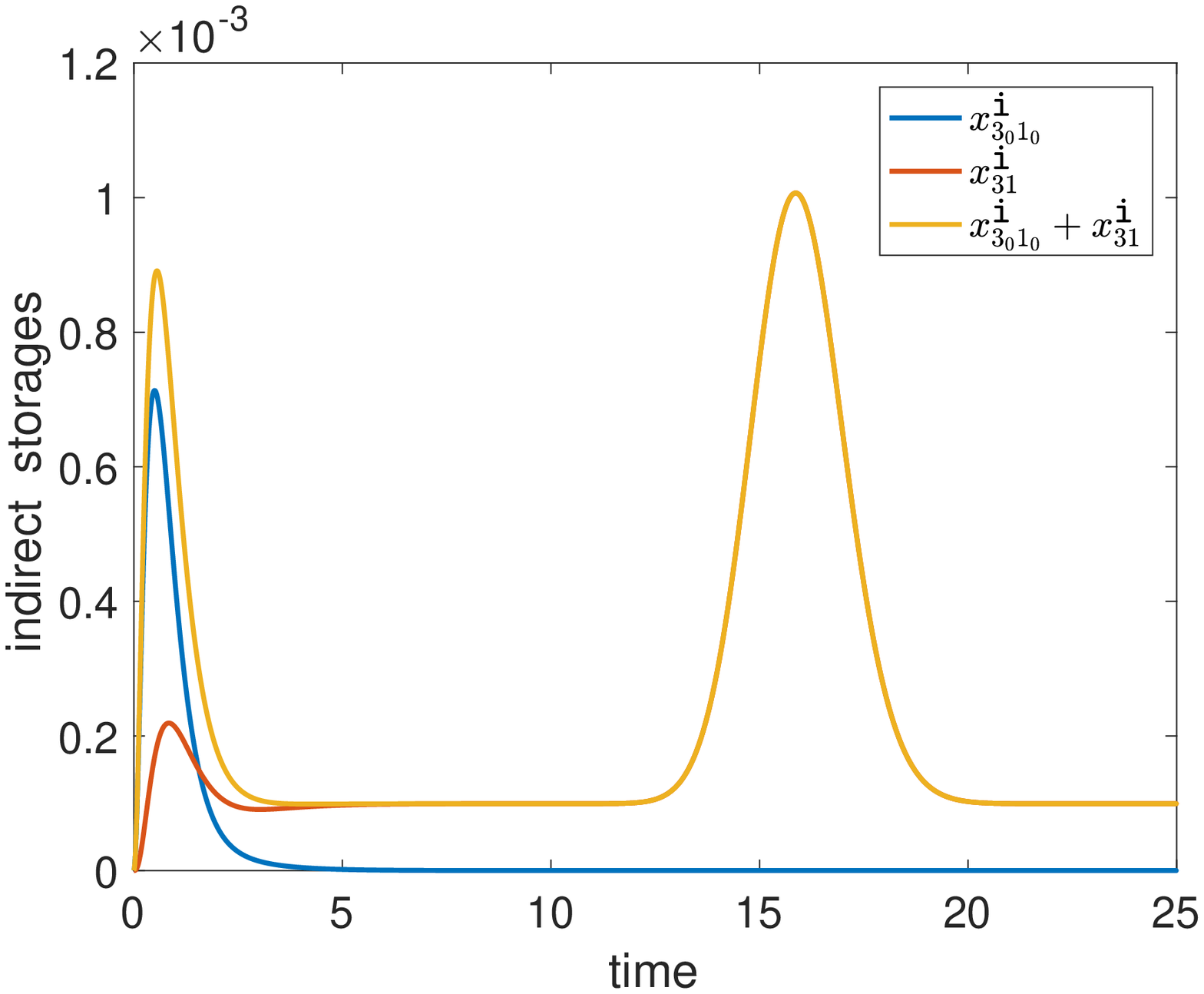}
        \caption{indirect (sub)storages}
        \label{fig:hallam_g_iX_31}
    \end{subfigure} 
    \begin{subfigure}[b]{0.32\textwidth}
        \includegraphics[width=\textwidth]{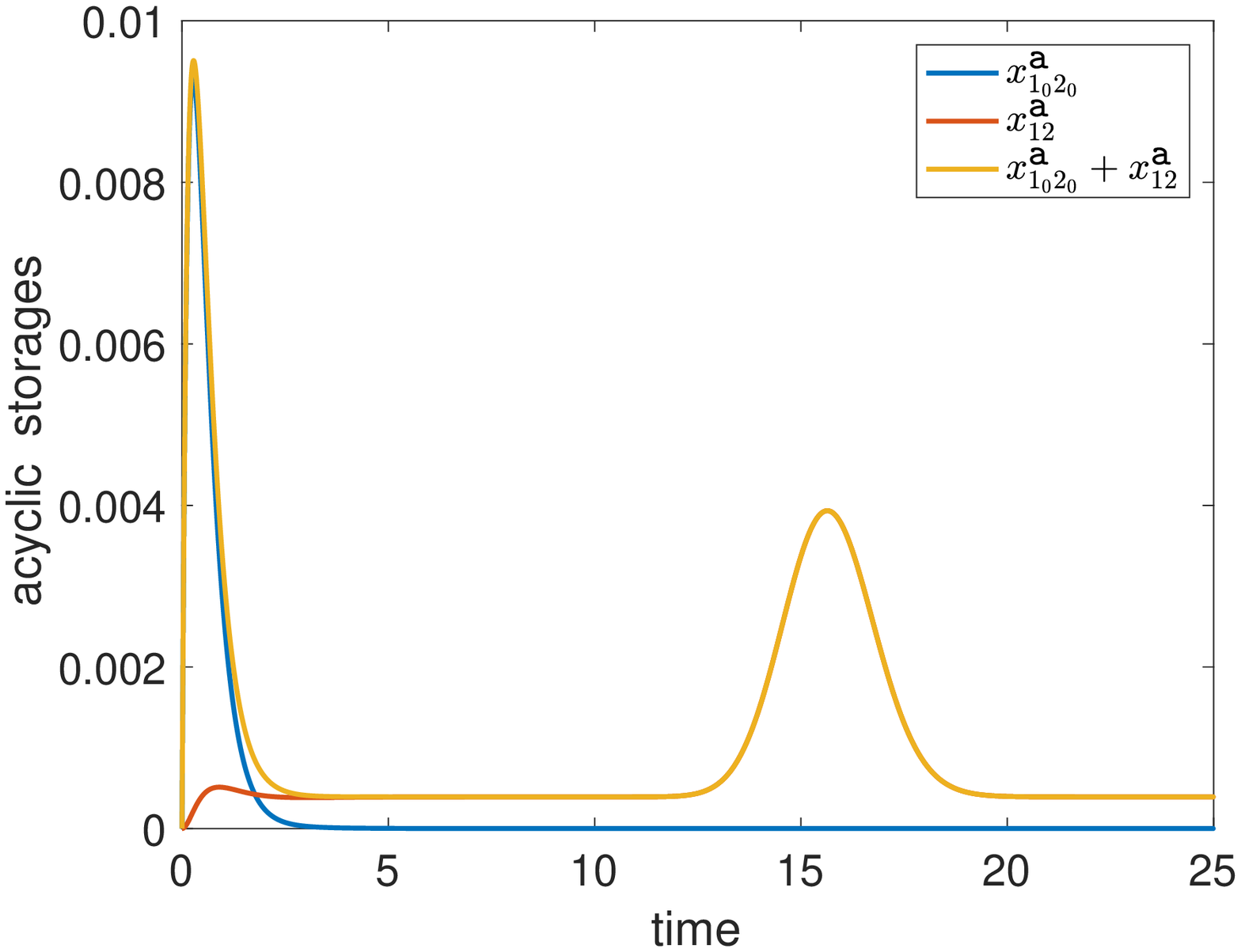}
        \caption{acyclic (sub)storages}
        \label{fig:hallam_g_aX_12}
    \end{subfigure} 
    \begin{subfigure}[b]{0.31\textwidth}
        \includegraphics[width=\textwidth]{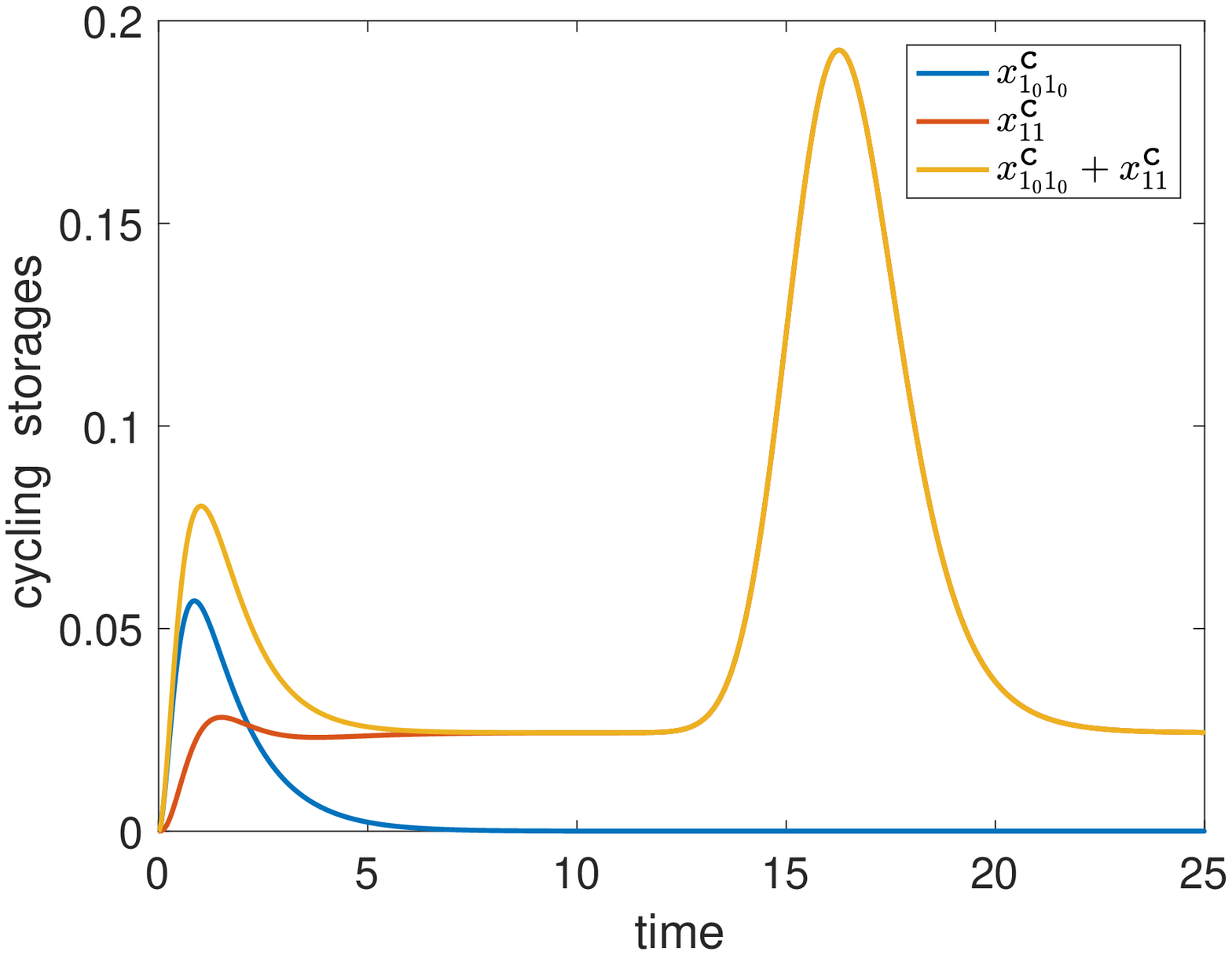}
        \caption{cycling (sub)storages}
        \label{fig:hallam_g_cX_11}
    \end{subfigure} 
    \caption{The graphical representation of some indirect, acyclic, and cycling flows and storages, as well as the corresponding initial subflows and substorages (Case study~\ref{appsec:ex_hallam}).}
    \label{fig:hallam_diact}
\end{figure}

The subflow path $p_{3_1 1_1}^1$ is extended to path 
\begin{equation}
\begin{aligned}
p_{0_1 1_1}^1 = 0_1 \mapsto 1_1 \to 2_1 \to 3_1 \to 1_1 \to 2_1 \to 1_1 \to 0_1 
\end{aligned}
\nonumber
\end{equation}
to compute the local output $f^1_{0_1 1_1 2_1}(t)$ (a segment of environmental output $y_1(t)$) derived from the local (and environmental) input $z_1(t)= 1$ along that particular path (see Fig.~\ref{fig:hallam_diag}). That is, the fate of $z_1(t)$ along path $p_{0_1 1_1}^1$ within the system is determined. The corresponding transient subflow and associated substorage functions at each step (subcompartment) along the path are also presented in Fig.~\ref{fig:tr_f_2ind} and~\ref{fig:tr_x_2ind}. Since $f^1_{0_1 1_1 2_1}(t) \leq 6.28 \times 10^{-5}$, at most about $\% 0.006$ of $z_1(t)$ exits the system through the given subflow path $p^1_{0_1 1_1}$ at any time $t$.

These results indicate that the proposed dynamic subsystem partitioning methodology enables dynamically tracking the fate of an arbitrary amount of nutrient flow and associated nutrient storage along a given flow path. Consequently, the spread of an arbitrary amount of nutrient from one compartment to the entire system can be monitored. Moreover, the effect of one compartment on any other in terms of the nutrient transfer, through not only direct but also indirect interactions, can be determined.

The \texttt{diact} flows and storages are introduced in Section~\ref{apxsec:flows}. The indirect flow and storage from compartment 1 to 3, $\tau^\texttt{i}_{31}(t)$ and $x^\texttt{i}_{31}(t)$, the acyclic flow and storage from compartment 2 to 1, $\tau^\texttt{a}_{12}(t)$ and $x^\texttt{a}_{12}(t)$, and the cycling flow and storage at compartment 1, $\tau^\texttt{c}_{11}(t)$ and $x^\texttt{c}_{11}(t)$, generated by the environmental inputs, as well as the corresponding initial subflows and substorages derived from the initial stocks transmitted in the same directions are depicted in Fig.~\ref{fig:hallam_diact}. As seen from the graphs, all initial \texttt{diact} subflows and substorages vanish as the system converges to a steady-state and, then, the system behavior is eventually dominated by the environmental inputs. Ecologically, the acyclic flow and storage, $\tau^\texttt{a}_{12}(t)$ and $x^\texttt{a}_{12}(t)$, represent the nutrient flow at time $t$ and the associated nutrient storage generated by this flow during $[t_1,t]$ that visit the resource compartment only once\textemdash do not return to this compartment for a second time later\textemdash after being directly or indirectly transmitted from the producer compartment. The initial acyclic subflow and substorage, $\tau^\texttt{a}_{1_02_0}(t)$ and $x^\texttt{a}_{1_02_0}(t)$, represent the same phenomena within the initial subsystem. Similarly, the indirect flow and storage, $\tau^\texttt{i}_{31}(t)$ and $x^\texttt{i}_{31}(t)$, represent the nutrient flow and storage transmitted indirectly from the resource compartment through the producer to the consumer compartment. The cycling flow and storage, $\tau^\texttt{c}_{11}(t)$ and $x^\texttt{c}_{11}(t)$, represent the nutrient flow and storage transmitted indirectly from the resource compartment through other compartments back into itself. The other \texttt{diact} (sub)flows and (sub)storages can be interpreted similarly, for both the subsystems and initial subsystem to analyze the intercompartmental dynamics generated respectively by the environmental inputs and initial stocks, individually and separately.
\begin{figure}[t]
    \centering
    \begin{subfigure}[b]{0.45\textwidth} 
        \includegraphics[width=\textwidth]{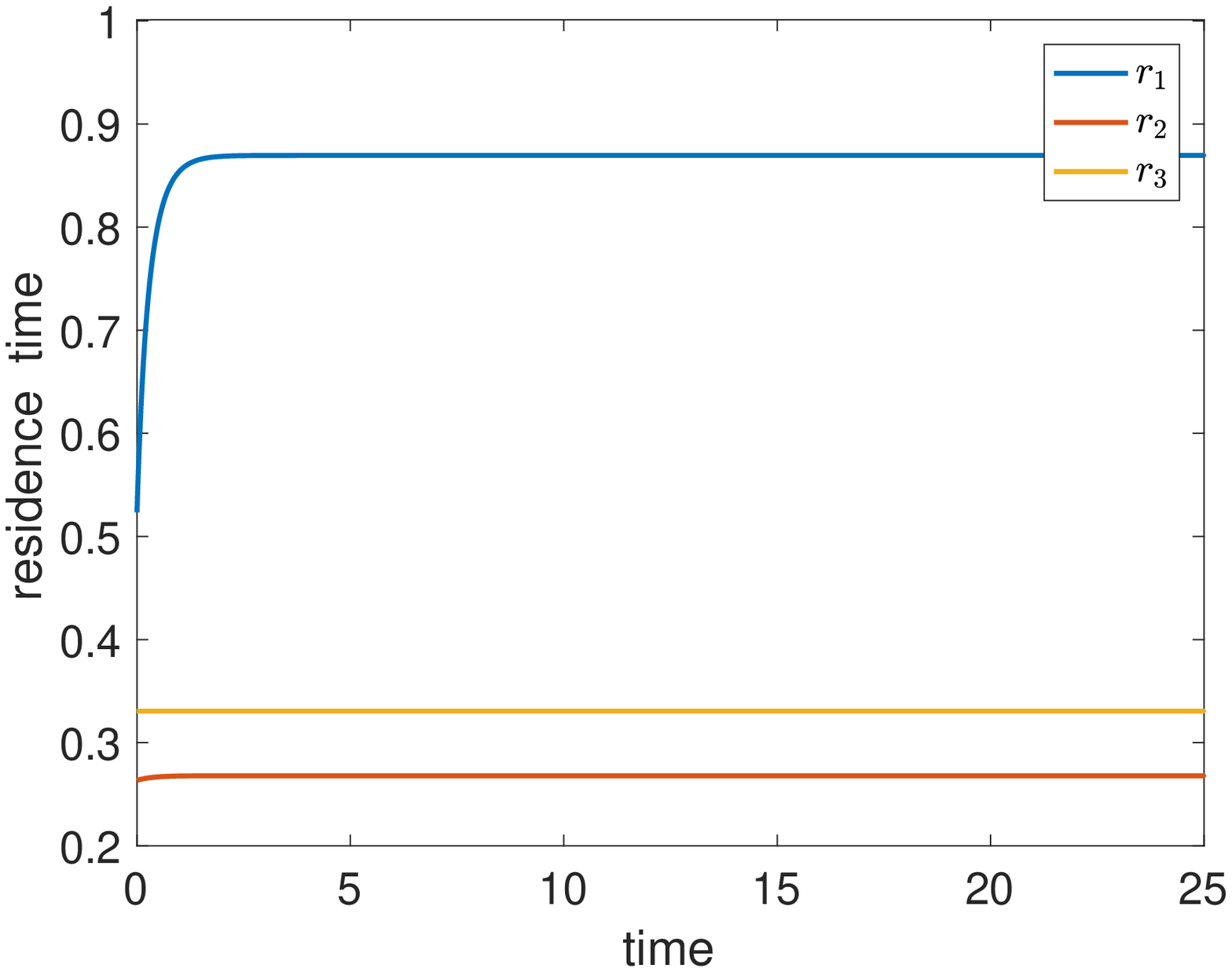}
        \caption{constant input} 
        \label{fig:hallam_RT}
    \end{subfigure}
    \begin{subfigure}[b]{0.45\textwidth}
        \includegraphics[width=\textwidth]{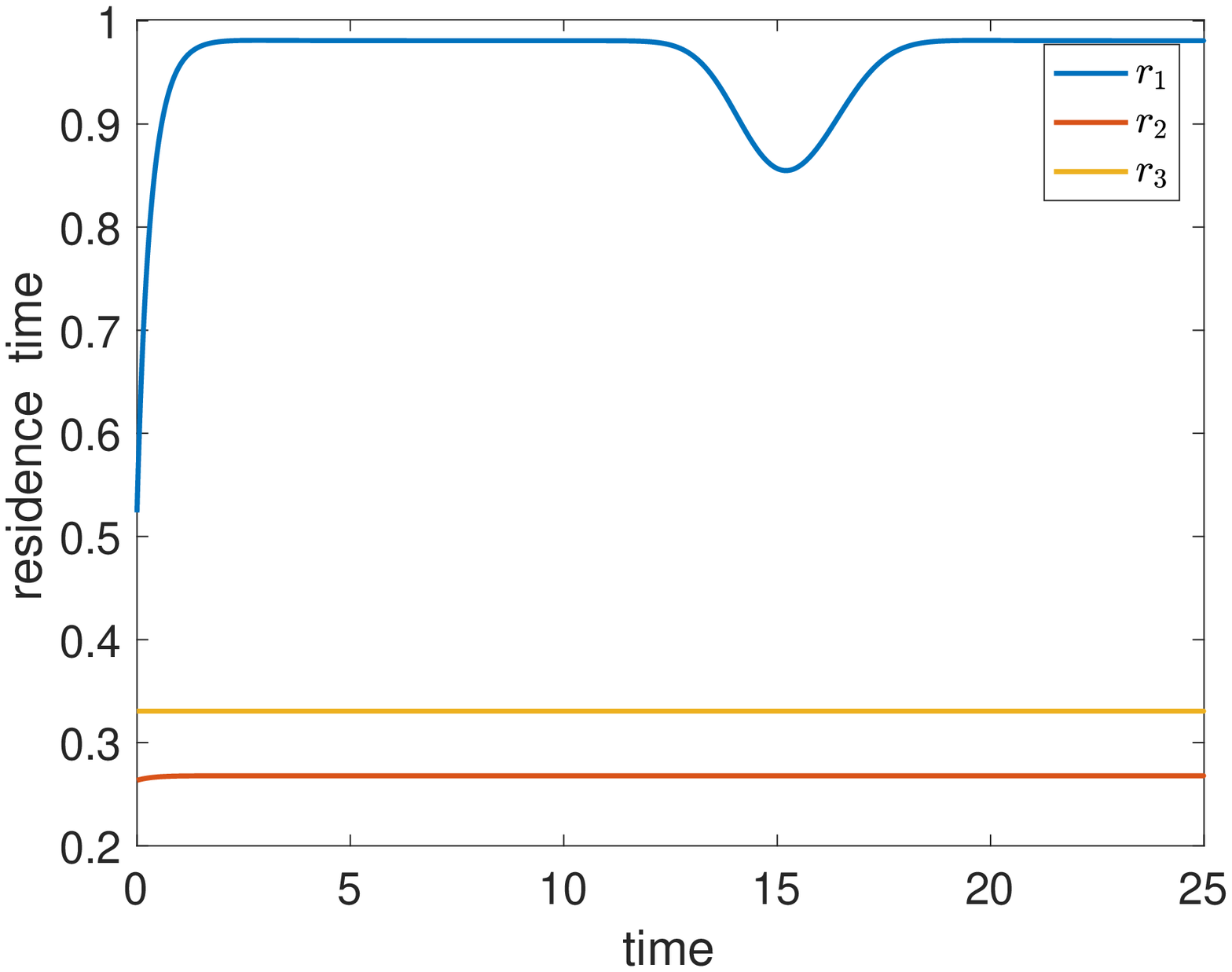}
        \caption{time-dependent input}
        \label{fig:hallam_g_RT}
    \end{subfigure} 
    \caption{The graphical representation for the residence times of the system compartments with both the constant and time-dependent environmental inputs, $z(t) = [1, 1, 1]^T$ and $z(t) = [1, {\mathrm{e}}^{\frac{-(t-15)^2}{2}}+0.1, 1]^T$ (Case study~\ref{appsec:ex_hallam}).}
    \label{fig:hallam_rt}
\end{figure}

The residence time matrix is another novel mathematical measure proposed for quantitative system analysis \cite{Coskun2017NDP,Coskun2017DESM}. The $i^{th}$ diagonal element of $\mathcal{R}(t,{x})$ at time $t_1$, ${r}_{i} (t_1,{x})$, can be interpreted as the time required for the outward throughflow, at the constant rate of $\hat{\tau}_i(t_1,{x})$, to completely empty compartment $i$ with the storage of $x_{i}(t_1)$. The diagonal structure of the residence time matrix indicates that all subcompartments of compartment $i$ vanish simultaneously. The residence times measure compartmental activity levels \cite{Coskun2017SCSA}. The smaller the residence time the more active the corresponding compartment. The derivative of the residence time matrix is called the {\em reverse activity rate matrix} \cite{Coskun2017DESM}. 

The residence time functions for this model with both the constant and time-dependent environmental inputs are depicted in Fig.~\ref{fig:hallam_rt}, for a comparison. The residence times of both the consumer and producer compartments are almost constant and the same in both cases. Interestingly, the Gaussian impulse at the producer compartment, $z_2(t)$, has no significant impact on the activity level of the consumer and even that of the producer compartment itself. However, the decrease in the input into the producer compartment from constant $z_2(t)=1$ to $z_2(t) = {\mathrm{e}}^{\frac{-(t-15)^2}{2}}+0.1$ results in an overall increase in the residence time of the resource compartment (and all of its subcompartments) from the steady state value of ${r}_{1} = 0.87$ days to ${r}_{1} = 0.98$ days. Moreover, the maximum impulse at $t=15$ decreases this residence time, $r_1(t)$, locally in time. That is, 
\[ \mathcal{R}(10) = \mathcal{R}(25) = \diag{([0.98,0.27,0.33])} 
\quad \mbox{but} \quad
\mathcal{R}(15) = \diag{([0.85,0.27,0.33])} . \]
Consequently, the residence time of the resource compartment adversely impacted by the environmental input into the producer compartment.
\begin{figure}[t]
    \centering
    \begin{subfigure}[b]{0.46\textwidth}
        \includegraphics[width=\textwidth]{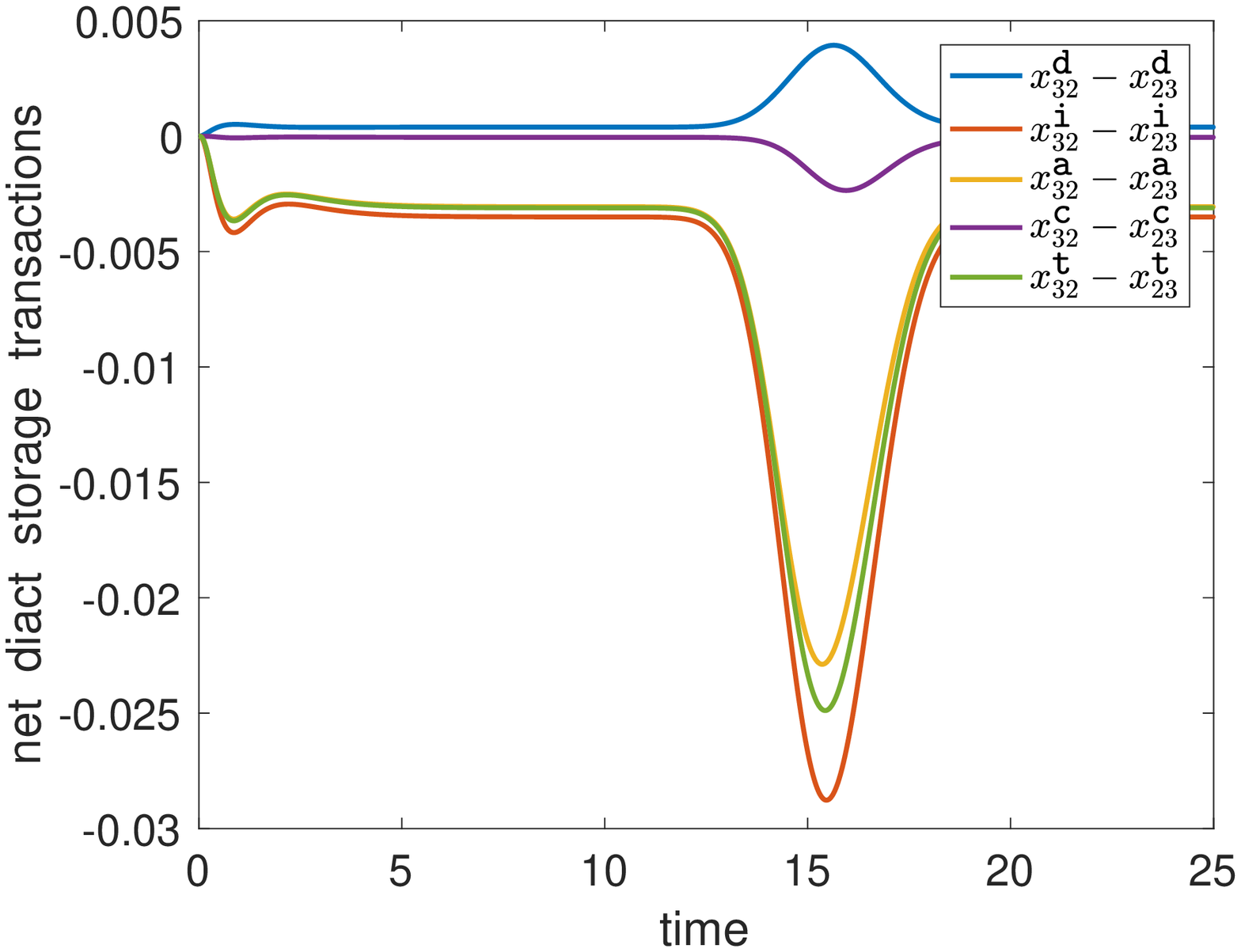}
        \caption{sign analysis}
        \label{fig:iint_delta}
    \end{subfigure} 
    \begin{subfigure}[b]{0.45\textwidth}
        \includegraphics[width=\textwidth]{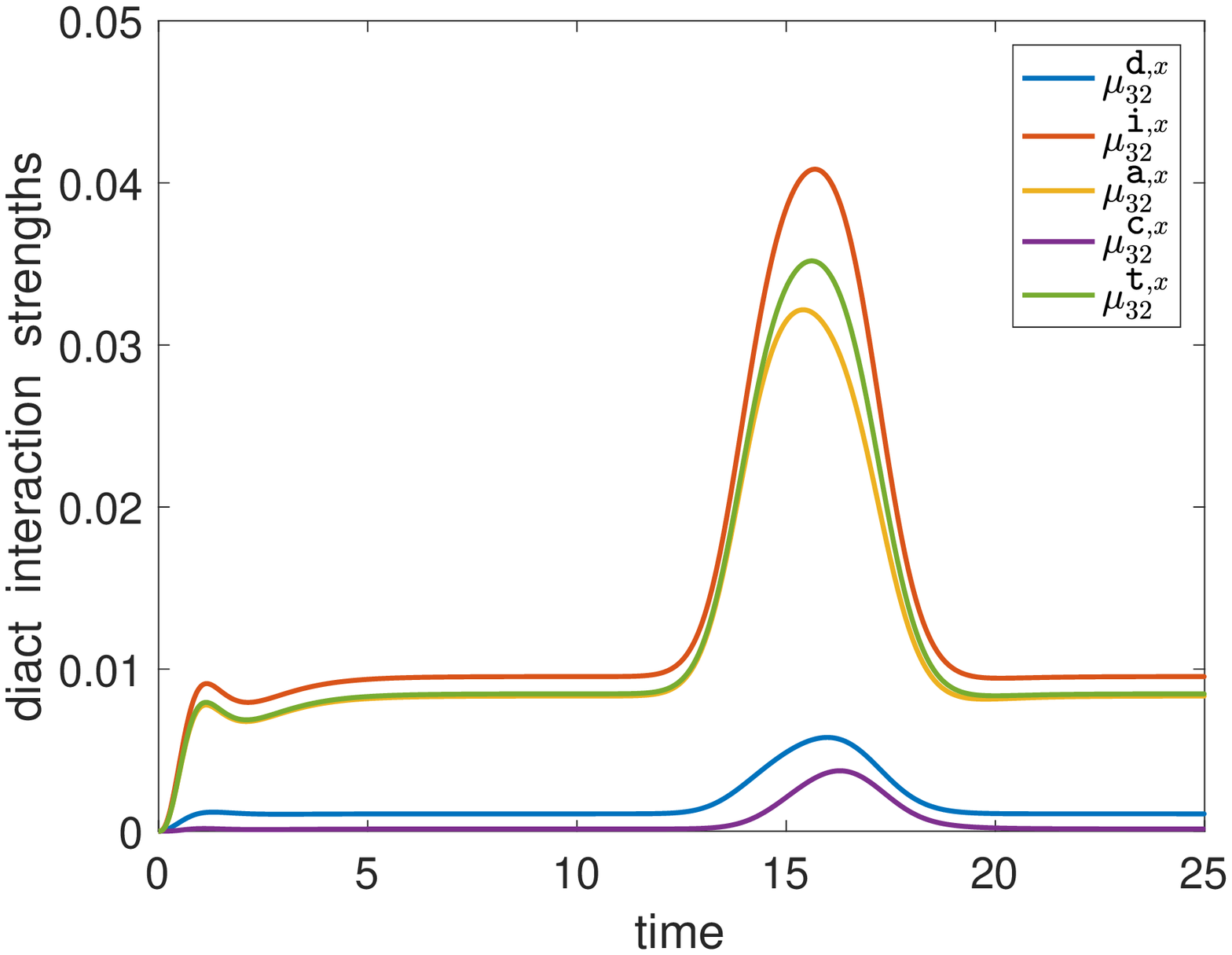}
        \caption{strength analysis}
        \label{fig:iint_mu}
    \end{subfigure} 
    \caption{The graphical representation for the net \texttt{diact} storage transactions and the strengths of the \texttt{diact} interspecific interactions between the producer ($i=2$) and consumer (3) compartments for the time-dependent input case (Case study~\ref{appsec:ex_hallam}).}
    \label{fig:hallam_iint}
\end{figure}

The mathematical classification of the \texttt{diact} interspecific interactions is also introduced in Section~\ref{sec:qdii}. The sign and strength of the \texttt{diact} interactions, induced by environmental inputs, between the producer and consumer compartments become
\begin{equation}
\label{eq:iidm_atwork}
\delta^{\texttt{*},x}_{32}(t) = \sgn{(x^\texttt{*}_{32}(t) - x^\texttt{*}_{23}(t) )} \quad \mbox{and} \quad \mu^{\texttt{*},x}_{32}(t) = 
\frac{| x^\texttt{*}_{32}(t) - x^\texttt{*}_{23}(t) | } { x_{2}(t) + x_{3}(t) } 
\end{equation} 
for the storage-based analysis. The numerical results for the net \texttt{diact} storage transactions and the strengths of the \texttt{diact} interactions are presented in Fig.~\ref{fig:hallam_iint}. 

As seen from the sign analysis in Fig.~\ref{fig:iint_delta},
\begin{equation}
\label{eq:iidm_dmX}
\begin{aligned}
\delta^{\texttt{d},x}_{32}(t) & = (+) , 
\quad
\delta^{\texttt{i},x}_{32}(t) = \delta^{\texttt{a},x}_{32}(t) = \delta^{\texttt{c},x}_{32}(t) = \delta^{\texttt{t},x}_{32}(t) =  (-) .
\end{aligned} 
\end{equation} 
These results indicate that the \texttt{diact} interactions induced by the environmental inputs between the producer and consumer compartments are all antagonistic. Although the consumer compartment directly benefits from the producer compartment as expected, interestingly, their indirect, cycling, acyclic, and total interactions are detrimental to the consumer compartment. The strengths of the \texttt{diact} iterations are ordered as follows:
\begin{equation}
\label{eq:iidm_strX}
\begin{aligned}
\mu^{\texttt{c},x}_{32}(t) < \mu^{\texttt{d},x}_{32}(t) < \mu^{\texttt{a},x}_{32}(t) < \mu^{\texttt{t},x}_{32}(t) < \mu^{\texttt{i},x}_{32}(t) 
\end{aligned} 
\end{equation} 
for $t \geq 0.39$. Therefore, the indirect interaction between the producer and consumer compartments is the strongest of all \texttt{diact} interactions. Since the indirect interaction dominates the direct interaction, their overall interactions is counterintuitively detrimental to the consumer compartment after $t = 0.39$.

The detailed information and inferences enabled by the proposed methodology cannot be obtained through the analysis of the original system by the state-of-the-art techniques, as demonstrated in these case studies.

\section{Discussion}
\label{sec:disc}

Environment is not an easy concept to define in general and, in particular, to analyze mathematically. One reason for this is that nature is always on the move and ecological systems struggle to adapt to constantly changing circumstances. Although sound rationales are offered in the literature for the analysis of natural system dynamics, they are only for special cases, such as linear models and static systems. In recent decades, there has been several attempts to analyze dynamic ecological networks, but each of them bears disadvantages. The need for dynamic and nonlinear methodologies has always existed. This manuscript proposes a novel mathematical methodology for the analysis of nonlinear dynamic compartmental systems to comprehensively address these shortcomings.

Considering a hypothetical ecosystem with several interacting species for which the effect of a specific pollutant needs to be investigated, monitoring the evolution of that pollutant within this food web would be critical to addressing the potential harm. Given the initial pollutant stocks in each of the species, current deterministic mathematical methods can analyze the composite throughflow and storage of the toxin in the species. The evolution of each environmental pollutant input separately within the web, however, cannot be determined through the current methodologies. In the case that multiple species exposed to the same pollutant from the environment, the proposed system partitioning methodology enables dynamically partitioning the composite pollutant flow at and storage in any species into subcompartmental segments based on their constituent sources from the initial pollutant stocks and environmental pollutant inputs. In other words, the system partitioning enables tracking the evolution of the initial pollutant stocks and environmental pollutant inputs in each species, as well as the associated pollutant storages derived from these inputs and stocks individually and separately within the food web. 

The proposed subsystem partitioning methodology can then dynamically track the fate of arbitrary intercompartmental pollutant flows and associated storages in each species along a given food chain in the web as well. Therefore, the spread of an arbitrary amount of toxin from one species to the entire web or along a specific food chain can be monitored. Such information can help, for example, to identify critical pathways in which intervention helps to alleviate the effects of the pollutant. The direct, indirect, acyclic, cycling, and transfer (\texttt{diact}) flows and associated storages of the pollutant from one species directly or indirectly to any other\textemdash including itself\textemdash can also be determined.

More technically, the proposed system partitioning methodology explicitly generates mutually exclusive and exhaustive subsystems. Except the initial subsystem\textemdash which is driven by the initial stocks\textemdash each subsystem is driven by a single environmental input. The subsystems are running within the original system and have the same structure and dynamics as the system itself, except for their initial stocks and environmental inputs. The system partitioning yields the subthroughflows and substorages that represent flows and storages derived from the initial stocks and generated by environmental inputs in each compartment. That is, the composite compartmental storages and throughflows, $x_i(t)$ and $\tau_i(t)$, are dynamically partitioned into the subcompartmental substorage and subthroughflow segments, $x_{i_k}(t)$ and $\tau_{i_k}(t)$, based on their constituent sources from the initial stocks and environmental inputs, $x_{i_0}$ and $z_k(t)$. Equipped with these measures, the system partitioning ascertains the dynamic distribution of the initial stocks and environmental inputs and the organization of the associated storages derived from these stocks and inputs within the system. In other words, the system partitioning enables dynamically tracking the evolution of the initial stocks and environmental inputs individually and separately within the system. The system partitioning methodology, therefore, refines system analysis from the current static, linear, compartmental to the dynamic, nonlinear, subcompartmental level to explore the full complexity of the ecological systems. 

The subsystems are then further decomposed into subflows and substorages along a set of mutually exclusive and exhaustive directed subflow paths. The subsystem partitioning methodology yields the transient subflows and substorages in each subcompartment along a given subflow path within a subsystem, generated by or derived from an arbitrary intercompartmental flow or storage. Therefore, arbitrary composite intercompartmental flows and associated storages can be apportioned dynamically into transient subflow and substorage segments along a given set of subflow paths. That is, the transient subflows and substorages determine the dynamic distribution of arbitrary intercompartmental flows and the organization of the associated storages generated by these flows within the subsystems. As a result, the spread of an arbitrary flow or storage segment from one compartment to the entire system can be monitored. Moreover, an archive of compartments visited by arbitrary system flows and storages can be also compiled. In brief, the subsystem partitioning methodology enables tracking the fate of arbitrary intercompartmental flows and associated storages within the subsystems. 

The proposed mathematical method, as a whole, enables the decomposition to the utmost level, or ``atomization,'' of the system flows and storages. In addition to the subthroughflows, substorages, transient flows and storages, the dynamic direct, indirect, acyclic, cycling, and transfer (\texttt{diact}) flows and storages from one compartment directly or indirectly to any other\textemdash including itself\textemdash are also systematically formulated for the quantification of intercompartmental flow and storage dynamics. The \texttt{diact} flows and storages are derived explicitly through the dynamic and path-based approaches, which are based on the system and subsystem partitioning methodologies, respectively. As an immediate application, a mathematical technique that characterizes and classifies the neutral and antagonistic nature of \texttt{diact} interspecific interactions in food webs and determines the strength of these interactions is also developed based on the \texttt{diact} flows and storages. The illustrative case studies in Section~\ref{sec:results} demonstrate the rigor and efficiency of these mathematical system analysis tools as ecological system indicators.

For a comparison of the proposed methodology with the state-of-the-art techniques, we first note that, at a steady state, the proposed dynamic methodology agrees with the current techniques for static ecological network analysis, as shown by \cite{Coskun2017SCSA}. In recent decades, there have been several attempts to analyze dynamic ecological networks. The first actual dynamic analysis was limited to linear systems with time-dependent input \cite{Hippe1983}. The proposed method is applied to the linear ecosystem model introduced by \cite{Hippe1983} as an illustrative case study in Section~\ref{sec:results}. It is shown that the analytic solutions obtained by the proposed methodology agree with those obtained by Hippe's approach. Further results that are not available through Hippe's approach, such as the \texttt{diact} flows and storages, are also presented for this linear system.

The dynamic approach is extended from linear to nonlinear systems by \cite{Hallam1985}. The authors provided, however, only closed-form, abstract formulations that are difficult to apply to real cases. The proposed methodology is also applied to the nonlinear ecosystem model analyzed by \cite{Hallam1985}, and the results, together with their ecological interpretations, are presented in Section~\ref{sec:results}. A comparison of our results with the ones provided by the authors was not possible because, unlike the comprehensive dynamic analysis enabled by the proposed methodology for nonlinear systems, the authors could only provide asymptotic solutions to the model at steady state through their methodology.

Individual-based algorithms that rely on particle tracking simulations are also proposed for dynamic nonlinear ecological models in the literature. A truncated infinite series formulation, for example, was proposed by \cite{Shevtsov2009}. However, the authors' approach was approximate, computationally resource-intensive, and offered no guarantees of series convergence. While a guarantee of convergence is added by \cite{Kazanci2012}, the computation remained resource-intensive due to the individual-based simulation technique \cite{Kazanci2009, Tollner2009}.

This is the first manuscript in the literature that comprehensively addresses all the previously identified problems and shortcomings. The method's primary limitation is that it is designed for the analysis of conservative models defined in Eq.~\ref{eq:consv}. Since the conservation principles are fundamental laws of nature, a large class of real-world problems are formulated based on conservation principles in many fields. On the other hand, there are still various non-conservative systems that cannot be analyzed by the proposed methodology in its current form.

The proposed methodology can easily be extended for similar analyses to systems of higher order ordinary and partial differential equations whose source terms governing the intercompartmental interactions are in the form of the conservative compartmental systems, as defined in Eq.~\ref{eq:consv}. Such an extension of the system partitioning methodology to partial differential equations enables spatiotemporal analysis of ecological systems.

\section{Conclusions}
\label{sec:conc}

In the present paper, we developed a comprehensive mathematical method for the analysis of nonlinear dynamic compartmental systems through the system decomposition theory. The proposed method is based on the novel analytical and explicit, mutually exclusive and exhaustive system and subsystem partitioning methodologies. While the proposed dynamic {\em system} partitioning provides the subthroughflow and substorage matrices to determine the distribution of the initial stocks and environmental inputs, as well as the organization of associated storages individually and separately within the system, the {\em subsystem} partitioning yields the transient flows and storages to determine the distribution of arbitrary intercompartmental flows and the organization of associated storages within the subsystems. Consequently, the evolution of the initial stocks, environmental inputs, and arbitrary intercompartmental system flows, as well as the associated storages derived from these stocks, inputs, and flows can be tracked individually and separately within the system. Moreover, the transient and the dynamic direct, indirect, acyclic, cycling, and transfer (\texttt{diact}) flows and associated storages transmitted along a given flow path or from one compartment directly or indirectly to any other within the system are systematically formulated to ascertain the intercompartmental dynamics.
  
Traditional ecology is still largely a descriptive empirical science. This narrows the field's scope of applicability and compromises its capacity to deal with complex ecological networks. The proposed dynamic method enhances the strength and extends the applicability of the state-of-the-art techniques and provides significant advancements in theory, methodology, and practicality. It serves as a quantitative platform for testing empirical hypotheses, ecological inferences, and, potentially, theoretical developments. Therefore, this method has the potential to lead the way to a more formalistic ecological science. We consider that the proposed methodology brings a novel complex system theory to the service of urgent and challenging environmental problems of the day. Several case studies from ecosystem ecology are presented to demonstrate the accuracy and efficiency of the method.

The proposed methodology also lays groundwork for the development of new mathematical system analysis tools as quantitative ecological indicators. The time dependent nature of these quantities enables also their time derivatives and integrals to be formulated as novel system measures. Multiple such dynamic \texttt{diact} measures and indices of matrix, vector, and scalar types which may prove useful for environmental assessment and management are systematically introduced in a separate paper by \cite{Coskun2017DESM}.

\section*{Acknowledgments}
The author would like to thank the members of the ecosystem ecology group in the Department of Mathematics at UGA, for useful discussions and valuable feedback on a prior draft of this paper during his visit in 2016. In particular, the author is indebted to Caner Kazanci for introducing the static environ theory to the author, identifying some open problems in the area, and reviewing a prior draft of the paper, and other group members, Bernard C. Patten for his detailed review of a prior draft of the paper and Malcolm Adams for useful discussions and comments. The author also thanks Hasan Coskun, Sangwon Suh, Stuart Borett, and Brian Fath for their review of a prior draft of the paper and helpful comments.

\bibliographystyle{siamplain}
\bibliography{ecology}

\end{document}